\documentclass[11pt,a4paper]{article}
\usepackage{jheppub}

\usepackage{subcaption}

\usepackage{braket}
\usepackage{dsfont}
\usepackage{comment}
\title{Krylov Localization and suppression of complexity}

\author[a]{E. Rabinovici,} 

\author[b]{A. S\'{a}nchez-Garrido,}

\author[a]{R. Shir}

\author[b]{and J. Sonner}

\affiliation[a]{Racah Institute of Physics, The Hebrew University, Jerusalem 9190401, Israel}
\affiliation[b]{Department of Theoretical Physics, University of Geneva, 24 quai Ernest-Ansermet, 1214 Gen\`eve 4, Switzerland}  

\emailAdd{eliezer@mail.huji.ac.il}
\emailAdd{Adrian.SanchezGarrido@unige.ch}
\emailAdd{ruth.shir@mail.huji.ac.il}
\emailAdd{Julian.Sonner@unige.ch}

\abstract{Quantum complexity, suitably defined, has been suggested as an important probe of late-time dynamics of black holes, particularly in the context of AdS/CFT. A notion of quantum complexity can be effectively captured by quantifying the spread of an operator in Krylov space as a consequence of time evolution. Complexity is expected to behave differently in chaotic many-body systems, as compared to integrable ones. In this paper we investigate Krylov complexity for the case of interacting integrable models at finite size and find that complexity saturation is suppressed as compared to chaotic systems. We associate this behavior with a novel localization phenomenon {\it on the Krylov chain} by mapping the theory of complexity growth and spread to an Anderson localization hopping model with off-diagonal disorder, and find that localization is enhanced in the integrable case due to a stronger disorder in the hopping amplitudes, inducing an effective suppression of Krylov complexity. We demonstrate this behavior for an interacting integrable model, the XXZ spin chain, and show that the same behavior results from a phenomenological model that we define: This model captures the essential features of our analysis and is able to reproduce the behaviors we observe for chaotic and integrable systems via an adjustable disorder parameter.}

\begin{document}

\maketitle

\section{Introduction}
The notion of complexity is playing an increasingly important role in our understanding of quantum systems, from the  concrete to the abstract. It is important to quantify the intuitive idea of how {\it hard} it is to simulate a given quantum system, either on a classical computer for practical purposes, or in fact on a quantum computer or a simulator in a putative laboratory setting \cite{nielsen2002quantum}. This set of questions has motivated the introduction of gate complexity \cite{nielsen2002quantum}, aiming to quantify how complicated a quantum circuit is needed to simulate a quantum system of interest. However, gate complexity appears to be hard to uniquely generalize  from discrete settings to continuum ones. This is required for investigating complexity in the context of the states in the continuous Hilbert space of the world of quantum field theory. These are of additional interest if one wishes to import complexity into the domain of gravitational physics\footnote{For results regarding the complexity of simulating holographic systems, see \cite{Garcia-Alvarez:2016wem,Babbush:2018mlj,Hartnoll:2021qyl,Xu:2020shn}.}. In this setting, complexity, appropriately defined, is being entertained as a fundamental quantity for our understanding of quantum gravity - for the most part via the AdS/CFT correspondence. Attempts are to concretely implement this by associating to it geometrical quantities such as the volume of the black-hole interior \cite{Stanford:2014jda,Susskind:2018pmk}, or momentum \cite{Susskind:2018tei,Barbon:2020olv}, or indeed one of a multitude of other options \cite{Belin:2021bga}. A challenge for this set of ideas is that while it is possible to quantitatively  calculate the behavior of the bulk  complexity, once defined,  the mathematical definition of the corresponding field-theory boundary complexity to be equated to these geometrical quantities is not well enough defined. In particular, all the definitions suggested in \cite{nielsen2002quantum, doi:10.1126/science.1121541} for the quantum boundary system require ad hoc add ins, be they arbitrary tolerance parameters for gate complexity or penalty functions for the Nielsen `complexity geometry' definition.  It therefore behooves us to explore the different options available, with particular regard to their possible extension to the continuum field theory setting.

At a very basic level, an appropriate notion of complexity should enable us to think of chaotic strongly correlated many-body systems, or chaotic strongly coupled quantum field theories, as being more complicated or complex than their free, or more generically interacting integrable cousins. In other words we expect chaotic theories to time evolve simple states and operators efficiently into more complex states and/or operators, while on the other hand integrable non-chaotic theories are expected not to increase complexity as fast. In this paper we will concretize this scenario for a particular notion of complexity, namely Krylov complexity, to which we will turn after a few more general remarks.

The notion of complexity has received attention in non-integrable theories \cite{Stanford:2014jda,Parker:2018yvk,Barbon:2019wsy,Rabinovici:2020ryf, Chapman:2021jbh} and much effort has been focused on exploring their chaotic properties which are supposed to parallel those of black holes, which are examples of maximally scrambling systems of finite entropy $S$ in the microcanonical sense. The comparison to black holes leads one to expect an early period of exponential growth, which gives way at times of order $t_s \sim \log S$ to a linear growth phase which in turn lasts until $t\sim \exp S$. At this time complexity reaches a plateau where it assumes values proportional to the size of the Hilbert space -- that is exponential in $S$. At very long time scales of order Poincar\'e time, $e^{e^S}$, we expect complexity to exhibit Poincar\'e recurrences (when applicable). At early times, the exponential rate of complexity growth has been linked to quantum versions of Lyapunov exponents \cite{Maldacena:2015waa, Parker:2018yvk} and their bounds\footnote{However, see also \cite{Dymarsky:2021bjq} for an alternative take on the relation to the Lyapunov bound.}. 

In order to put our arguments on a more concrete footing, in this paper we focus our attention on the notion of Krylov complexity \cite{Parker:2018yvk}, $C_K(t)$, (also referred to as `K-complexity'), which depends both on the system Hamiltonian $H$ and an operator of interest ${\cal O}$. Intuitively $C_K(t)$ measures the spread of the operator ${\cal O}$ in operator space by characterizing its growth under Heisenberg evolution with respect to the Hamiltonian $H$. Specifically it does so by quantifying the average position of ${\cal O}(t)$ in a judiciously chosen basis, obtained by orthonormalizing operators obtained from successive commutations with the Hamiltonian, $\left[H, \ldots , \left[H , {\cal O} \right]\right]$. The output of this orthogonalization procedure is an ordered orthonormal basis and a set of normalization coefficients, the Lanczos coefficients $b_n$. As we shall explain further below, the spread of the operator over this basis can be understood in terms of a quantum-mechanical hopping model of a particle on an auxiliary space, the so-called Krylov chain, with hopping amplitudes determined by $b_n$, and with K-complexity playing the role of the position expectation value on that chain. It should be noted that K-complexity is bounded from above by the Krylov space dimension, which is always smaller than the dimension of the Hilbert space of operators, through its construction. It also does not depend on a tolerance parameter as gate complexity does \cite{nielsen2002quantum}, or on a penalty metric as geometric complexity does \cite{doi:10.1126/science.1121541}.
Other recent work related to the study of K-complexity includes \cite{Jian:2020qpp, Kar:2021nbm, Kim:2021okd, Caputa:2021sib, Dymarsky:2021bjq,Caputa:2021ori,Patramanis:2021lkx}. In particular, works like \cite{Caputa:2021ori,Dymarsky:2021bjq,Kar:2021nbm,Dymarsky:2019elm} have applied it to the realm of holographic conformal field theory.

In a previous paper we developed highly accurate and stable numerical methods, which in conjunction with analytical bounds, allowed us to determine the complete $b-$sequences and resulting Krylov dynamics up to late time for a class of strongly interacting chaotic many-body systems \cite{Rabinovici:2020ryf}. We found that the Krylov space in SYK$_4$ \cite{Kitaev_SYK,Sachdev_SYK,MaldacenaSYK} for ETH operators \cite{Peres_1984, Deutsch_1991, Srednicki_ETH} is maximal with a dimension that scales exponentially with system size, and that the associated Lanczos coefficients $b_n$ display an initial quasi-linear growth up to $n\sim S$ followed by a very slow nearly linear non-perturbative descent to zero at $n\sim e^S$. Note that SYK$_4$ is rare example of a strongly interacting system, in which ETH can be shown to apply both numerically and analytically, \cite{ Sonner:2017hxc, Nayak:2019khe, Nayak:2019evx}. The aforementioned behavior of the Lanczos sequence for ETH operators then induces a K-complexity profile that starts growing exponentially up to the scrambling time $t_s\sim \log S$ and then transitions to a long period of linear growth up to eventual saturation at a value exponentially big in system size at times of the order of the Heisenberg time $t_H\sim e^S$. The qualitative shape of the profile obtained was in agreement with that of the time dependent profile expected from bulk considerations \cite{Susskind:2014rva,Brown:2015bva,Brown:2015lvg,Susskind:2018pmk}. While the emphasis of that work was put on the behavior of chaotic models, it is important to contrast their behavior with that of integrable systems. This line of inquiry was initiated in \cite{Rabinovici:2020ryf} by studying the SYK$_2$ model, and we found that its $b-$sequence terminates far below the bound $D^2  - D +1$ (here $D$ is the dimension of Hilbert space) which in turn leads to K-complexity saturating far below the value of the chaotic SYK$_4$ model. In this paper we investigate whether under-saturation of K-complexity is a ubiquitous phenomenon in more generic integrable models, allowing thus to distinguish their complexity signature from chaotic ones. The main thrust of the argument follows from studying the $b-$sequence and its associated one-particle hopping problem in integrable theories, where we argue that the presence of near-degeneracies resulting from the Poissonian nature of the spectrum of these systems implies an erratic structure akin to that encountered in random hopping models that show wave function localization. More precisely, we will map the Krylov chain dynamics to an Anderson random hopping model with off-diagonal disorder, rather than the diagonal disorder of the original Anderson problem  \cite{Anderson_AbsDiff}. Our main finding is the observation of a disordered structure in the Lanczos sequence of a local operator in the XXZ spin chain, a canonical example of strongly-interacting integrable system solvable via the Bethe Ansatz, which induces some degree of localization that prevents maximally efficient propagation of the particle in Krylov space, resulting in a K-complexity saturation value at late times that is lower than that obtained in previous studies for SYK$_4$ \cite{Rabinovici:2020ryf}. We confirmed that this effect was due to disorder by studying a phenomenologically-inspired model with a disordered Lanczos sequence with adjustable fluctuation strength, yielding results that reproduce closely those obtained from XXZ.

%... \js{fill in what we think is most pertinent. I suggest: a) idea of `weak localization' on Krylov chain. b) Localization of center of band, c) quantitative agreement of XXZ as example of nontrivial interacting theory with our disorered toy model. }

This paper is organised as follows: Section \ref{Sect_Dynamics_Krylov} summarizes the notions of Krylov space and K-complexity, as well as the mapping between Krylov chain dynamics and a one-dimensional hopping problem, provides a heuristic argument for the expected disorder in the Lanczos coefficients of integrable systems, and presents the framework of the Anderson problem with off-diagonal disorder, introducing some tools that we will use along the way. Section \ref{Section_XXZ} presents and discusses the XXZ spin chain, which will be used for the main computations in this paper. Section \ref{Sect_Numerical_Results} presents the numerical results obtained from XXZ and a comparison with SYK$_4$. Section \ref{Sect_Toys} shows results obtained from phenomenological models of disordered Lanczos sequences that display similar results to XXZ. Finally, our results raise some open questions in the realm of quantum many body physics, as well as some interesting puzzles within the framework of the AdS/CFT correspondence. These are discussed in Section \ref{Sect_conclusions}. 

\section{Dynamics in the Krylov chain}\label{Sect_Dynamics_Krylov} 

\subsection{Krylov space and the Lanczos algorithm}

During its time evolution, an operator $\mathcal{O}$ gradually explores a subspace of the space of operators of a given system. This subspace is referred to as the Krylov space, and its structure as well as the rate at which typical operators explore it can be elucidated by constructing a suitable orthonormal basis, the Krylov basis, according to the Lanczos algorithm \cite{Parker:2018yvk, Recursion_Method, Lanczos1950AnIM}.
For convenience, we shall use the \textit{smooth-ket} notation $|\mathcal{O})$ to denote the element of operator space corresponding to an operator $\mathcal{O}$. In the Heisenberg picture, time evolution is generated by the Liouvillian \textit{operator}\footnote{Operators acting on operator space are sometimes referred to as \textit{superoperators} (see e.g. \cite{Parker:2018yvk}), a terminology that has no relation with supersymmetry and which we shall not adopt in this work.} $\mathcal{L}$, which satisfies the relation $\mathcal{L}|\mathcal{O})=\Big{|} \left[H,\mathcal{O}\right]\Big{)}$, where $H$ is the system Hamiltonian, so that:
\begin{equation}
    \centering
    \label{Heisenberg-evol}
    \Big|\mathcal{O}(t)\Big)=e^{it\mathcal{L}}|\mathcal{O})=\sum_{n=0}^{+\infty} \frac{(it)^n}{n!}\mathcal{L}^n |\mathcal{O})\,.
\end{equation}
This motivates the definition of the Krylov space associated to the operator $\mathcal{O}$, $\mathcal{H}_{\mathcal{O}}$, as the linear span of all powers of the Liouvillian acting on it or, in other words, the linear span of all nested commutators of $H$ with $\mathcal{O}$:

\begin{equation}
    \centering
    \label{Krylov_space}
    \mathcal{H}_{\mathcal{O}}=\text{span}\left\{\mathcal{L}^n|\mathcal{O})\right\}_{n=0}^{+\infty}~.
\end{equation}
For a physical system with a Hilbert space of finite dimension $D$ the operator space has dimension $D^2$. Therefore, even if the set of all powers of the Liouvillian acting on $\mathcal{O}$ contains infinitely many elements, only a finite number of them can be linearly independent, yielding a finite dimension for the Krylov space defined in (\ref{Krylov_space}), which we denote $K$. This can be shown explicitly using a spectral decomposition of $\mathcal{O}$ in terms of the energy basis:

\begin{equation}
    \centering
    \label{O-spectral-decomp}
    |\mathcal{O}) \equiv \sum_{a,b=1}^D O_{ab} \ket{E_a}\bra{E_b}\,.
\end{equation}
where $\left\{E_a,\;\ket{E_a}\right\}_{a=1}^D$ are the eigenvalues and eigenstates (respectively) of $H$. In fact, one can show that the elements appearing inside of the sum in (\ref{O-spectral-decomp}) are eigenstates of the Liouvillian: Denoting $\omega_{ab}\equiv E_a-E_b$ and $|\omega_{ab})\equiv\ket{E_a}\bra{E_b}$, we see that

\begin{equation}
    \centering
    \label{Liouvillian-eigenstates}
    \mathcal{L}|\omega_{ab}) = \omega_{ab}|\omega_{ab})\,.
\end{equation}
Therefore, as discussed in \cite{Rabinovici:2020ryf}, the Krylov dimension $K$ is equal to the number of non-vanishing projections of $\mathcal{O}$ over eigenspaces of the Liouvillian, since each application of $\mathcal{L}$ does not mix the projections of $\mathcal{O}$ over each eigenspace. This implies the upper bound
\begin{equation}
    \centering
    \label{K_bound}
    K\leq D^2-D+1\,,
\end{equation}
 which arises from the fact that the zero eigenvalue $\omega_{aa}=0$ (or zero \textit{phase}, as the eigenvalues of $\mathcal{L}$ appear as phases in the spectral decomposition of the two-point function of $\mathcal{O}$) is always at least $D$ times degenerate.
 
 Given an inner product in operator space, we can build a basis for Krylov space fulfilling simultaneously the requirements of orthonormality and of retaining some notion of ordering suitable for time evolution. In this work we shall make use of the infinite-temperature inner product, or Frobenius inner product:
 \begin{equation}
     \centering
     \label{Inner-product}
     \left(\mathcal{A}|\mathcal{B}\right)=\frac{1}{D}\textrm{Tr}\left[\mathcal{A}^\dagger \mathcal{B}\right], \quad \|\mathcal{A}\|^2 = (\mathcal{A}|\mathcal{A}) = \frac{1}{D}\textrm{Tr}\left[\mathcal{A}^\dagger \mathcal{A}\right],
 \end{equation}
 where $\|\cdot \|$ is the induced norm. An orthonormal basis for the space (\ref{Krylov_space}) is constructed by performing the \textit{Lanczos algorithm}, for an operator $\mathcal{O}$ time-evolving under a Hamiltonian $H$, given the inner product (\ref{Inner-product}). This basis, in which the Liouvillian is tridiagonal, is known as the \textit{Krylov basis}.  The general structure of the Lanczos algorithm is as follows:

 \begin{enumerate}
     \item $|\mathcal{O}_0)=\frac{1}{\|\mathcal{O}\|}|\mathcal{O})$.
     \item For $n=1$:
     \begin{itemize}
        \item[2.1.]  $|\mathcal{A}_1) = \mathcal{L}|\mathcal{O}_0)$.
        \item[2.2.] $b_1 = \|\mathcal{A}_1\|$.
     \item[2.3.] \textbf{If} $b_1\neq 0$: $|\mathcal{O}_1)=\frac{1}{b_1}|\mathcal{A}_1) $.
     \textbf{Otherwise} stop.
     \end{itemize}
     \item For $n>1$:
     
     \begin{itemize}
         \item[3.1.] $|\mathcal{A}_n) =\mathcal{L}|\mathcal{O}_{n-1}) - b_{n-1}|\mathcal{O}_{n-2})$.
         \item[3.2.] $b_n = \|\mathcal{A}_n\|$.
         \item[3.3.] \textbf{If} $b_n \neq 0$: 
    $|\mathcal{O}_n)=\frac{1}{b_n}|\mathcal{A}_n) $.
    \textbf{Otherwise}: stop.
     \end{itemize}
    
 \end{enumerate}
 The algorithm terminates once all independent directions  in $\mathcal{H}_\mathcal{O}$ are exhausted, and its output consists of an orthonormal basis for $\mathcal{H}_\mathcal{O}$, $\{|\mathcal{O}_n)\}_{n=0}^{K-1}$ known as the Krylov basis, and a set of coefficients, $\{b_n\}_{n=1}^{K-1}$, known as Lanczos coefficients. As announced, the Krylov elements are orthonormal and retain a notion of ordering suitable for time evolution, since $|\mathcal{O}_n)$ is a linear combination of up to $n$ nested commutators of $H$ with $\mathcal{O}$, and therefore they are effectively probed gradually one after the other thanks to the structure of (\ref{Heisenberg-evol}). Note that this basis can have an infinite number of elements in a Hilbert space with an unbounded spectrum, given an operator with projections over an infinite number of energy differences.  However, we will be interested in long-time behaviour of operator time evolution, and consider Hilbert spaces of finite dimension $D\sim e^S$, where $S$ denotes generically system size or entropy.
 
 In the Krylov basis, the restriction of the Liouvillian over Krylov space admits a matrix representation $L_{mn}\equiv (\mathcal{O}_m| \mathcal{L}|\mathcal{O}_n)$ which takes a tridiagonal form:

\begin{comment} 
\begin{equation}
    \centering
    \label{L-tridiagonal}
    \big( L_{mn} \big) = \begin{pmatrix} 0 & b_1 & 0 & 0&\dots &0&0&0&0 \\ b_1 & 0 & b_2 & 0 & \dots & 0&0 &0&0\\ 0 & b_2 & 0 & b_3 & \dots & 0&0&0&0\\0&0& b_3 &0&\dots& 0& 0& 0& 0 \\ \vdots &\vdots & \vdots & \vdots & \ddots & \vdots&\vdots&\vdots&\vdots \\ 0&0&0&0&\dots & 0&b_{K-3} &0&0\\0&0&0&0&\dots&b_{K-3}&0 &b_{K-2}&0\\0 & 0 & 0 & 0 & \dots &0& b_{K-2}&0&b_{K-1} \\0 & 0 & 0 & 0& \dots&0& 0&b_{K-1}&0\end{pmatrix}
\end{equation}

\end{comment}

\begin{equation}
    \centering
    \label{L-tridiagonal}
    \big( L_{mn} \big) = \begin{pmatrix} 0 & b_1 & 0 &\dots &0&0&0 \\ b_1 & 0 & b_2 & \dots & 0&0 &0\\ 0 & b_2 & 0 & \dots & 0&0&0 \\ \vdots &\vdots & \vdots & \ddots & \vdots&\vdots&\vdots \\0&0&0&\dots&0 &b_{K-2}&0\\0 & 0 & 0 & \dots & b_{K-2}&0&b_{K-1} \\ 0 & 0 & 0& \dots& 0&b_{K-1}&0\end{pmatrix}\,.
\end{equation}
This feature can be exploited to connect the dynamics over Krylov space to a one-dimensional hopping problem, as we shall discuss in Section \ref{Subsection_Krylov_hopping}.

The tridiagonal Liouvillian matrix (\ref{L-tridiagonal}) can be diagonalized with eigenvectors $|\omega_i)$ and associated eigenfrequencies $\omega_i$,
\begin{equation}\label{Liouvillian_eigenvectors}
    \mathcal{L}|\omega_i) = \omega_i |\omega_i), \quad i = 0,\dots, K-1 ~.
\end{equation}
Note that these eigenvectors and eigenfrequencies are not all the eigenvectors and eigenfrequencies of the full Liouvillian introduced in (\ref{Liouvillian-eigenstates}), but the particular direction of each eigenspace (i.e. the space spanned by eigenvectors with the same eigenvalue) selected by the operator by a non-zero projection. In particular there are only $K$ of the $|\omega_i)$ instead of $D^2$ which is the number of $|\omega_{ab})$'s. For a precise description of the relationship between $|\omega_i) $ and $|\omega_{ab})$ see \cite{Rabinovici:2020ryf}.
 
\subsection{Krylov space dynamics as a hopping problem} \label{Subsection_Krylov_hopping}

By construction, Krylov space is the minimal subspace of operator space that contains the time-evolving operator $\mathcal{O}(t)$ at all times. Thus, in order to keep track of this time evolution,  we can decompose $\Big|\mathcal{O}(t)\Big)$ in terms of the Krylov elements:

\begin{equation} \label{phi_def}
    |\mathcal{O}(t)) = e^{it\mathcal{L}} |\mathcal{O}) = \sum_{n=0}^{K-1} \phi_n(t)|\mathcal{O}_n)\,,
\end{equation}
where $\phi_n(t)$ can be viewed as wave functions over the Krylov basis or, as we shall refer to it, the Krylov \textit{chain}. This terminology is justified from the fact that, as stated in (\ref{L-tridiagonal}), $\mathcal{L}$ has a tridiagonal form in the Krylov basis, which one can write more suggestively as:

\begin{equation}
    \centering
    \label{L-tight-binding}
    \mathcal{L}=\sum_{n=0}^{K-2}b_{n+1}\Bigg( \big|\mathcal{O}_{n}\big)\big(\mathcal{O}_{n+1}\big| + \big|\mathcal{O}_{n+1}\big)\big(\mathcal{O}_{n}\big|\Bigg)~.
\end{equation}
We note that (\ref{L-tight-binding}) resembles the Hamiltonian of a one-dimensional tight-binding model on a finite chain with $K$ sites, with zero potential energies and hopping amplitudes given by the Lanczos coefficients. The statement is that the time evolution of the operator can be mapped to such a one-dimensional quantum-mechanical problem just by constructing its Krylov basis and studying its dynamics with respect to it \cite{Parker:2018yvk}, by solving the Schrödinger-like equation for the wave functions:

\begin{equation}
    -i \dot{\phi}_n(t) = \sum_{m=0}^{K-1} L_{nm}\phi_m(t)=b_{n+1}\phi_{n+1}(t)+b_n\phi_{n-1}(t)\,,
\end{equation}
with the initial condition $\phi_n(0)=\delta_{n0}$ (we assume that the operator is normalized, for simplicity) and the boundary conditions $b_0=b_K=0$ which ensure finiteness of the chain. 

We can thus understand the label $n$ as a position on the Krylov chain, and the Krylov elements $|\mathcal{O}_n)$ as localized states on that chain. In general, we always begin with a localized state $|\mathcal{O}_0)$ which over time can become delocalized depending on the dynamics of the Liouvillian dictated by the hopping amplitudes given by the Lanczos coefficients $b_n$.  In \cite{Dymarsky:2019elm} this perspective was used to show a relation between non-integrability and de-localization of the operator on the Krylov chain.

Two enlightening probes of the dynamics on the Krylov chain are K-complexity, $C_K(t)$ \cite{Parker:2018yvk} and K-entropy, $S_K(t)$ \cite{Barbon:2019wsy}. The former gives the average position of the propagating packet $\phi_n(t)$ over the Krylov chain,

\begin{equation}
    \centering
    \label{KC-definition}
    C_K(t)=\sum_{n=0}^{K-1}n|\phi_n(t)|^2,
\end{equation}
while the latter is nothing but another name for the Shannon entropy of the wave packet, and measures how spread or localized it is:

\begin{equation}
    \centering
    \label{KS-definition}
    S_K(t)=-\sum_{n=0}^{K-1}|\phi_n(t)|^2 \log |\phi_n(t)|^2.
\end{equation}

In \cite{Parker:2018yvk} it was proposed, exploiting the relation between the asymptotics of the Lanczos coefficients and other quantities such as the two-point function, that maximally chaotic systems should display, in the thermodynamic limit, an asymptotically linear profile for the Lanczos coefficients, $b_n\sim \alpha n$, which in turn would imply an exponential growth for K-complexity $C_K(t)\sim e^{2\alpha t}$. Subsequent work \cite{Barbon:2019wsy} proposed that, for operators in finite systems satisfying the eigenstate thermalization hypothesis, ETH \cite{Srednicki_ETH,Srednicki_1999}, the linear growth of the Lanczos sequence should transition to a constant plateau phase around $n\sim S$, where $S$ denotes entropy or system size; this implies a transition from exponential to linear growth of K-complexity at times of order of the scrambling time $t_{s}\sim \log S$, and this linear growth should persist up to times of the order of the Heisenberg time $t_H\sim e^S$, when all the directions of the Krylov space have been explored and K-complexity would therefore saturate. 

Performing numerical simulations, in \cite{Rabinovici:2020ryf} we were able to verify these facts for the Sachdev-Ye-Kitaev model \cite{Kitaev_SYK,Sachdev_SYK,MaldacenaSYK} with $4$-site interactions, encountering the additional feature of a non-perturbative descent that corrects the plateau of the Lanczos sequence making it terminate at $b_K=0$, as it should. This work also observed that typical operators\footnote{For chaotic systems, the notion of typicality amounts to satisfying the eigenstate thermalization hypothesis.} in maximally chaotic systems should saturate the Krylov dimension upper bound (\ref{K_bound}) due to the absence of degeneracies in the spectrum of the Hamiltonian, whose level-spacing statistics satisfy distributions of the Wigner-Dyson type and therefore feature level repulsion \cite{BGS_conjecture,Mehta}, and to the fact that these operators are dense (in the sense of not having any zero elements) in the energy basis. In these systems, the wave packet $\phi_n(t)$ spreads in time efficiently over the Krylov chain and at times of the order of $t_H$ it tends to a uniform distribution $|\phi_n(t>t_H)|^2\sim\frac{1}{K}$, implying that K-complexity saturates at $C_K(t>t_H)\sim\frac{K}{2}$, a value wich is exponentially big in system size because $K$ saturates the bound (\ref{K_bound}), and thus $K\sim D^2\sim e^{2S}$. Conversely, free or quadratic systems present in general a large number of degeneracies in the spectrum of the Hamiltonian, as well as potentially rational relations, implying degeneracies in the spectrum of the Liouvillian; this, together with abundant selection rules due to symmetries imposing the vanishing of operator matrix elements in the energy basis, reduces the Krylov dimension and therefore the upper bound for K-complexity. This was exemplified in \cite{Rabinovici:2020ryf} taking the case of SYK with 2-site interactions, where the Krylov space of a single Majorana scales only linearly with system size, and hence so does the K-complexity upper bound.

This paper intends to explore the remaining middle ground: there exists a large class of strongly-interacting integrable systems which feature no exact degeneracies but still exhibit Poissonian level-spacing statistics \cite{Berry_Tabor} and are usually solvable by the means of the Bethe Ansatz \cite{1931ZPhy...71..205B, Samaj_bajnok_2013}.
In these systems, typical operators\footnote{In an integrable system, the notion of \textit{typicality} might be more ambiguous. In this work, we reserve this designation for local operators for the case of local systems, or non-extensive operators for the case of non-local systems with $q$-body interactions.} need not be sparse in the energy basis \cite{Rigol_XXZ}, thus admitting a Krylov space whose dimension can potentially be still exponential in system size. The difference between these systems and maximally chaotic ones should therefore appear at the dynamical level and, in this spirit, this work intends to explore whether the statistical properties of the Lanczos coefficients of the former systems may induce some form of localization preventing the effective diffusion of the wave packet through the Krylov chain, thus lowering the long-time value of the expectation value of its position, which is nothing but K-complexity.
 
\begin{comment}

In the Krylov basis, the Liouvillian has a tridiagonal form,
\begin{equation} \label{Off-diagonal_Liouvillian}
    \mathcal{L}|\mathcal{O}_n) = b_{n+1} |\mathcal{O}_{n+1}) + b_{n-1} |\mathcal{O}_{n-2}) ~.
\end{equation}

\end{comment}

\begin{comment}

Note that in a general Hamiltonian problem, the equivalent of (\ref{Off-diagonal_Liouvillian}) would include potential energies associated with each site via a diagonal term of the form $V_n |\mathcal{O}_n)$.

The dynamics of the time-evolving operator $|\mathcal{O}(t))$ are given by
\begin{equation} \label{phi_def}
    |\mathcal{O}_0(t)) = e^{i\mathcal{L}t} |\mathcal{O}_0) = \sum_{n=0}^{K-1} \phi_n(t)|\mathcal{O}_n)
\end{equation}
where $\phi_n(t)$ satisfy the Schr\"odinger-like equation:
\begin{equation}
    -i \dot{\phi}_n(t) = \sum_{m=0}^{K-1} L_{nm}\phi_m(t),\quad \phi_n(0)=\delta_{n0}
\end{equation}
with $L_{mn}= (\mathcal{O}_m|\mathcal{L}|\mathcal{O}_n)$, and $\mathcal{L}|\mathcal{O}_n)$ is given in (\ref{Off-diagonal_Liouvillian}).

\end{comment}

\subsection{Disorder in Lanczos-sequences of integrable systems}
As a hopping model over the Krylov chain, the properties of the Lanczos coefficients will determine the dynamics.  As we shall show below, the Lanczos coefficients depend on the Liouvillian frequencies and on the spectral decomposition of the initial operator. We will argue heuristically that the abundance of energy differences smaller than the mean level spacing $\Delta$ in integrable systems induces an erratic behavior in their Lanczos coefficients; this is a result of the Poissonian nature of their spectrum.  Conversely, chaotic systems have fewer energy differences below $\Delta$ due to level repulsion, and we will argue that one would expect a less erratic behavior in the $b$-sequence of typical operators in these systems. We have observed this effect comparing the Lanczos sequences obtained from typical operators in SYK$_4$ \cite{Rabinovici:2020ryf} with those obtained from local operators in the XXZ spin chain, see subsection \ref{Subs_Anderson_loc} and Figures \ref{fig:Lanczos_XXZ} and \ref{fig:sigma}.

The elements $b_n$ of the Lanczos-sequence can be written as ratios of Hankel determinants of the moments of the operator two-point function (see for example  \cite{SANCHEZDEHESA1978275}). The moments are given by: 
\begin{eqnarray} \label{moments}
    \mu_n = \sum_{i=0}^{K-1} |O_i|^2 \omega_i^n\,, 
\end{eqnarray}
where $O_i=(\mathcal{O}_0|\omega_i)$ is the spectral decomposition of the initial operator, and it can be proved that the Lanczos coefficients satisfy the following recursion:
\begin{equation} \label{bn_Hankel}
    b_n^2 = \frac{D_{n-2} D_n}{D_{n-1}^2} \, , \quad n\geq 1\,,   
\end{equation}
where 
\begin{equation} \label{Hankel_Det}
    D_n = \begin{vmatrix} 
    \mu_0 & \mu_1 & \mu_2 & \dots & \mu_{n} \\
    \mu_1 & \mu_2 & \mu_3 & \dots & \mu_{n+1}  \\
    \vdots & \vdots & \vdots & \dots & \vdots \\
    \mu_{n} & \mu_{n+1} & \mu_{n+2} & \dots & \mu_{2n}
    \end{vmatrix} 
\end{equation}
are the so-called Hankel determinants, with $D_{-1}=1$ and $D_0=\mu_0=\sum_{i=0}^{K-1}|O_i|^2=1$ for a normalized initial operator. Note that all odd moments are zero since the operator is Hermitian and the Liouvillian spectrum is symmetric around zero. In  Appendix \ref{Appx_Hankel_Det} we show that such a Hankel determinant can be expressed directly in terms of the spectral decomposition $O_i$ and the Liouvillian frequencies $\omega_i$,
\begin{equation} \label{Dn}
    D_{n-1} = \sum_{\{i_1,i_2,\dots i_n \} \subset \{0, 1, \dots , K-1 \} } \prod_{i \in \{i_1,i_2,\dots i_n \}} |O_i|^2\prod_{a,b \in \{i_1,i_2,\dots i_n \}} (\omega_a-\omega_b)^2\,,
\end{equation}
where the sum is over all ${{K}\choose{n}}$ possible ways of choosing $n$ frequencies out of the $K$ frequencies in the spectrum of the Liouvillian, and the final product is the square of a Vandermonde determinant constructed out of these $n$ frequencies.  Given Poisson statistics in the level spacing of the Hamiltonian, one expects to find in the Liouvillian frequencies which are smaller than the mean level spacing, see Section \ref{Section_XXZ}. With enough such frequencies, for some $n$ every choice of $n$ frequencies out of $K$ will include some very small $\omega_a-\omega_b$ in the Vandermonde determinants, reducing the value of $D_{n-1}$ compared with $D_{n-2}$ (this will also depend on the projections $O_i$ over the eigenspaces corresponding to those particular sets of frequencies). Specifically near the center of the Liouvillian spectrum, for integrable systems, we expect smaller frequencies and small frequency differences than for chaotic systems. 
Accordingly, $b_n^2$ given by the ratio (\ref{bn_Hankel}), will undergo a fluctuation, becoming either larger or smaller than the previous one, depending on whether $D_{n-1}$ is in the numerator or the denominator. Conversely in chaotic systems the smaller amount of very small frequencies, would result in a less erratic $b_n$-sequence.

\subsection{Some results from Anderson localization}\label{Subs_Anderson_loc}
As argued around equation (\ref{L-tight-binding}), dynamics on the Krylov chain constitute a one-dimensional hopping problem where the hopping amplitudes are given by the Lanczos coefficients. Thus, the erratic nature of the $b$-sequence induced by the abundance of small energy differences in integrable models can be effectively described by off-diagonal disorder in the Anderson sense. We will therefore now collect some useful results pertaining to Anderson localization that will inform our subsequent discussion. Such problems were studied as an electron localization problem (together with diagonal disorder) in \cite{Thouless_1972}; as a pure off-diagonal system in \cite{Fleishman_1977}, where it was established that the center of the band state was localized with a localization length increasing as $\sqrt{K}$, where $K$ is the chain size; also in \cite{PhysRevB.24.5698}, where localization was shown by studying the transmission coefficient; and including long-range correlations in \cite{PhysRevB.72.174207}.  For a review, see for example \cite{IZRAILEV2012125}.  

For a maximal Krylov dimension (which is attained whenever no exact degeneracies are present in the Hamiltonian spectrum and the operator is fully dense in the energy basis), the Liouvillian always has a zero frequency eigenstate.  This eigenstate, which is the center-of-the-band state, can be reconstructed from the Lanczos-sequence.  In particular, expanding the Liouvillian eigenstates in the Krylov basis as

\begin{equation}
    \centering
    \label{L_eigenstate_krylov}
    |\omega ) = \sum_{n=0}^{K-1}\psi_n|\mathcal{O}_n)\,,
\end{equation}
the eigenvalue problem 
\begin{eqnarray} \label{EV_problem}
    \omega \psi_n = b_{n+1}\psi_{n+1}+b_n\psi_{n-1},\quad n=0,...,K-1
\end{eqnarray}
can be solved exactly for a given $b$-sequence for $\omega=0$.  The even elements of the eigenvector are given by:
\begin{eqnarray}
     \frac{\psi_{2n}}{\psi_0} &=& \prod_{i=1}^{n}\frac{ b_{2i-1}}{b_{2i}}, \quad n = 0,\dots, \frac{K-1}{2}
\end{eqnarray}
and all odd elements are zero.

\begin{figure}
    \centering
    \includegraphics[scale=0.6]{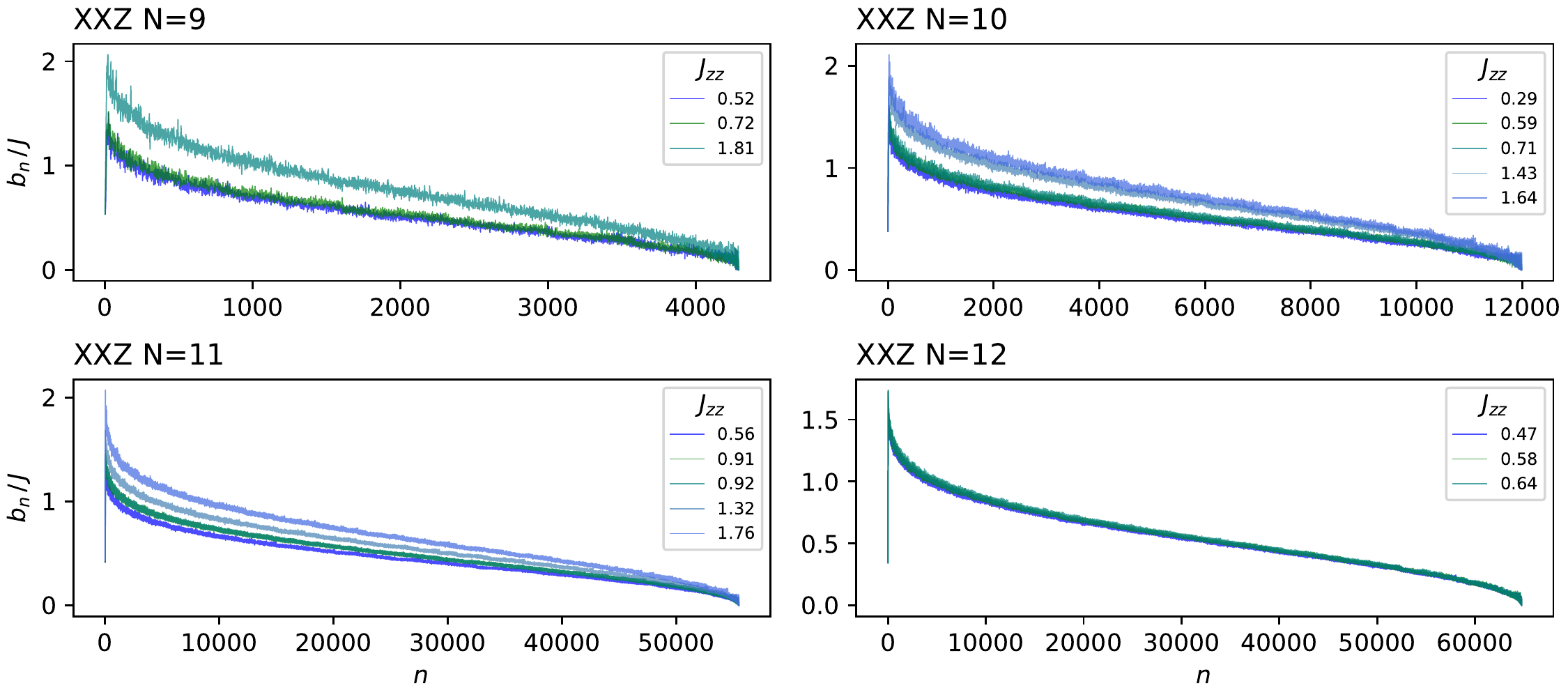}
    \caption{The Lanczos sequences for the various XXZ cases studied in this paper. For the details of the XXZ model see Section \ref{Section_XXZ}. For details on the one-site operator used and further numerical results, see Section \ref{Sect_Numerical_Results}.}
    \label{fig:Lanczos_XXZ}
\end{figure}

\begin{figure}
    \centering
    \includegraphics[scale=0.5]{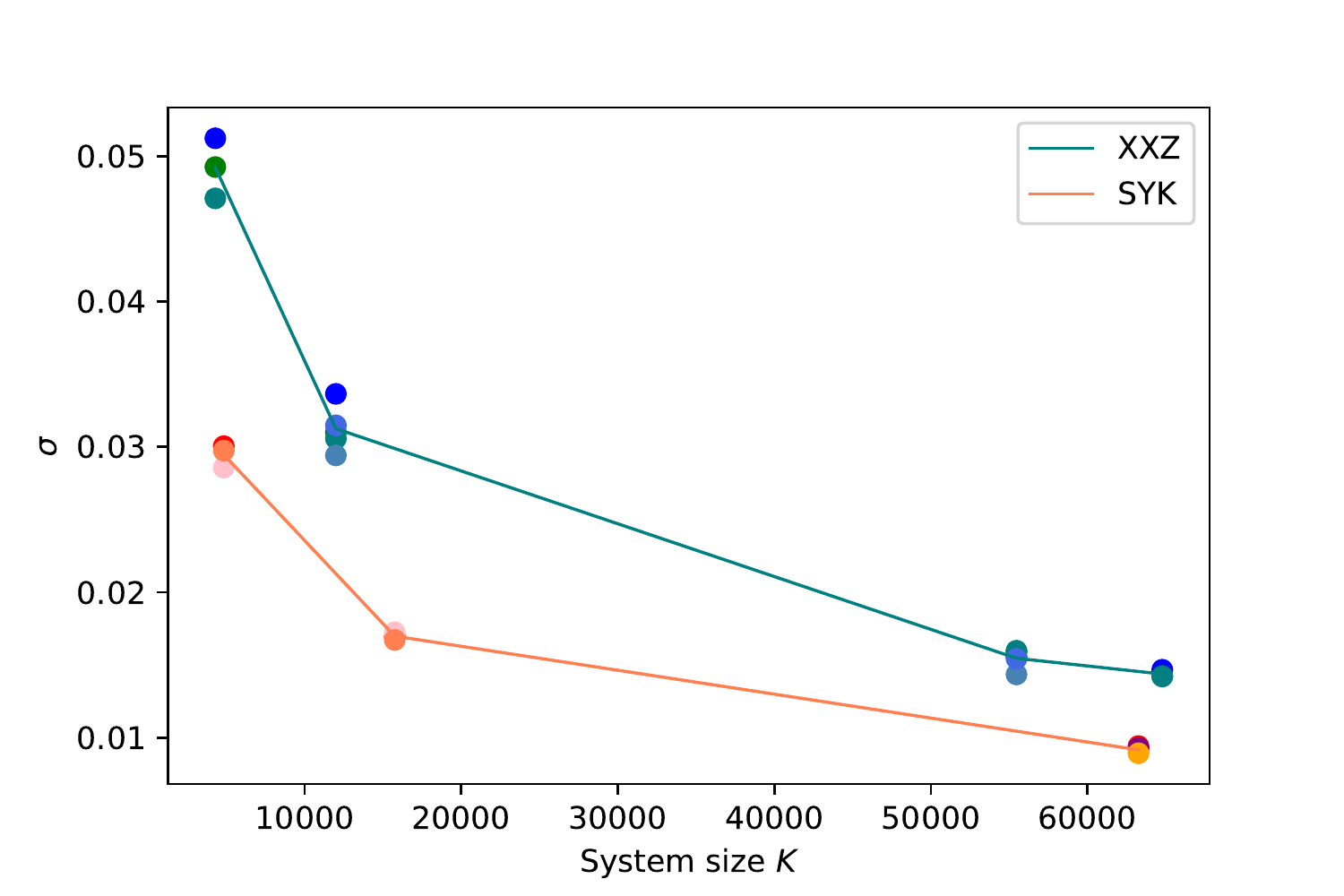}
    \caption{Using Lanczos-sequence data for various XXZ and complex SYK$_4$ systems of different Krylov dimensions, the standard deviation of the logarithm of ratios of consecutive Lanczos coefficients is represented by $\sigma$ on the $y$-axis, while the $x$-axis represents the Krylov dimension.  The smoother the Lanczos-sequence is, the closer $\sigma$ is to zero. The lines connect the average values of each set of points for a given system size and are shown to help the reader.
    For each system size, each point represents a single realization of SYK and a single $J_{zz}$ value of XXZ. In particular we present the following data.  \textbf{SYK}: 3 realizations for $L=8$ fermions, 3 realizations for $L=9$ and 5 realizations for $L=10$ (their Lanczos sequences can be found in \cite{Rabinovici:2020ryf}); \textbf{XXZ}: 3 different $J_{zz}$ coefficients for $N=9$ spins, 5 for $N=10$, 5 for $N=11$ and 3 for $N=12$. The specific values chosen for $J_{zz}$ for each system size can be found in Figure \ref{fig:Lanczos_XXZ}.
    }
    \label{fig:sigma}
\end{figure}

In the absence of long-range correlations between the hopping amplitudes, it was first proposed in \cite{Fleishman_1977} that in the case of disordered $b$'s this state is localized.  The localization length was shown to be proportional to $\sqrt{K}$ and inversely proportional to $\sigma$, $l_{loc}\propto \sqrt{K}/\sigma$, where $\sigma^2$ is the variance of the logarithm of ratios of consecutive hopping amplitudes:
\begin{eqnarray} \label{disorder_strength}
    \sigma^2 = \textrm{Var}(x_i), \quad x_i \equiv \ln \Big| \frac{b_{2i-1}}{b_{2i}} \Big| ~.
\end{eqnarray}
Since $b_n$ are fluctuating, the values of $x_i$ will fluctuate around zero, and will be randomly distributed with mean 0. The variance $\sigma^2$ will depend on the details of the $b_n$ distribution. In Figure \ref{fig:Lanczos_XXZ} we depict the Lanczos sequences of a one-site operator for various instances of the XXZ spin chain (for more details, see Sections \ref{Section_XXZ} and \ref{Sect_Numerical_Results}). The statistics of their fluctuations were analysed by computing (\ref{disorder_strength}) in each case\footnote{Only the first half of each sequence was used for this computation, as the statistics of the very small Lanczos coefficients towards the end of the descent are numerically less reliable due to the amplifying effect of the logarithm and the ratio in the definition of the variables $x_i$.}. The results for $\sigma$ are depicted in Figure \ref{fig:sigma}, where they are also compared to SYK$_4$ systems of similar size using data from \cite{Rabinovici:2020ryf}. Although disorder strength is seen to decrease with the system size for both XXZ and SYK$_4$, our results indicate that indeed XXZ exhibits larger disorder in its Lanczos sequences. In particular, for comparable system sizes the disorder parameter $\sigma$ is bigger in XXZ than in SYK$_4$

\subsection{Probes of late-time behavior} \label{Sec:late-time_behavior}
In this section we introduce some of the quantities used to study and observe localization in \cite{luck:cea-01485001}, and relate them to K-complexity and its long-time-average. 

The time-dependent coefficients $\phi_n(t)$ given in (\ref{phi_def}) can be written in the basis of Liouvillian eigenvectors, 
\begin{eqnarray}
    |\mathcal{O}(t)) &=& \sum_{j=0}^{K-1} e^{i \omega_j t} |\omega_j)(\omega_j|\mathcal{O}_0) 
    =\sum_{m=0}^{K-1} \sum_{j=0}^{K-1} e^{i \omega_j t} (\omega_j|\mathcal{O}_0) (\mathcal{O}_m|\omega_j) |\mathcal{O}_m)\,,
\end{eqnarray}
where we expanded the Liouvillian eigenstates in the Krylov basis 
$|\omega_j) = \sum_{m=0}^{K-1} |\mathcal{O}_m)(\mathcal{O}_m|\omega_j)$. 
From (\ref{phi_def}), using the orthonormality of the Krylov elements $(\mathcal{O}_m|\mathcal{O}_n)=\delta_{mn}$,  we recognize
\begin{eqnarray}
    \phi_n(t) = (\mathcal{O}_n|\mathcal{O}(t)) =  \sum_{j=0}^{K-1} e^{i \omega_j t} (\omega_j|\mathcal{O}_0) (\mathcal{O}_n|\omega_j)~.
\end{eqnarray}
We now give $|\phi_n(t)|^2$ the interpretation of a transition probability from $|\mathcal{O}_0)$ to $|\mathcal{O}_n)$ at time $t$:
\begin{eqnarray}
\label{P0n_def}
    P_{0n}(t) \equiv |\phi_n(t)|^2  = \sum_{i,j=0}^{K-1} e^{i(\omega_j-\omega_i)t} (\omega_j|\mathcal{O}_0) (\mathcal{O}_n|\omega_j) (\omega_i|\mathcal{O}_n)(\mathcal{O}_0|\omega_i) ~.
\end{eqnarray}
This expression suggests that we interpret K-complexity as the average location on the Krylov chain to which the operator will transition at each time slice $t$:
\begin{eqnarray}
    C_K(t) = \sum_{n=0}^{K-1} n |\phi_n(t)|^2 = \sum_{n=0}^{K-1} n P_{0n}(t)\,.
\end{eqnarray}
To study the long-time behaviour of K-complexity, which is the main goal of this paper, we perform a long-time-average over $|\phi_n(t)|^2$:
\begin{eqnarray}
\label{Q0n_longtimeavg}
    \overline{|\phi_n|^2} = \lim_{T\to \infty} \frac{1}{T}\int_0^T |\phi_n(t)|^2 dt\,.
\end{eqnarray}
By construction, the spectrum of the restriction of the Liouvillian to Krylov space has no degeneracies, and hence the phase-differences average out so that only the diagonal terms with $i=j$ in (\ref{P0n_def}) contribute to (\ref{Q0n_longtimeavg}):
\begin{eqnarray} \label{Transition_Probability}
    \overline{|\phi_n|^2} \equiv Q_{0n}= \sum_{i=0}^{K-1}   |(\mathcal{O}_0|\omega_i)|^2 |(\mathcal{O}_n|\omega_i)|^2 ~.
\end{eqnarray}
The long-time-averaged K-complexity $\overline{C_K}$ can now be interpreted in two ways:
\begin{enumerate}
    \item The long-time-averaged expectation value of the position to which $|\mathcal{O}_0)$ can ``hop'': 
    \begin{eqnarray} \label{KC_Q0n}
        \overline{C_K} = \sum_{n=0}^{K-1} n\, Q_{0n}\,.
    \end{eqnarray}
    \item The spectral average of the K-complexities of the Liouvillian eigenstates, weighted by the overlap of such eigenstates with the initial condition:
    \begin{eqnarray} \label{KC_KCi}
        \overline{C_K} = \sum_{i=0}^{K-1} |(\mathcal{O}_0 |\omega_i)|^2   \sum_{n=0}^{K-1} n\,  |( \mathcal{O}_n |\omega_i)|^2\equiv \sum_{i=0}^{K-1}|(\mathcal{O}_0 |\omega_i)|^2\, C_K^{(i)}\,, 
    \end{eqnarray}
    where $C_K^{(i)}$ is the K-complexity of the eigenstate of phase $\omega_i$, that is, the average position on the Krylov chain of this stationary state. Eigenstates with smaller $C_K^{(i)}$ will be more localized towards the left of the chain and in general will have more significant overlap with the initial condition.
\end{enumerate}
The transition probability, and hence also $\overline{C_K}$, are influenced by localization effects in the eigenstates. The following two limiting cases are instructive:
\begin{itemize}
\item \textit{Full delocalization}: for which the eigenstates $|\omega_i)$ are fully delocalized over $|\mathcal{O}_n)$, and $|(\mathcal{O}_n |\omega_i)|^2\sim \frac{1}{K}$ for every $i,n$, implying both $Q_{0n}=\frac{1}{K}$ for all $n$ and $C_K^{(i)}=\frac{K}{2}$ for all $i$. Applying either (\ref{KC_Q0n}) or (\ref{KC_KCi}) we obtain, consistently:
    \begin{equation}
        \overline{C_K}=\frac{1}{K}\sum_{n=0}^{K-1} n \approx \frac{K}{2}\,.
    \end{equation}
\item \textit{Full localization}: for which the eigenstates $|\omega_i)$ are fully localized over $|\mathcal{O}_n)$, and $|( \mathcal{O}_n |\omega_i)|^2= \delta_{ni}$ for every $i,n$. In this case $Q_{0n}=\delta_{0n}$ for any $n$ and $C_K^{(i)}=i$ for all $i$, and we obtain:
    \begin{equation}
        \overline{C_K}= \sum_{i=0}^{K-1}\delta_{0i}\sum_{n=0}^{K-1} n\, \delta_{ni} = 0 ~.
    \end{equation}
Notice that in this extreme case $C_K(t)=0$ for all times because $|(\mathcal{O}_n | \omega_i )|^2 = \delta_{ni}$ implies that $|\omega_0)=|\mathcal{O}_0)$ (up to a phase), so that the initial condition is itself a stationary state and therefore it doesn't propagate.
\end{itemize}

For the case of a constant Lanczos-sequence, the eigenstates and eigenvalues of the Liouvillian as well as $\overline{C_K}$ can be computed analytically. This case is shown in detail in Appendix \ref{appx:Constant_b_analytics}.

We shall now turn to discuss the Hamiltonian of the XXZ spin chain, an interacting model whose integrable nature makes it an ideal candidate for studying the erratic structure of the Lanczos coefficients in this kind of systems and the consequent imprints that localization leaves on Krylov space dynamics.

\section{Complexity in integrable models: the XXZ spin chain}\label{Section_XXZ}

Motivated by the arguments in Subsection \ref{Subsection_Krylov_hopping}, we intend to study operator dynamics in systems with an exponentially big Krylov dimension for typical operators but which still fall within the category of integrable systems. These conditions are fulfilled by strongly-interacting systems that are integrable via the Bethe Ansatz. In fact, in the framework of the \textit{algebraic} Bethe Ansatz it is possible to show that there exist extensively many commuting conserved charges that can be retrieved from the series expansion in the spectral parameter of the transfer matrix \cite{Samaj_bajnok_2013,slavnov2019algebraic,Reffert_Bethe}. However, these symmetries do not necessarily imply an extensive number of degeneracies in the spectrum of the Hamiltonian: rather, they enforce an intricate block-diagonal structure in which eigenstates carry several quantum numbers corresponding to different conserved charges (even though the underlying global symmetry might not be immediately apparent and the charges might not even be local). As a result of this, energy eigenvalues are effectively uncorrelated and therefore the nearest-neighbor level spacing statistics become Poissonian, implying that eigenvalues in the spectrum don't feature level repulsion: They can be arbitrarily close to each other, even though not exactly degenerate. Additionally, local operators in these systems can be dense in the energy basis, even though the statistical properties of their matrix elements don't follow the ETH, as exemplified in \cite{Rigol_XXZ}. These features have the effect that the Krylov space dimension for typical operators in this kind of systems is close to the upper bound (\ref{K_bound}) or, at least, scales exponentially in system size as it would do for an ETH operator in a chaotic system. The distinguishing features of interacting integrable systems should be found in the impact on the actual dynamical evolution along the Krylov chain of the underlying spectral statistics and of the structure of the particular operator. The XXZ spin chain \cite{Heisenberg:1928mqa,YangXXZ_I,YangXXZ_II,Samaj_bajnok_2013} is a canonical example for this kind of models and we shall adopt it as the main workhorse of this project. Its Hamiltonian is given by:

\begin{equation}
    \centering
    \label{XXZ}
    H_{XXZ} = -\frac{J}{2} \sum_{n=1}^{N-1}\left[ \sigma_n^x \sigma_{n+1}^x+\sigma_n^y \sigma_{n+1}^y + J_{zz}\left( \sigma_n^z \sigma_{n+1}^z -1\right) \right],
\end{equation}
where $J$ is an overall dimensionful energy scale\footnote{In the numerics it was set to $1$, so it can be thought of as setting the units of the problem, as there is no other dimensionful parameter.} and $J_{zz}$ is the dimensionless coupling strength. Note that the special values $J_{zz}=0,1$ correspond to the XY and XXX models, respectively, and that the summation limits signal implicitly open boundary conditions (OBC), as opposed to periodic (PBC). Unless stated otherwise, we will use OBC in our numerical analysis.

The Hamiltonian (\ref{XXZ}) is integrable via Bethe Ansatz. In order to unveil the Poissonian nature of its spectrum, it is necessary to first mod out the discrete symmetries that do cause exact degeneracies and restrict to a Hilbert space sector whose dimension will still be exponential in system size. $H_{XXZ}$ commutes with the two following global symmetries (see for example \cite{Santos2013}):

\begin{itemize}
    \item Total spin projection along the $z$-direction (or total magnetization):
    \begin{equation}
        \centering
        \label{Total-Spin}
        S^z = \sum_{n=1}^N S_n^z\,,
    \end{equation}
    where $S_n^z=\frac{1}{2}\sigma_n^z$ denotes the spin projection along the $z$-axis at site $n$.
    \item Parity, defined as the transformation that exchanges sites as $n\longleftrightarrow{ N+1-n}$:
    \begin{equation}
        \centering
        \label{Parity}
        P = \prod_{n=1}^{\lfloor \frac{N}{2}\rfloor}\mathcal{P}_{n,N+1-n}\,,
    \end{equation}
    where $\mathcal{P}_{nm}$ is the permutation operator between sites $n$ and $m$. As defined, parity is a space inversion with respect to the chain center.
\end{itemize}

\begin{figure}
    \centering
    \includegraphics[width=10cm]{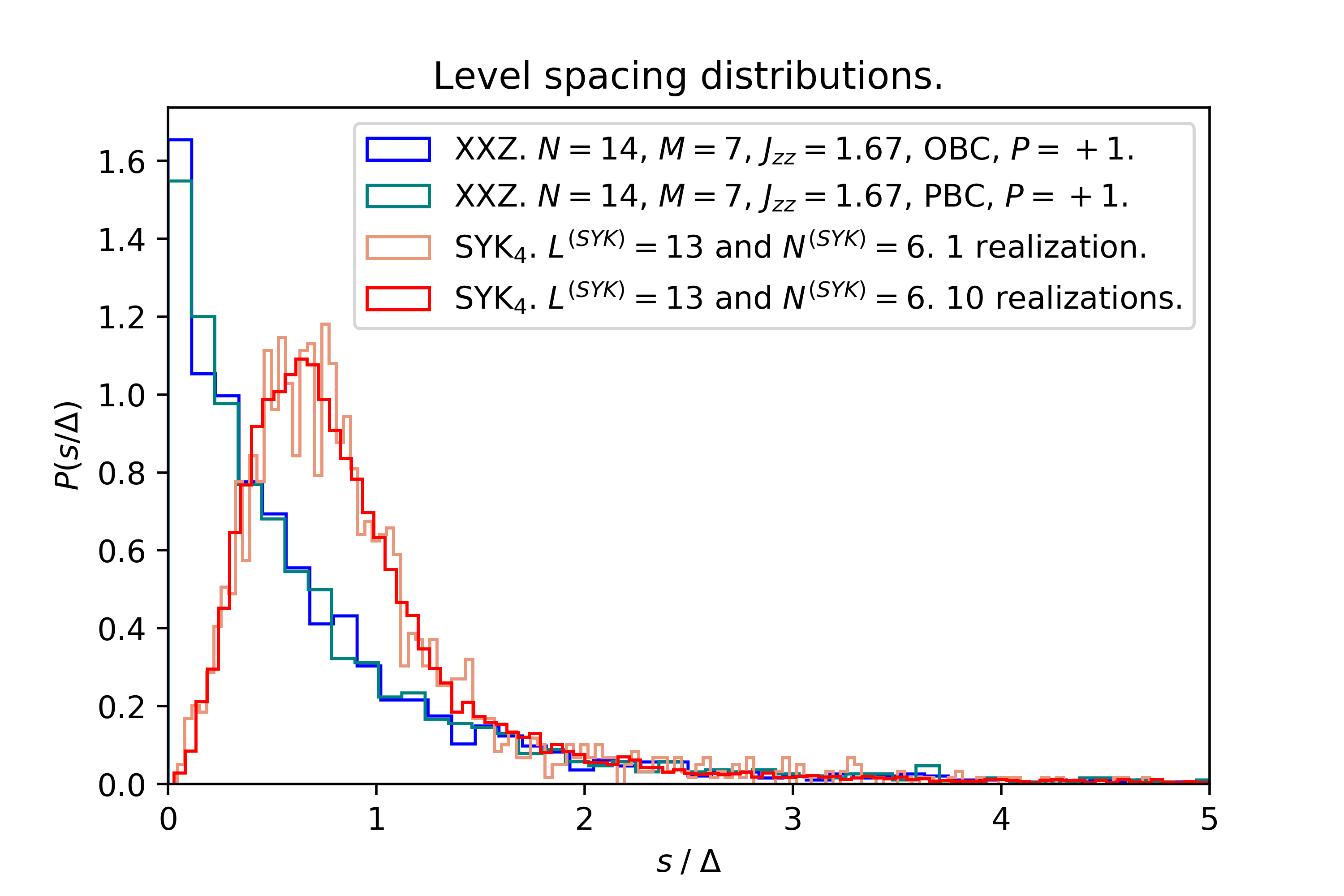}
    \caption{Nearest-neighbor level spacing distributions for XXZ with a specific $J_{zz}$ coupling with $N=14$ spins in the sector $M=7$, $P=+1$, both with open and periodic boundary conditions. For comparison, we include the corresponding distributions for complex SYK$_4$ with $L^{(SYK)}=13$ sites and occupation number $N^{(SYK)}=6$, whose Hilbert space dimension is equal to that of the positive parity sector of the above-mentioned XXZ instance. All distributions are normalized so that the area under each curve is equal to one, and all energy differences have been normalized by the mean level spacing $\Delta$ in each system.}
    \label{fig:Histogram}
\end{figure}

There are other global symmetries that can be deduced from the algebraic Bethe Ansatz, but numerical checks indicate that only parity induces exact degeneracies when OBC are chosen. For instance, we can consider \textit{charge conjugation}, which takes the form of a spin flip on each site or, up to a global phase, a rotation of angle $\pi$:

    \begin{equation}
        \centering
        \label{R-operator-charge-conj}
        R = \prod_{n=1}^{N}\sigma_n^x = i^N \exp \big( -i \pi S^x \big)~.
    \end{equation}
It is possible to show that $[H_{XXZ},R]=0=[P, R]$, however, $[R,S^z]\neq 0$. Since $\left[S^z,P\right]=0$, we choose the set of commuting observables $\left\{H_{XXZ},S^z,P \right\}$, with which it is possible to block-diagonalize $H_{XXZ}$ and restrict ourselves to a Hilbert space sector of fixed parity and total magnetization. The total Hilbert space of the system, $\mathcal{H}$, has dimension $2^N$ and, thanks to $S^z$-symmetry, it can be split as the direct sum of sectors with a fixed number $M$ of spins up (i.e. fixed magnetization sectors):

\begin{equation}
    \centering
    \label{Hilbert-M-sectors}
    \mathcal{H}=\bigoplus_{M=0}^N\mathcal{H}_M ~.
\end{equation}
In each of these sectors, whose dimension is $D_M=\binom{N}{M}$, the total magnetization operator is proportional to the identity, $S^z = M-\frac{N}{2}$ and the restriction of the Hamiltonian on each of them is closed. In future discussions, we will consider without loss of generality only sectors with $M\leq \lfloor\frac{N}{2}\rfloor$, since the $M$-sector is mapped to the $N-M$-sector by $R$ symmetry\footnote{In fact, all non-trivial physics come from \textit{antiparallel} pairs of neighbouring spins \cite{Samaj_bajnok_2013}, as can be deduced from (\ref{XXZ}).}. Furthermore, the analysis based on Bethe Ansatz shows \cite{Samaj_bajnok_2013} that, in each sector, the spectral width scales with $M$ for fixed interaction strength $J_{zz}$, since the energy of each eigenvalue can be expressed as the sum of $M$ magnon dispersion relations, which are bounded. Therefore, in order to have a spectrum whose width doesn't scale with system size, we choose to normalize the Hamiltonian in a fixed-magnetization sector by the number of up spins; that is, we will work with $H^{(M)}$ given by:

\begin{equation}
    \centering
    \label{XXZ-sector-normalized}
    H^{(M)} := \frac{1}{M}H_{XXZ}^{(M)},
\end{equation}
where $H_{XXZ}^{(M)}$ denotes the restriction of the XXZ Hamiltonian $H_{XXZ}$ to a fixed magnetization sector with $M$ spins up. The spectrum of $H^{(M)}$ can present a number of exact degeneracies due to parity symmetry. In fact, each magnetization sector can be further split into positive and negative parity subsectors:

\begin{equation}
    \centering
    \label{Hilbert-M-P-sectors}
    \mathcal{H}_M=\mathcal{H}_M^+\oplus \mathcal{H}_M^{-},
\end{equation}
whose dimensions $D_M^\pm$ are computed in Appendix \ref{Appx_Sectors} and summarized in Table \ref{Table_Dim_ParitySectors}. Importantly, they scale asymptotically exponentially in system size. As numerical checks indicate, once we restrict ourselves to a sector $\mathcal{H}_M^P$, the spectrum of the Hamiltonian doesn't feature exact degeneracies due to any other discrete symmetries, and we can extract the universal Poissonian behavior of the spectral statistics, as ilustrated in Figure \ref{fig:Histogram}, where the level spacing distribution of the positive-parity sector in XXZ with both OBC and PBC are compared to that of an instance of complex SYK$_4$ with the same Hilbert space dimension. This serves to illustrate a point that could be misunderstood: the integrable character of the system reflects itself in the \textit{lack of correlation} between the energy eigenvalues, yielding a Poissonian distribution characterized by a high probability of finding energy differences smaller than the mean level spacing, as opposed to the characteristic level repulsion displayed by chaotic systems. Exact degeneracies are not necessarily a defining feature of integrable systems.

As a word of caution, we shall note that, for chains with an even number of sites $N$, there exists a zero-magnetization sector ($M=\frac{N}{2}\Longrightarrow S^z = 0$) in which accidentally $S^z$ and $R$ \textit{do} commute, since $R$ doesn't change the magnetization sector because there are as many spins up as spins down. Numerically (see for instance Figure \ref{fig:Histogram}) we don't observe extra degeneracies produced by this symmetry enhancement. In fact, it is conjectured in the literature \cite{Doikou:1998jh} that eigenstates within this magnetization sector are themselves eigenstates of $R$. In conclusion, for chains of even length, in the Hilbert space sector with $M=\frac{N}{2}$ and fixed $P$ we can classify eigenstates according to their $R$-eigenvalue ($\pm 1$), and the parity sectors can be split into $R$-subsectors:

\begin{equation}
    \centering
    \label{H-N/2-P-R}
    \mathcal{H}_{\frac{N}{2}}^{P}= \mathcal{H}_{\frac{N}{2}}^{P,R=+}\oplus \mathcal{H}_{\frac{N}{2}}^{P,R=-},\quad\quad\text{for }N\text{ even}
\end{equation}
whose dimensions $D_{\frac{N}{2}}^{PR}$ are also computed in Appendix \ref{Appx_Sectors} and can be written compactly as:

\begin{equation}
    \centering
    \label{R_sector_dimension}
    D_{\frac{N}{2}}^{PR} = \frac{1}{2}D_{\frac{N}{2}}^P + P\cdot R\cdot2^{\frac{N-4}{2}},\quad\quad\quad\text{for }N \text{ even.}
\end{equation}
Despite not granting exact degeneracies in the spectrum of the Hamiltonian, this symmetry enhancement can yield some selection rules for operators charged under $R$, which impose constraints on their matrix elements in the energy basis and hence can reduce the dimension of their associated Krylov space, as also discussed in Appendix \ref{Appx_Sectors}.

\begin{table}[ht]
\centering

\begin{tabular}{|c||c|c|}
\hline
       & $M$ odd & $M$ even \\ \hline && \\[-10pt] \hline
       && \\
$N$ odd  & $\begin{pmatrix}[1.5] \frac{N-1}{2} \\\frac{M-1}{2} \end{pmatrix}$  & $\begin{pmatrix}[1.5] \frac{N-1}{2} \\\frac{M}{2} \end{pmatrix}$      \\[20pt] \hline
&&\\
$N$ even & $0$     & $\begin{pmatrix}[1.5] \frac{N}{2} \\\frac{M}{2} \end{pmatrix}$      \\[20pt] \hline
\end{tabular}
\caption{The dimension of each Hilbert space sector of fixed parity and total magnetization is $D_M^{\pm}=\frac{D_M\pm A}{2}$, where $D_M=\binom{N}{M}$ and the expression for $A$, which depends on whether $N$ and $M$ are even or odd, is shown in this Table.} \label{Table_Dim_ParitySectors}
\end{table}

\section{Numerical results}\label{Sect_Numerical_Results}
In this section we present evidence that in the case of $H_{XXZ}$ discussed in the previous section and for a single-site operator which we introduce below, K-complexity indeed saturates at values below $\sim K/2$, which is the expected saturation value for chaotic systems such as SYK$_4$ studied previously in \cite{Rabinovici:2020ryf}, where the Lanczos sequences featured less disorder as illustrated in Figure \ref{fig:sigma}. 

We will work with $H^{(M)}$ defined in (\ref{XXZ-sector-normalized}) in the sector of fixed magnetization and parity $\mathcal{H}_M^{+}$ and will study the time evolution of the linear operator 
\begin{equation} \label{Study_Operator}
    \mathcal{O} = \sigma^z_{m} + \sigma^z_{N-m+1}\,,
\end{equation}
where $m$ can be any site on the spin chain $1\leq m \leq N$.  This operator was chosen because it respects the sector symmetries, namely total magnetization conservation and parity, in addition to being a local one-site observable.
We choose $m$ to be near the center of the spin chain, so that boundary effects will not play a role. In general, all the matrix elements of this operator in the energy basis (\ref{O-spectral-decomp}) are non-zero and therefore, given that we work in sectors with no degeneracies, the Krylov space dimension for such an operator is expected to saturate the upper bound of (\ref{K_bound}).  We tested this assertion by studying the matrix elements of this operator in the energy basis to very high precision. It would be interesting to also study other local operators; preliminary checks performed at small system sizes for linear combinations of one-site operators, two-site operators and linear combinations of two-site operators display a similar phenomenology to what will be described in this section for the operator (\ref{Study_Operator}).  Some of the Hamiltonians and operators were generated using the open-source package \cite{SciPostPhys.2.1.003}, while others were produced with programs of our own implementation, yielding coincident results.

\begin{figure}[ht]
    \centering
         \begin{subfigure}[t]{0.45\textwidth}
         \centering
         \includegraphics[width=\textwidth]{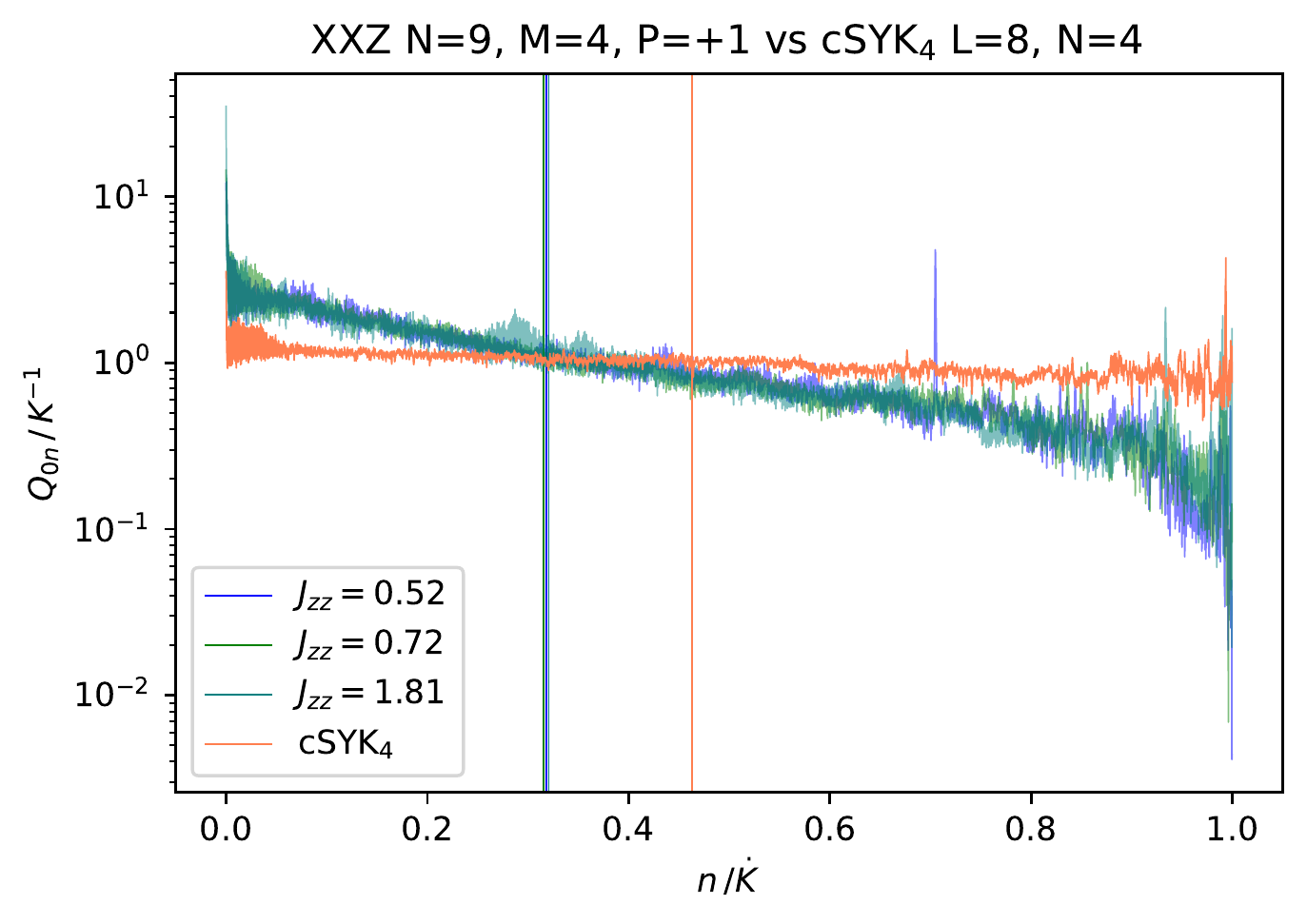}
         \caption{The Hilbert space dimension of XXZ with $N=9,\, M=4$, in the $P=+1$ sector is, according to Table \ref{Table_Dim_ParitySectors}, $D=66$.  For the operator $\mathcal{O}=\sigma_5^z$ the Krylov space dimension equals its upper bound $K=D^2-D+1=4291$. We find $\overline{C_K}\sim 0.318K$.  An SYK$_4$ system of Krylov space dimension $K_{SYK}=4831$ has $\overline{C_K}\approx 0.463 K_{SYK}$, much closer to the value expected for chaotic systems.}
         \label{fig:XXZ9}
        \end{subfigure}
     \hfill
     \begin{subfigure}[t]{0.45\textwidth}
         \centering
         \includegraphics[width=\textwidth]{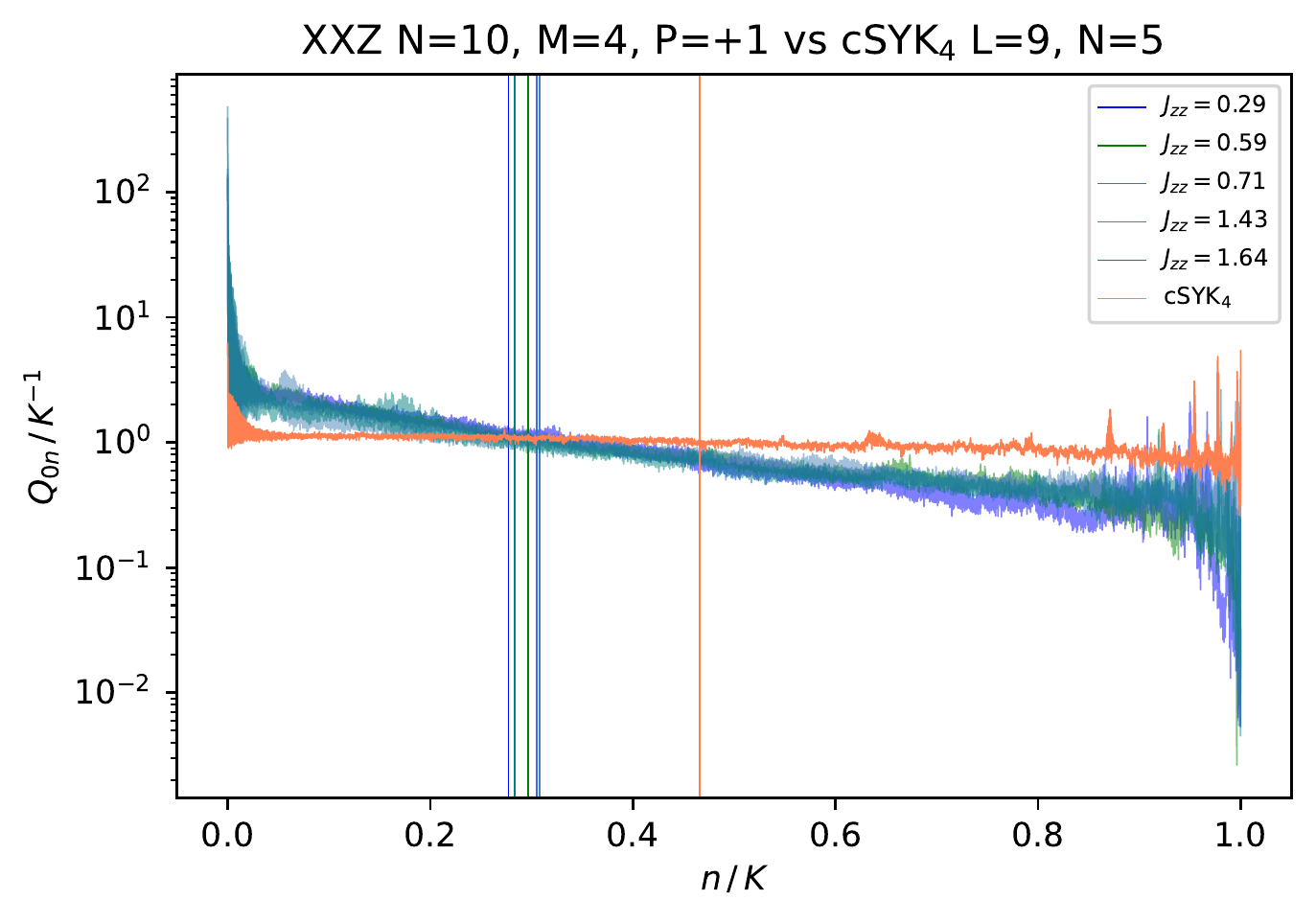}
         \caption{The Hilbert space dimension of XXZ with $N=10,\, M=4,\, P=+1$ is $D=110$ according to Table \ref{Table_Dim_ParitySectors}. For the operator $\mathcal{O}=\sigma_5^z+\sigma_6^z$ the Krylov space dimension saturates its upper bound and  $K=D^2-D+1=11991$. If the system were chaotic we would expect $\overline{C_K}\sim 0.5K$. We find $\overline{C_K}\approx 0.294K$.  For an SYK$_4$ system of Krylov space dimension $K_{SYK}=15751$, we find $\overline{C_K}\approx 0.466K_{SYK}$.}
         \label{fig:XXZ10}
        \end{subfigure} 
        \begin{subfigure}[t]{0.45\textwidth}
         \centering
         \includegraphics[width=\textwidth]{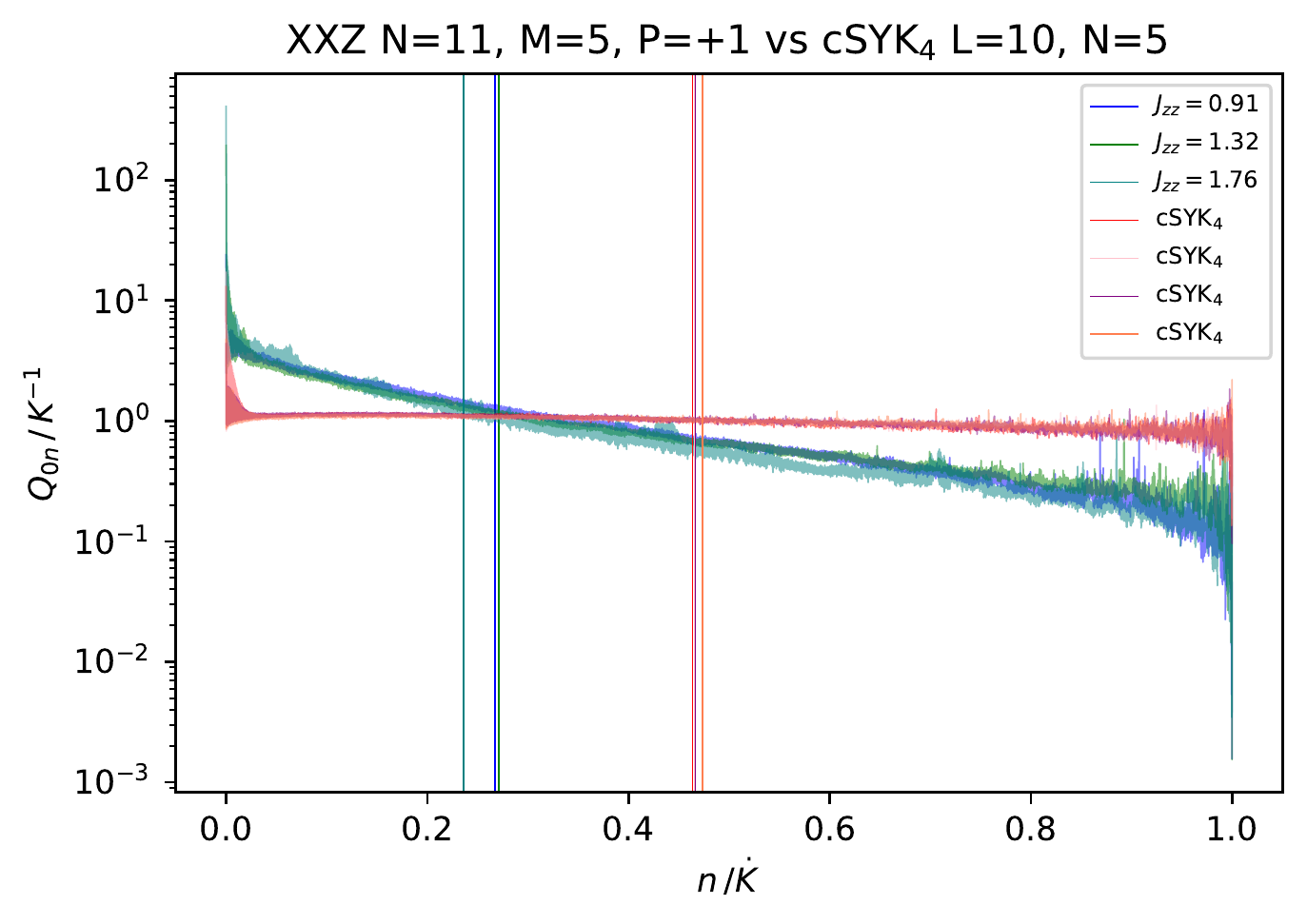}
         \caption{For XXZ with $N=11$ in the sector $M=5,\, P=+1$, the Hilbert space is of dimension $D=236$ (Table \ref{Table_Dim_ParitySectors}). Results for operators $\mathcal{O}=\sigma_6^z$ ($J_{zz}=0.91$) and $\mathcal{O}=\sigma_5^z+\sigma_7^z$ are shown; both saturate the upper bound on the Krylov space dimension $K=55461$. Nevertheless, the long-time average of K-complexity is below $0.5K$ and is (on average) $\overline{C_K}\sim 0.258K$. For SYK$_4$ systems of Krylov space dimension $K_{SYK}=63253$ we find (on average) $\overline{C_K}\approx 0.4675 K_{SYK}$.}
         \label{fig:XXZ11}
        \end{subfigure} 
      \hfill
     \begin{subfigure}[t]{0.45\textwidth}
         \centering
         \includegraphics[width=\textwidth]{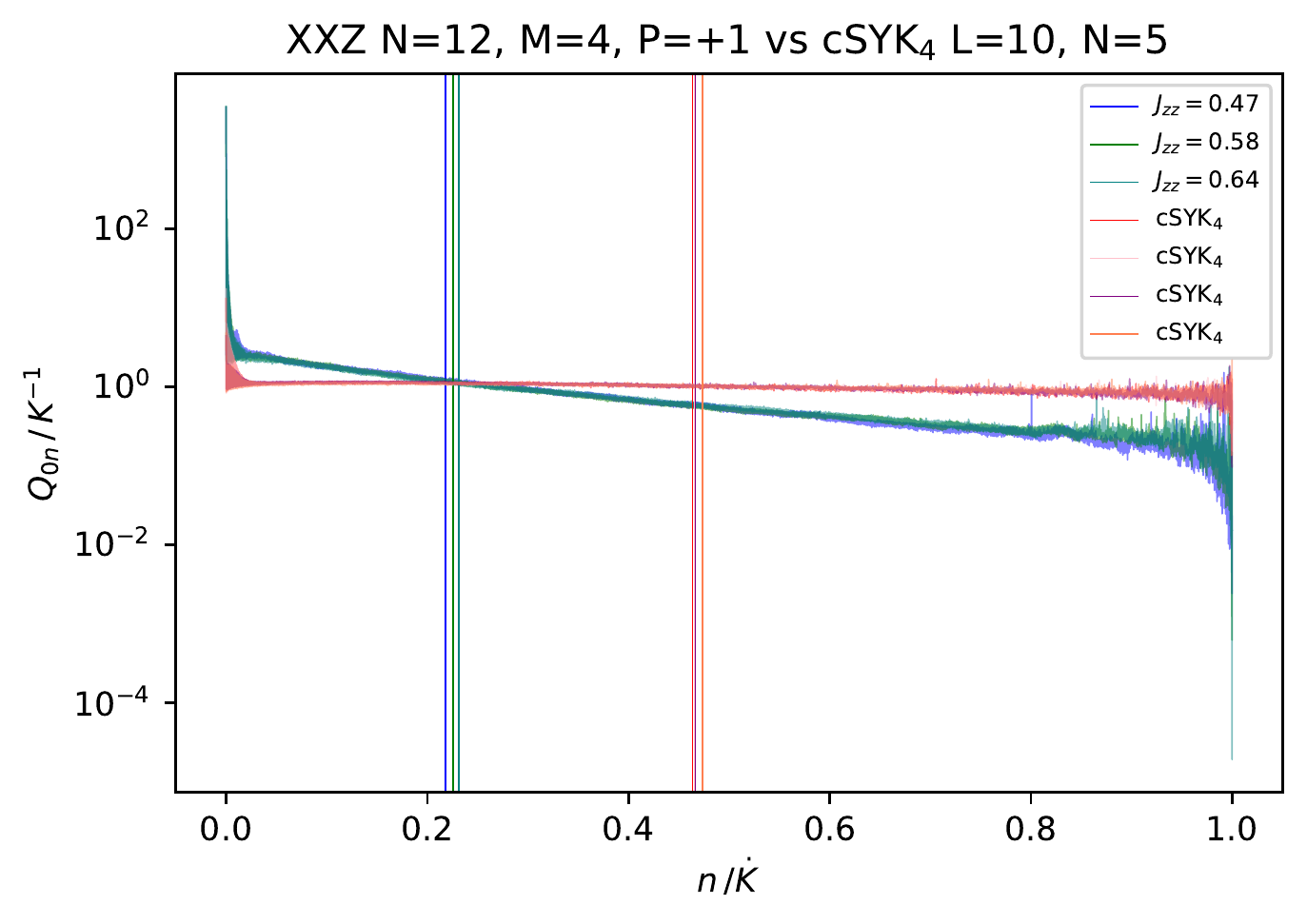}
         \caption{The Hilbert space dimension for XXZ with $N=12,\, M=4$ and $P=+1$ is $D=255$ (Table \ref{Table_Dim_ParitySectors}).  The operator $\mathcal{O}=\sigma_6^z+\sigma_7^z$ saturates the upper bound on the Krylov space dimension $K=D^2-D+1=64771$. The long-time average of K-complexity is below $0.5K$ and its value is (on average) $\overline{C_K}\approx 0.225 K$.  We plot again results for the same SYK$_4$ systems shown in Figure \ref{fig:XXZ11}.}
         \label{fig:XXZ12}
        \end{subfigure} 
    \caption{Numerical results for the transition probabilities $Q_{0n}$ defined in (\ref{Transition_Probability}) and $\overline{C_K}$ of XXZ systems with 9, 10, 11 and 12 spins and comparison with complex SYK$_4$ systems of 8, 9 and 10 fermions.  The vertical lines represent $\overline{C_K}$ in units of $K$ for each system.}
    \label{fig:Qn0_XXZ_vs_SYK}
\end{figure}

Using high precision arithmetic and the re-orthogonalization algorithms we developed for the study of SYK and which are described in \cite{Rabinovici:2020ryf}, we computed the Lanczos coefficients associated to operator (\ref{Study_Operator}) for various instances of XXZ, see Figure \ref{fig:Lanczos_XXZ}. Then, following the discussion in Section~\ref{Sec:late-time_behavior}, we plot in Figure \ref{fig:Qn0_XXZ_vs_SYK} the time-averaged transition probabilities $Q_{0n}$ for several XXZ systems with different numbers of spins $N$, different magnetization sectors $M$ and various choices of the $J_{zz}$ coupling. The weighted average of $Q_{0n}$, which is equivalent to the long-time average of K-complexity via equation (\ref{KC_Q0n}), is plotted as a vertical line for each system. In each case, we plotted also the transition probability and $\overline{C_K}$ for a similar sized complex SYK$_4$ system studied in \cite{Rabinovici:2020ryf}. To make the results clearer we normalized the $x$-axis with respect to the Krylov dimension $K$ for each system separately, as well as normalizing the transition probability in the $y$-axis with respect to $K^{-1}$.
We find that for XXZ the transition probability is biased towards the left side of the chain, whereas for SYK it is almost horizontal and of the order of $K^{-1}$. The left-biased transition probability for XXZ moves the saturation value of $C_K$ to a value smaller than $K/2$, while the almost flat profile of $Q_{0n}$ for SYK gives $\overline{C_K}$ a value close to $K/2$.

\begin{figure}[ht]
    \centering
    \includegraphics[scale=0.7]{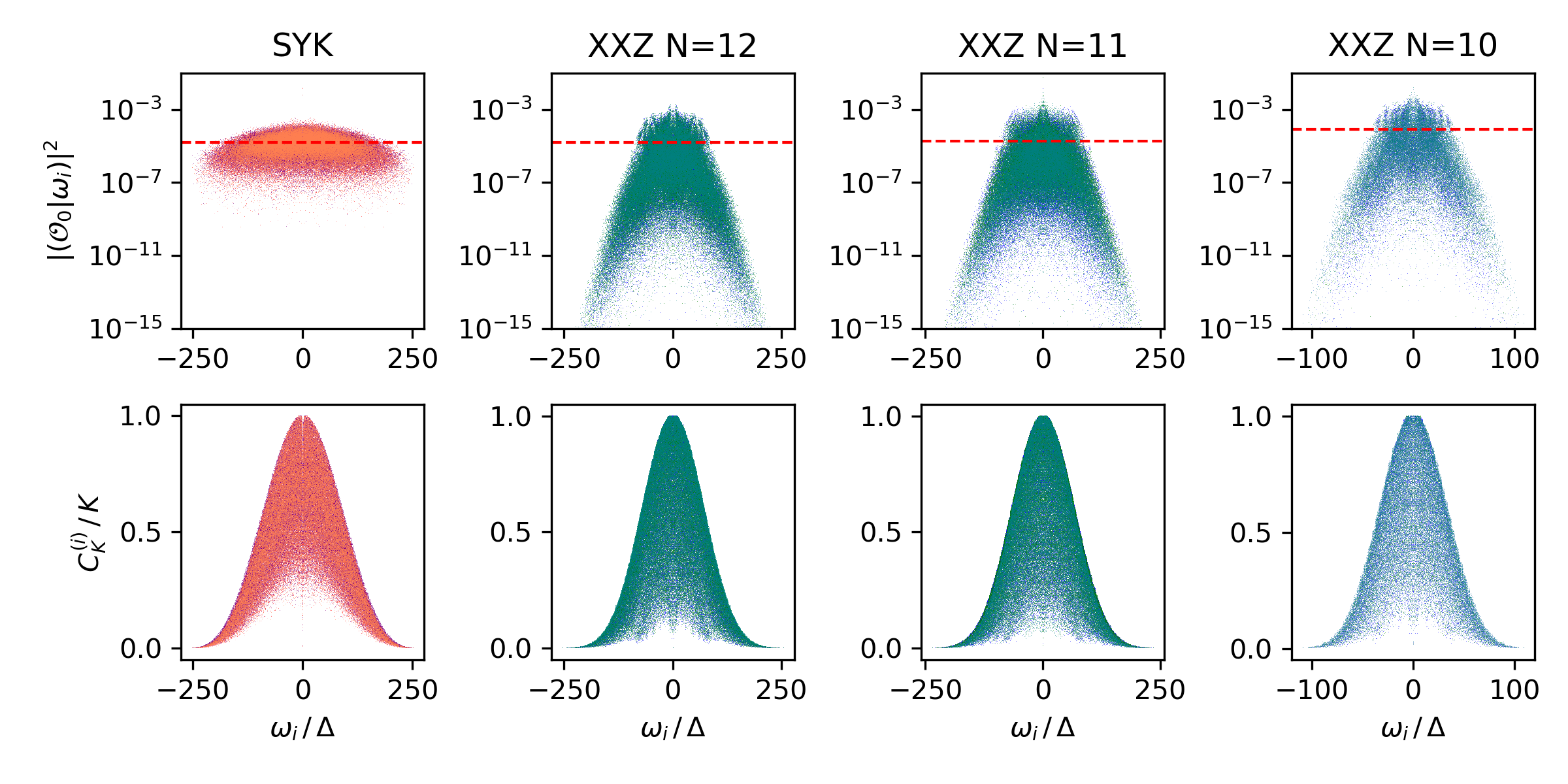}
    \caption{The spectral decomposition of the initial operator (top) and K-complexities of the individual Liouvillian eigenstates (bottom). Data for 4 random realization of cSYK$_4$ with 10 fermions are superimposed in the leftmost column.  Data for 3 choices of $J_{zz}$ coupling for XXZ with $N=12,\, M=4,\, P=+1$ are superimposed in the second column from the left. Data for 3 different choices of $J_{zz}$ couplings for XXZ with $N=11,\, M=5,\, P=+1$ are shown in the third column from the left; and data for 5 different choices of $J_{zz}$ couplings for XXZ with $N=10,\, M=4,\, P=+1$ are shown superimposed in the right column.  The frequencies are normalized according to $\Delta$ which is the mean level spacing computed for for each system separately. The dashed red line represents the value of $K^{-1}$ for each system. }
    \label{fig:KC_KCi}
\end{figure}

\begin{figure}[ht]
    \centering
    \includegraphics[scale=0.7]{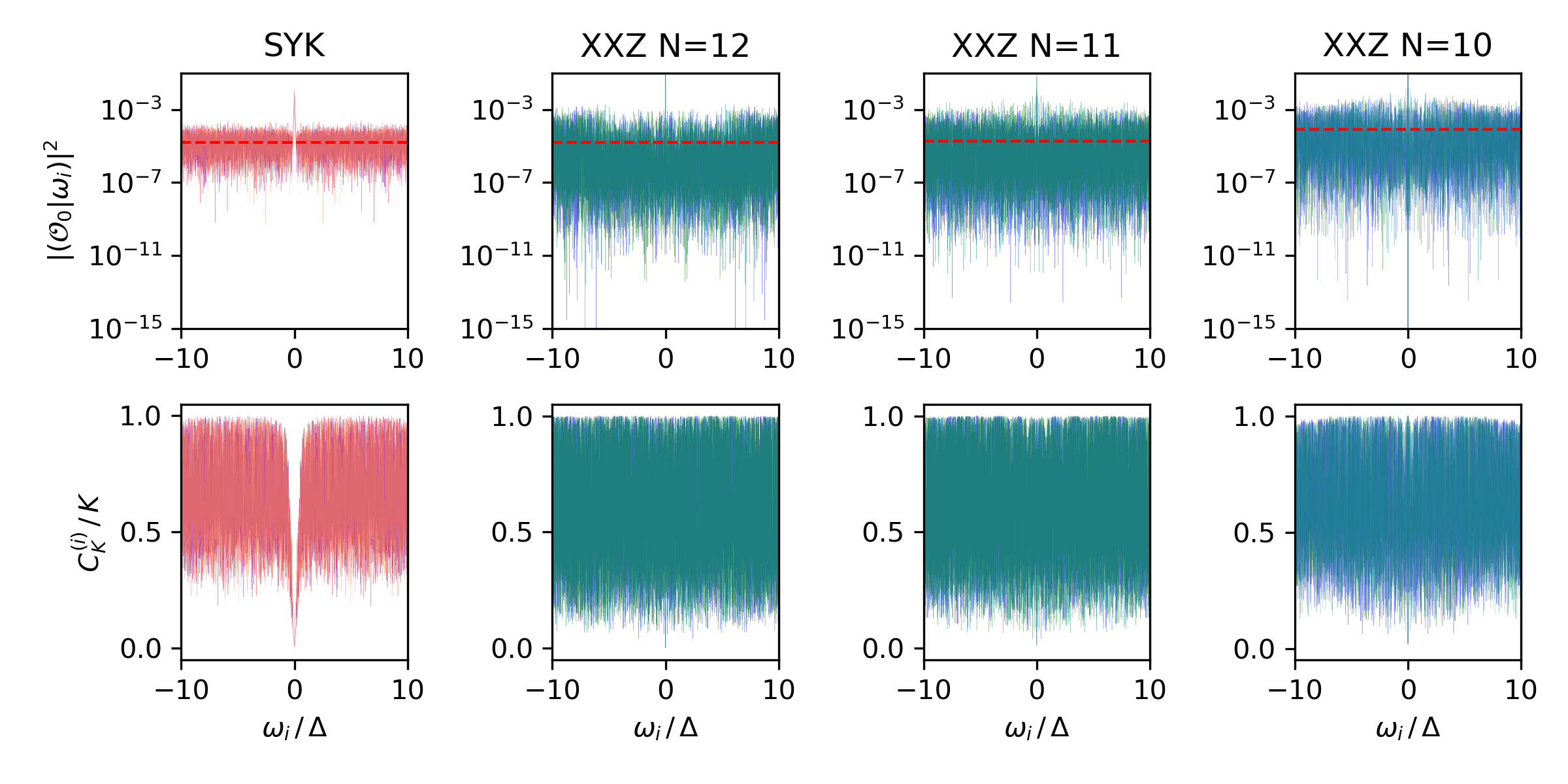}
    \caption{Same results as in Figure \ref{fig:KC_KCi} focused at the band-center. Note that the values of $C_K^{(i)}$ for XXZ reach smaller values than those for SYK.}
    \label{fig:KC_KCi_band_center}
\end{figure}

Equation (\ref{KC_KCi}) expresses the long-time average of K-complexity, $\overline{C_K}$, in terms the overlap of the Liouvillian eigenstates with the initial condition $|(\omega_i|\mathcal{O}_0)|^2$ and the spectral complexities $C_K^{(i)}$. The corresponding quantities computed numerically from XXZ and SYK are presented in Figures \ref{fig:KC_KCi} and \ref{fig:KC_KCi_band_center}. 
We find that two factors cooperate to decrease the value of K-complexity in XXZ compared with SYK:
(1)  In XXZ the spectral profile of the initial operator is peaked towards the band center\footnote{In this work, the term \textit{band center} is taken as a synonym of the \textit{center of the spectrum}.}, while for SYK the profile is flatter and of order $K^{-1}$ as expected from ETH \cite{Srednicki_1999}.  (2) In XXZ, K-complexities of the individual eigenstates near the band center reach smaller values (Figure \ref{fig:KC_KCi_band_center}) due to enhanced localization, compared with those of SYK where most eigenstates have K-complexities of around $K/2$. These eigenstates are particularly important, since the initial condition is localized to the left of the chain and therefore will have a larger overlap with them.  In general, the eigenstates in XXZ which have a stronger overlap with the initial condition, $|(\mathcal{O}_0|\omega_i)|^2$, have smaller $C_K^{(i)}$'s than those of of SYK, thus reducing the value of $\overline{C_K}$ computed via (\ref{KC_KCi}).

\begin{figure}[ht]
    \centering
    \includegraphics[scale=0.7]{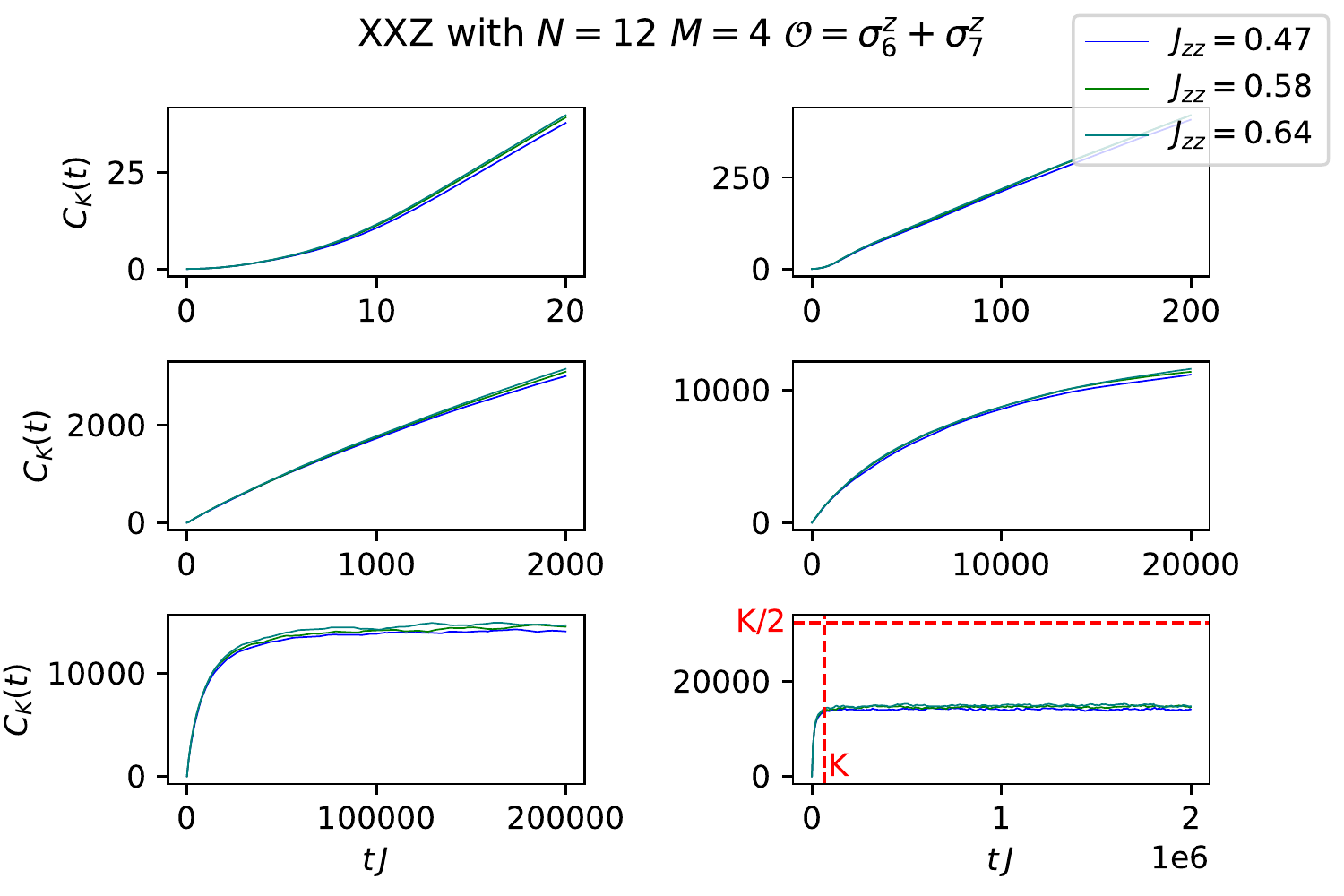}
    \caption{Time-dependent results for K-complexity. Shown here are results for XXZ with $N=12$ in the $M=4,\, P=+1$ sector at 3 different $J_{zz}$ couplings.  Each subplot shows a different time scale.  K-complexity is seen to saturate below $K/2$ at time scales of order $K$ (up to some dimensionful prefactor). }
    \label{fig:KC_time_dependent}
\end{figure}

\begin{figure}[ht]
    \centering
    \includegraphics[scale=0.7]{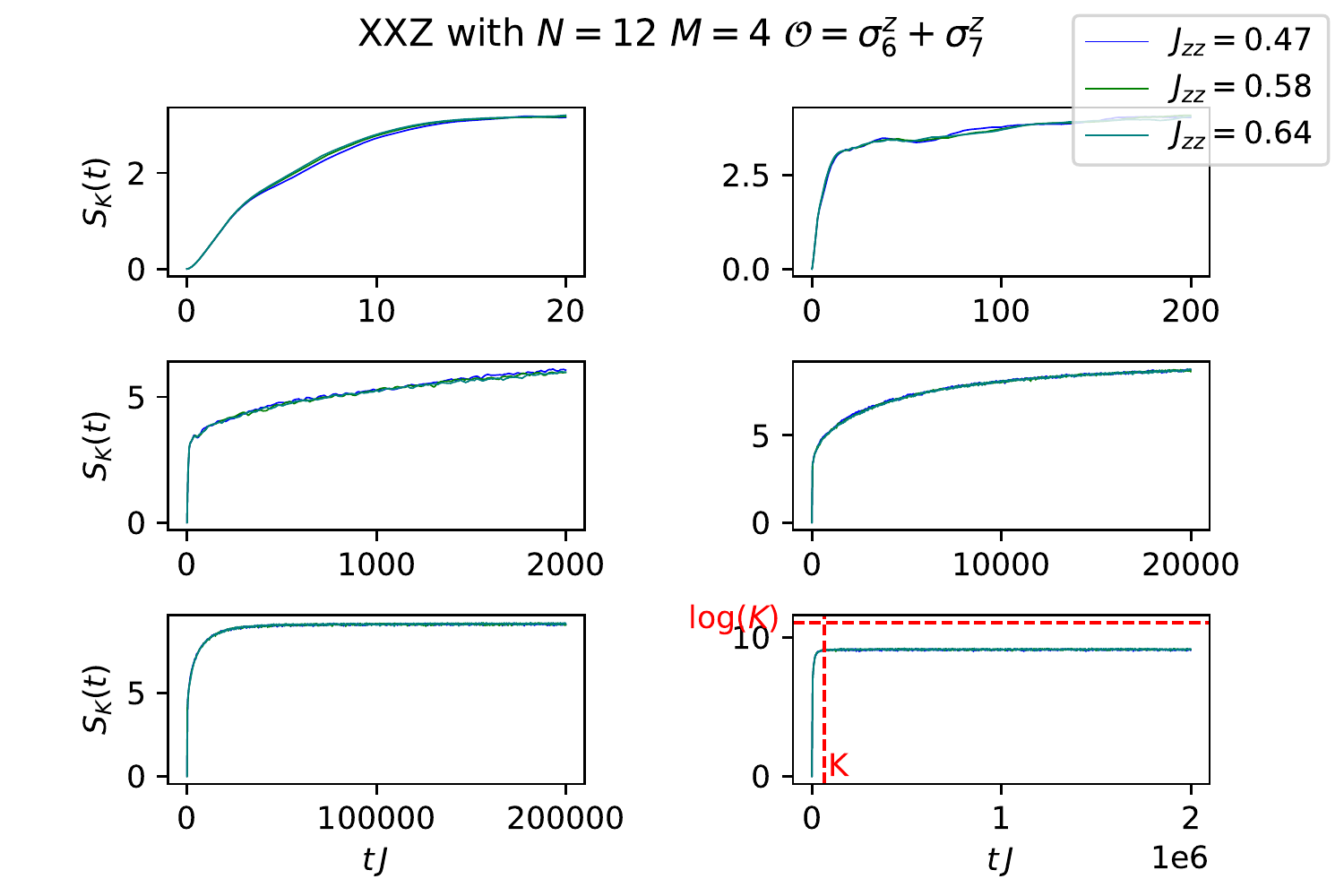}
    \caption{Time-dependent results for K-entropy.  Results for XXZ with $N=12$ in the $M=4,\, P=+1$ sector at 3 different $J_{zz}$ couplings are shown. K-entropy is seen to saturate below $\log(K)\sim N$ at time scales of order $K$. The general lower value of K-entropy is also a sign of localization.}
    \label{fig:KS_time_dependent}
\end{figure}

\begin{figure}[ht]
    \centering
    \includegraphics[scale=0.7]{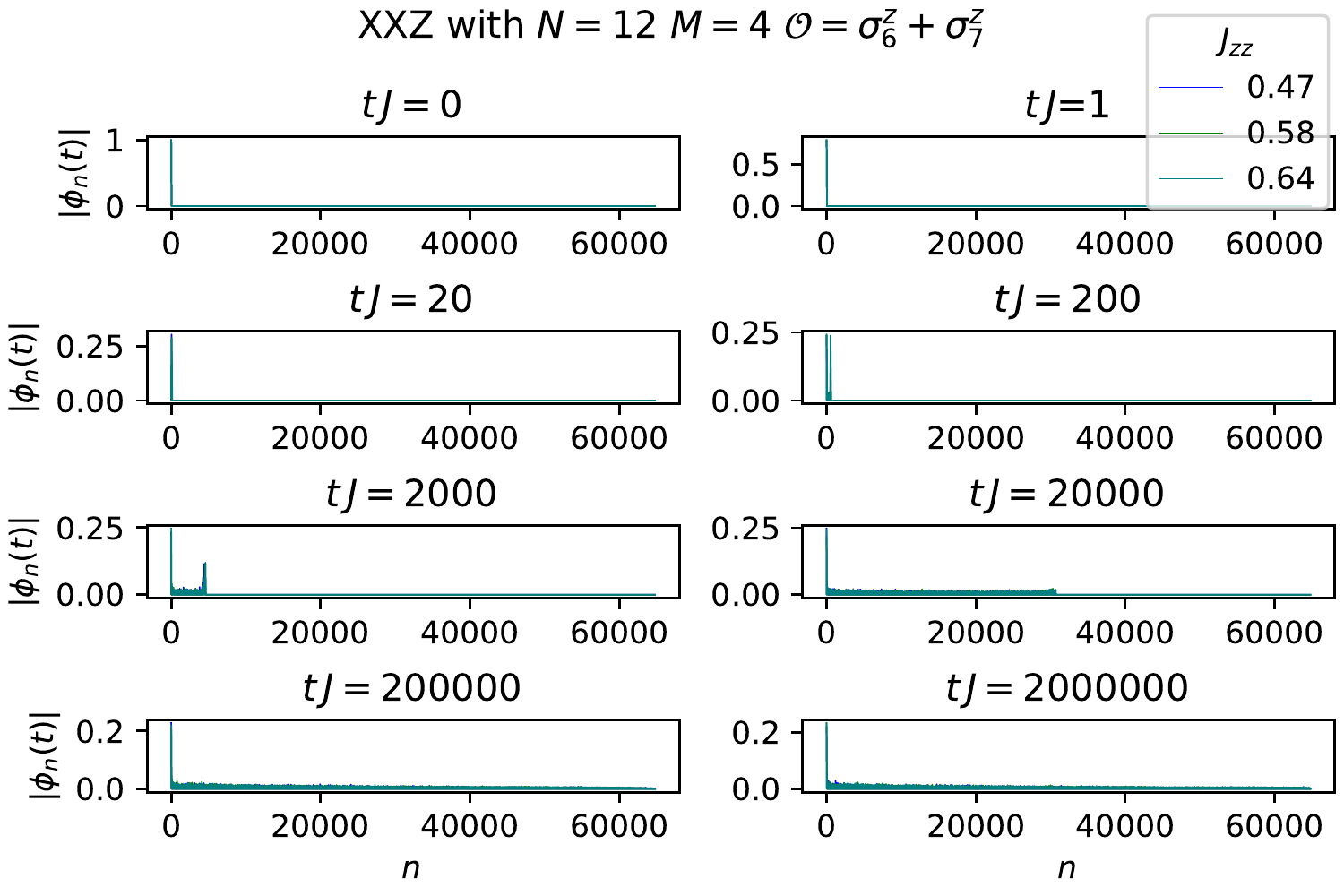}
    \caption{Snapshots of the absolute value of the wave-function at selected times for XXZ with $N=12$ in the $M=4,\, P=+1$ sector at 3 different $J_{zz}$. At $t=0$ the wave-function is a delta function on the first site on the Krylov chain. At late times the wave function  shows clear signs of localization on the first site (i.e. the initial condition), and is in general biased towards the left side of the Krylov chain. }
    \label{fig:Phi_time_dependent}
\end{figure}

Time-dependent results of K-complexity, K-entropy and snapshots of the absolute value of the wavefunction at various time scales are displayed in Figures \ref{fig:KC_time_dependent}, \ref{fig:KS_time_dependent} and \ref{fig:Phi_time_dependent}.  K-complexity is seen to saturate below $K/2$ as predicted by its late-time average. The time scale at which it saturates is of order $K$. Since $K\sim e^{2S}$ we find that the saturation \textit{time} for XXZ is similar to that observed in chaotic models such as SYK.
K-entropy saturates as well at time scales of order $K$, at a value smaller than $\log(K)\sim N$ -- the value signaling total delocalization of the operator wave function as seen in SYK in \cite{Rabinovici:2020ryf} -- which is another indication that the wave function is more localized than in SYK. Again, the saturation \textit{time} is of order $e^{2S}$ as seen in SYK. The snapshots of the wave function at increasing time scales show that, although the wave packet is spreading towards the right of the chain, there is always a stronger support on the initial site of the Krylov chain, and in general some bias towards the left region.

\textbf{Note:} In the case of an even number of spins in the zero magnetization sector $M=N/2$, both $P$ and $R$ preserve total magnetization and in Appendix \ref{Appx_selection_rules_Krylov} we argue that $\mathcal{O}$ of the form (\ref{Study_Operator}) is charged under $R$, which creates a selection rule and reduces the Krylov space dimension. Numerical results and discussion for $N=10,\, M=5$ are given at the end of Appendix \ref{Appx_selection_rules_Krylov}.

\section{Phenomenology of erratic Lanczos sequences}\label{Sect_Toys}

We have computed Lanczos sequences for various instances of XXZ and verified, in agreement with our initial expectation, that they feature disorder superimposed on the ascent-descent profile. Such Lanczos sequences play the role of disordered hopping amplitudes in the Krylov chain, and we observed that for those systems K-complexity saturated at late times at values smaller than half the Krylov space dimension, $\frac{K}{2}$, as a consequence of the long-time averaged transition probability being biased towards the left side of the chain (i.e. the side closer to the localized initial condition). In the framework of Anderson localization due to off-diagonal disorder, it is usually difficult to establish analytically features of time evolution, and customarily one focuses on properties of stationary states \cite{PhysRevB.24.5698,Fleishman_1977,IZRAILEV2012125}. For this reason, in this section we will present numerical computations with heuristically motivated Lanczos sequences whose main ingredients will be an ascent-descent profile with fluctuations on top, and we will show that they reproduce very closely the XXZ results, backing up the claim that the K-complexity phenomenology observed in this model is due to the disorder of its Lanczos coefficients. We will construct such Lanczos sequences for Krylov chains of size $K$ in successive levels of sophistication: First and for reference, we compute transition probabilities and K-complexity resulting from a flat $b$-sequence with fluctuations on top, which is nothing but the canonical setup for off-diagonal disorder; then, we will use two different Ansätze for a more realistic underlying profile displaying ascent and descent regimes. 

\subsection{Pure off-diagonal disorder}\label{Subs_Toy_FLAT}

\begin{figure}
    \centering
    \includegraphics[width=7.5cm]{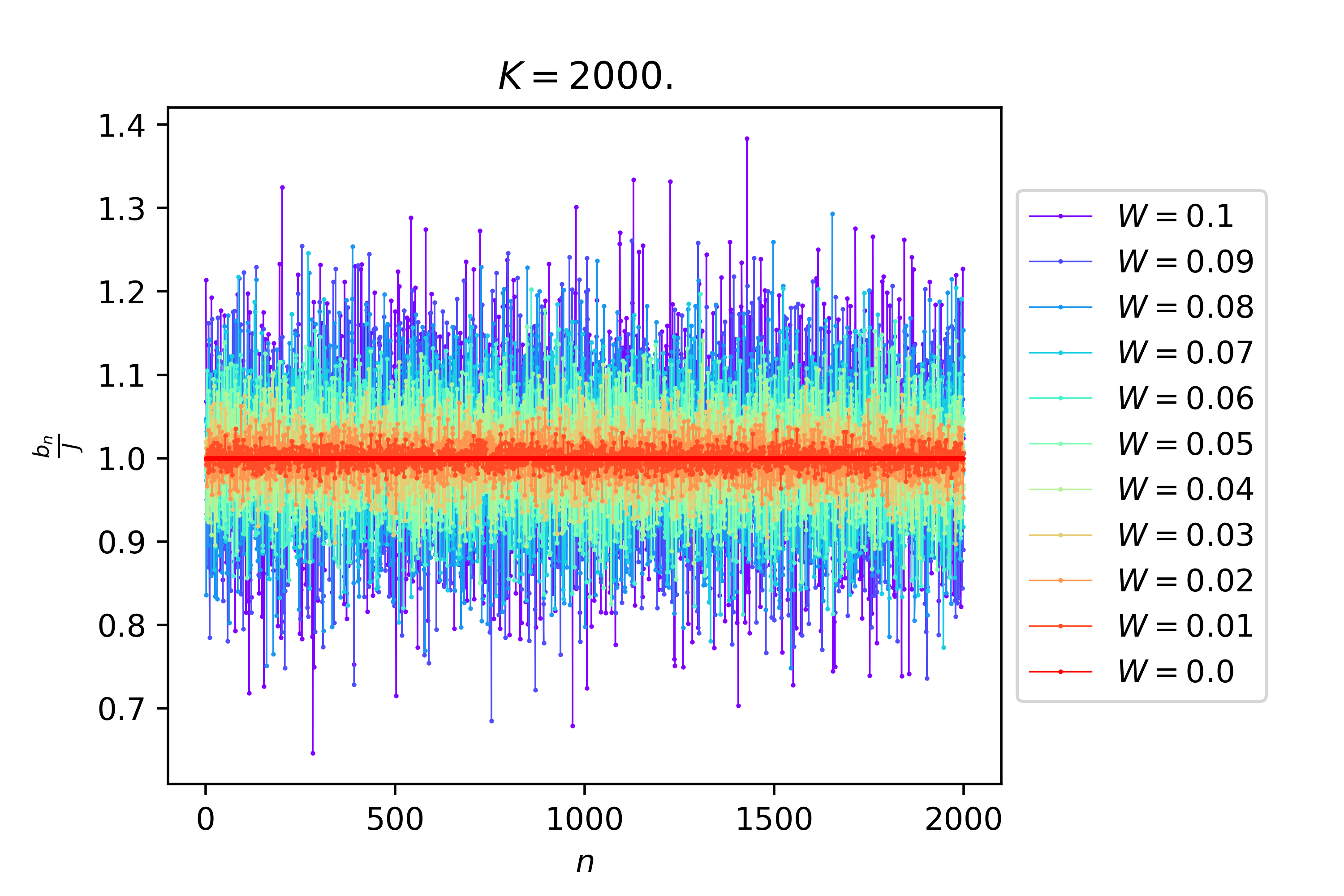} \\
    
    \includegraphics[width=7.5cm]{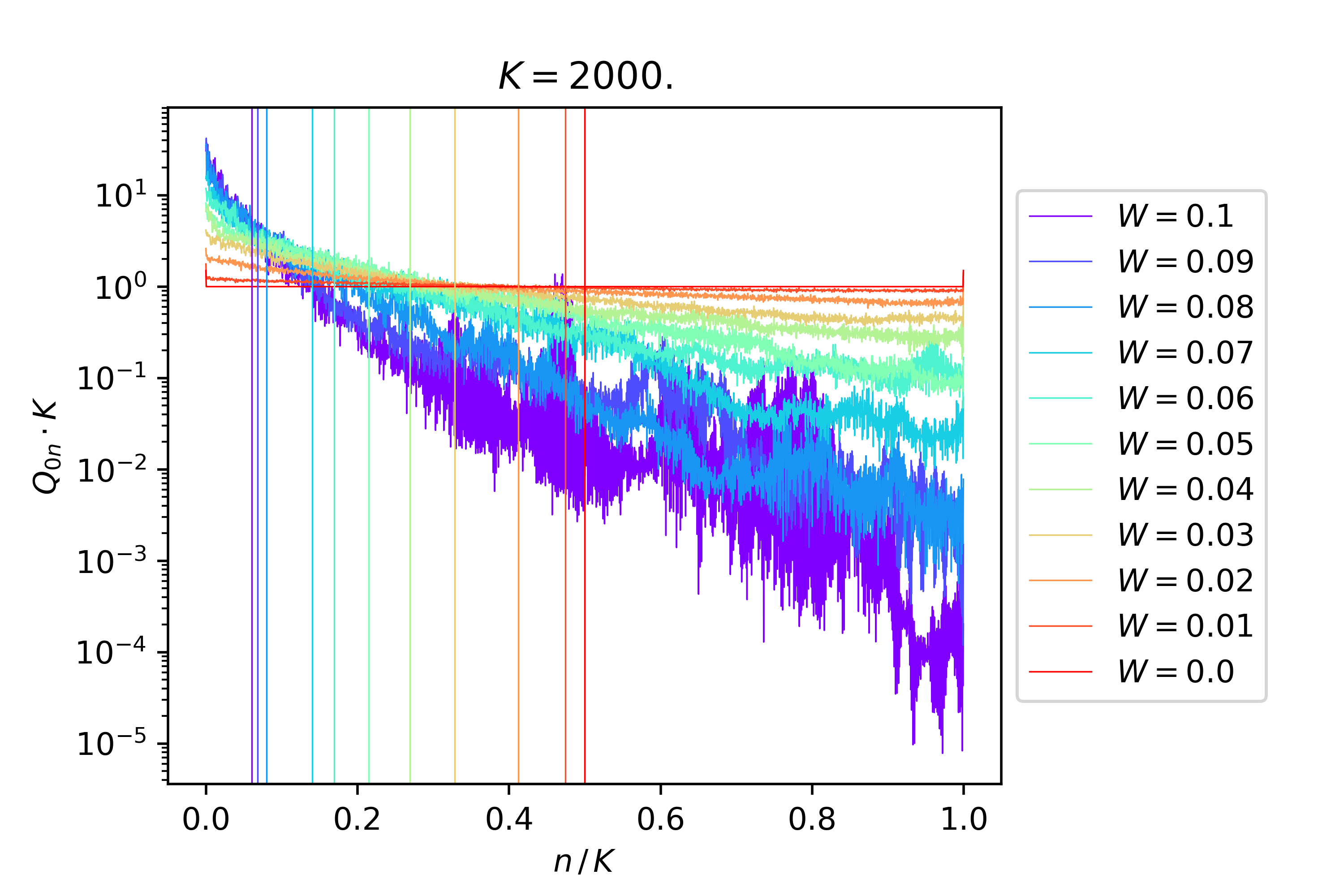} \includegraphics[width=7.5cm]{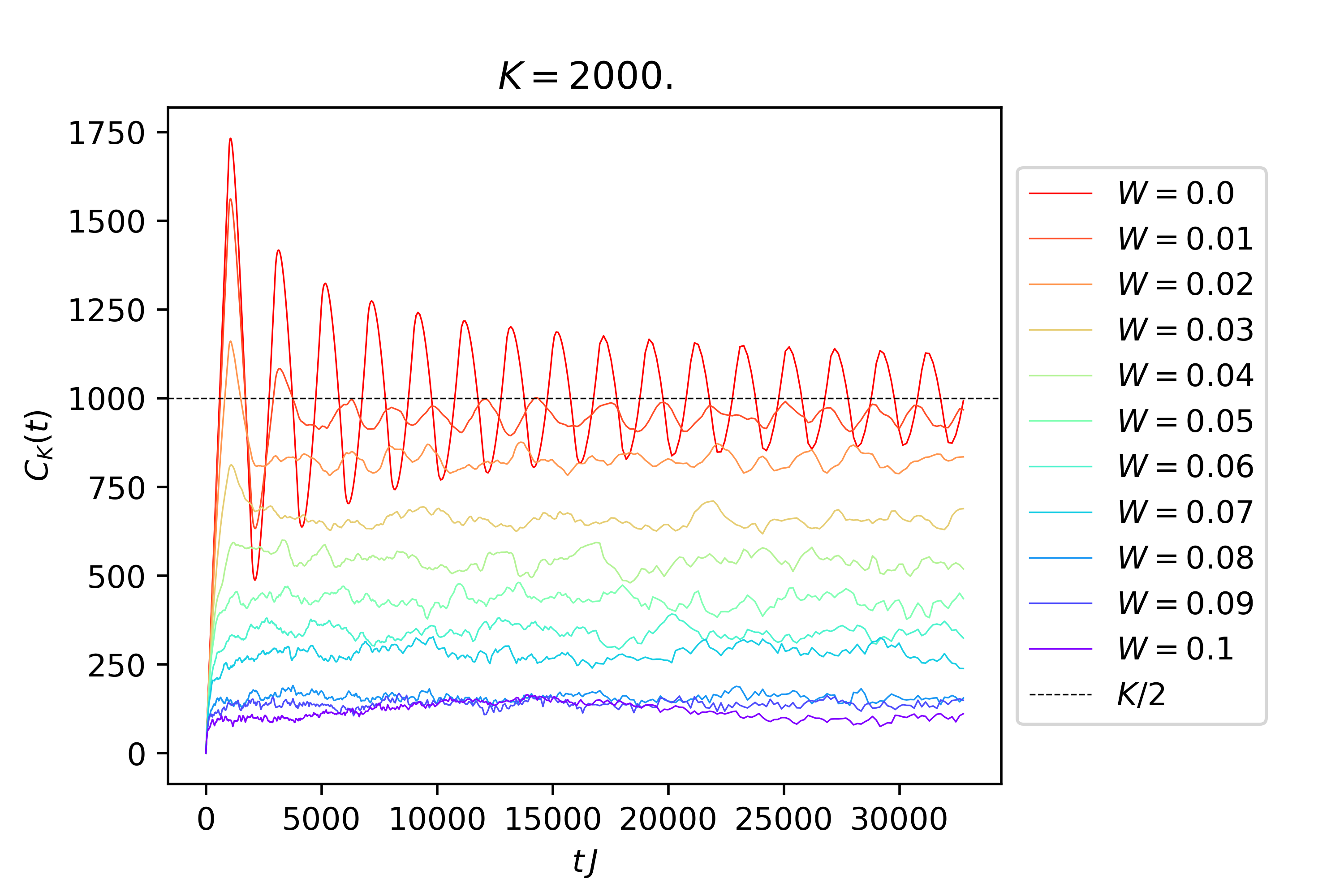}
    
    \caption{\textbf{Top:} Lanczos sequences generated by drawing each Lanczos coefficient from a normal distribution with unit mean and standard deviation $W$, following Subsection \ref{Subs_Toy_FLAT}. Each color corresponds to a single random realization for a fixed disorder strength, as indicated in the legend. In all cases we studied a Krylov chain of length $K=2000$. \textbf{Bottom left:} Transition probabilities computed out of each Lanczos sequence. \textbf{Bottom right:} K-complexity as a function of time.}
    \label{fig:Toy_Disorder_FLAT}
\end{figure}

In this canonical model for off-diagonal disorder, we generate a random sequence of Lanczos coefficients (hopping amplitudes) $\left\{b_n\right\}_{n=1}^{K-1}$ taking all the coefficients to be independent and identically distributed (i.i.d.) Gaussian random variables with mean $J$ and standard deviation $WJ$, where $J$ is to be thought of as a dimensionful parameter setting the energy scale\footnote{Again, it was set to $1$ in the numerics, so that all dimensionful quantities computed are to be thought of as normalized by $J$.}, and $W$ is a dimensionless parameter controlling the disorder strength. Figure \ref{fig:Toy_Disorder_FLAT} depicts various random realizations of Lanczos sequences with different disorder strengths, as well as the corresponding long-time averaged transition probabilities and K-complexities as a function of time computed out of them. We observe that transition probabilities are more and more biased to the left as disorder increases, and in fact decay roughly exponentially, yielding a K-complexity profile that saturates at late times at values that decrease with disorder strength. We note that, in the disorder-free case, K-complexity shows persistent oscillations before saturating at $\frac{K}{2}$, consistent with the wave packet bouncing back and forth between the edges of the Krylov chain before diffusing efficiently into a uniform distribution; in Appendix \ref{appx:Constant_b_analytics} we show analytically that the time-averaged transition probability is indeed a constant (up to edge effects), yielding $\sim \frac{K}{2}$ as the K-complexity long-time average. The addition of disorder contributes efficiently to destroying the coherence of the propagating packet, thus washing out complexity oscillations apart from reducing its saturation value (c.f. Figure \ref{fig:Toy_Disorder_FLAT} - bottom right).

\subsection{Disordered sequence with ascent and linear descent}

\begin{figure}
    \centering
    \includegraphics[width=7.5cm]{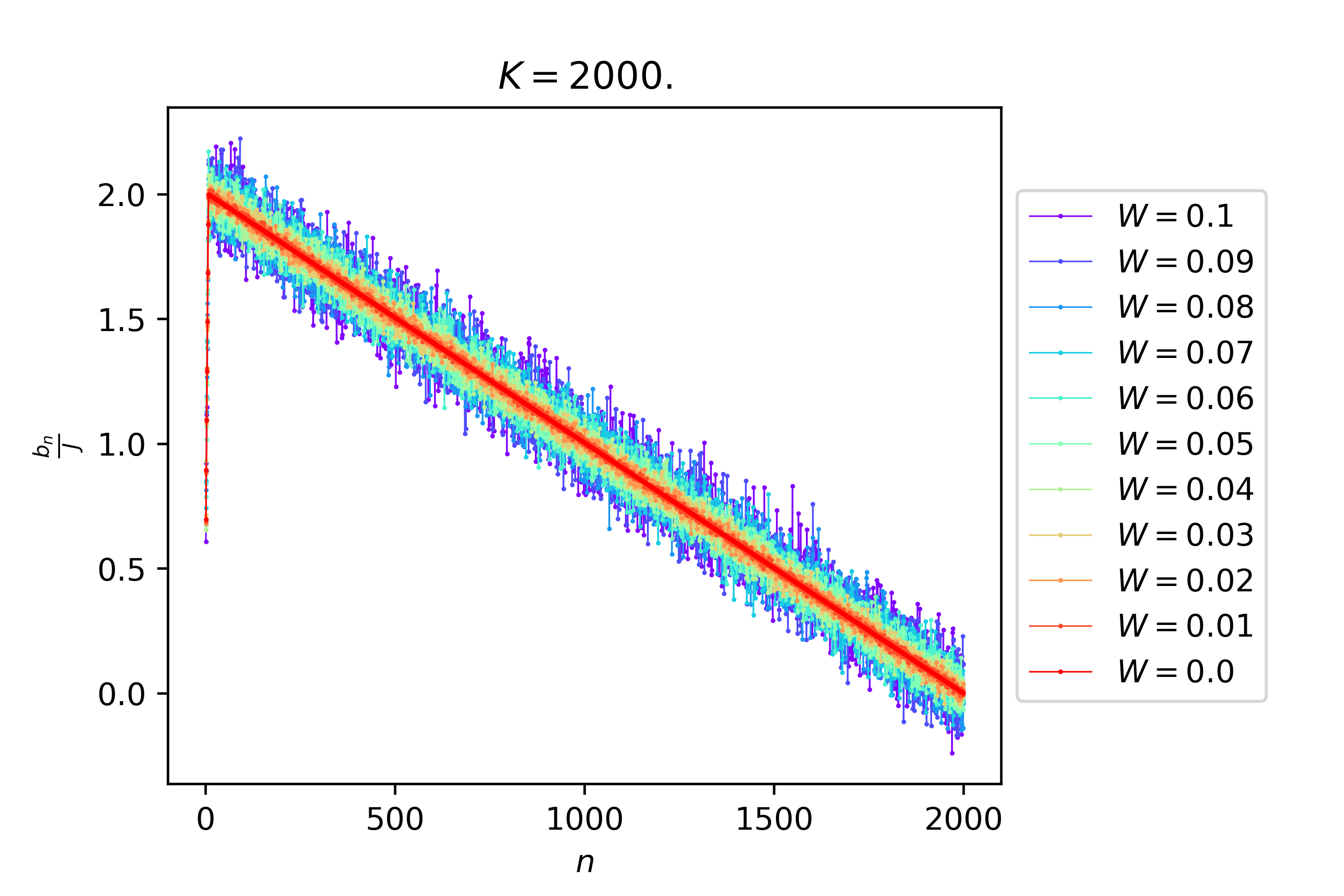} \includegraphics[width=7.5cm]{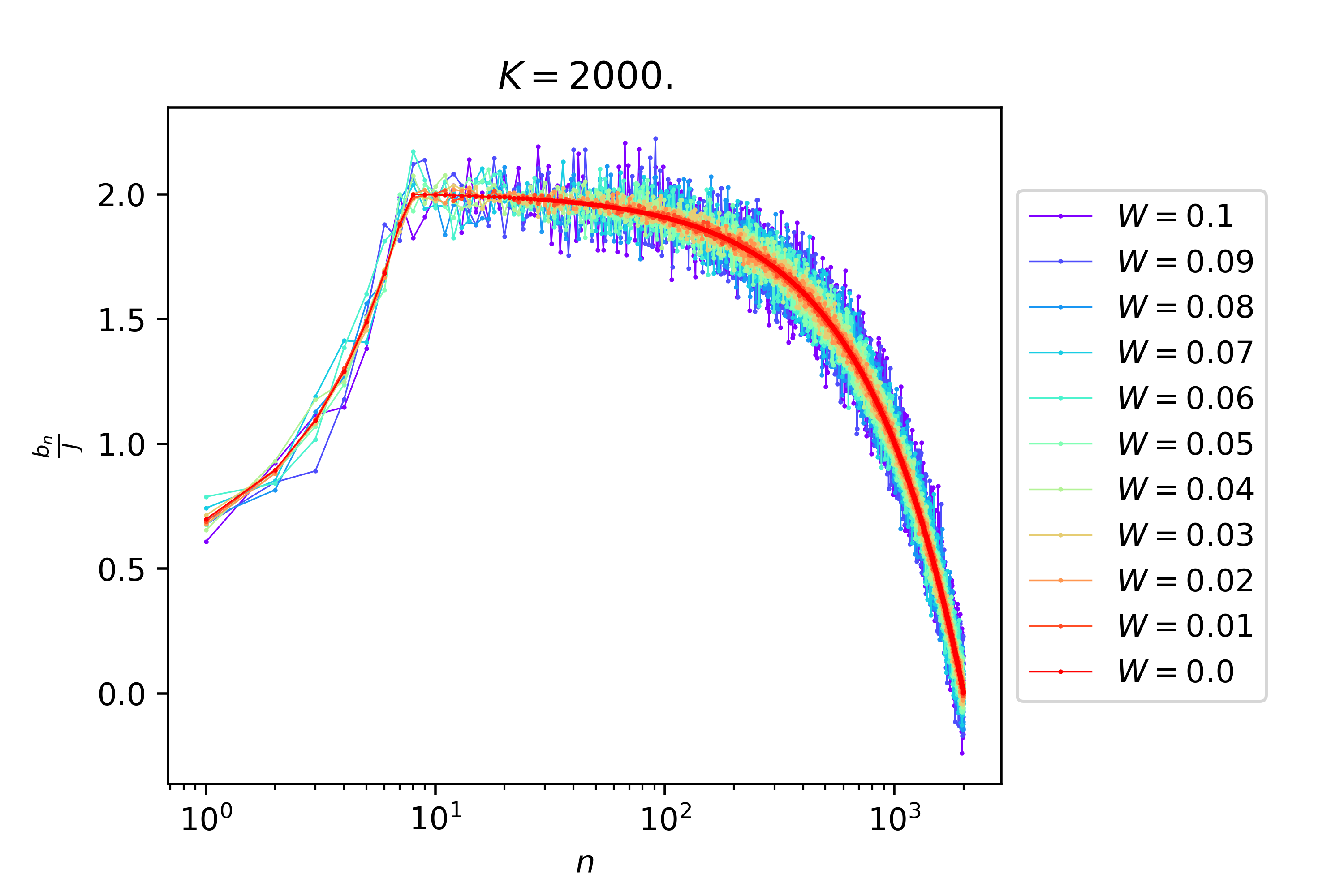} \\
    \includegraphics[width = 7.5cm]{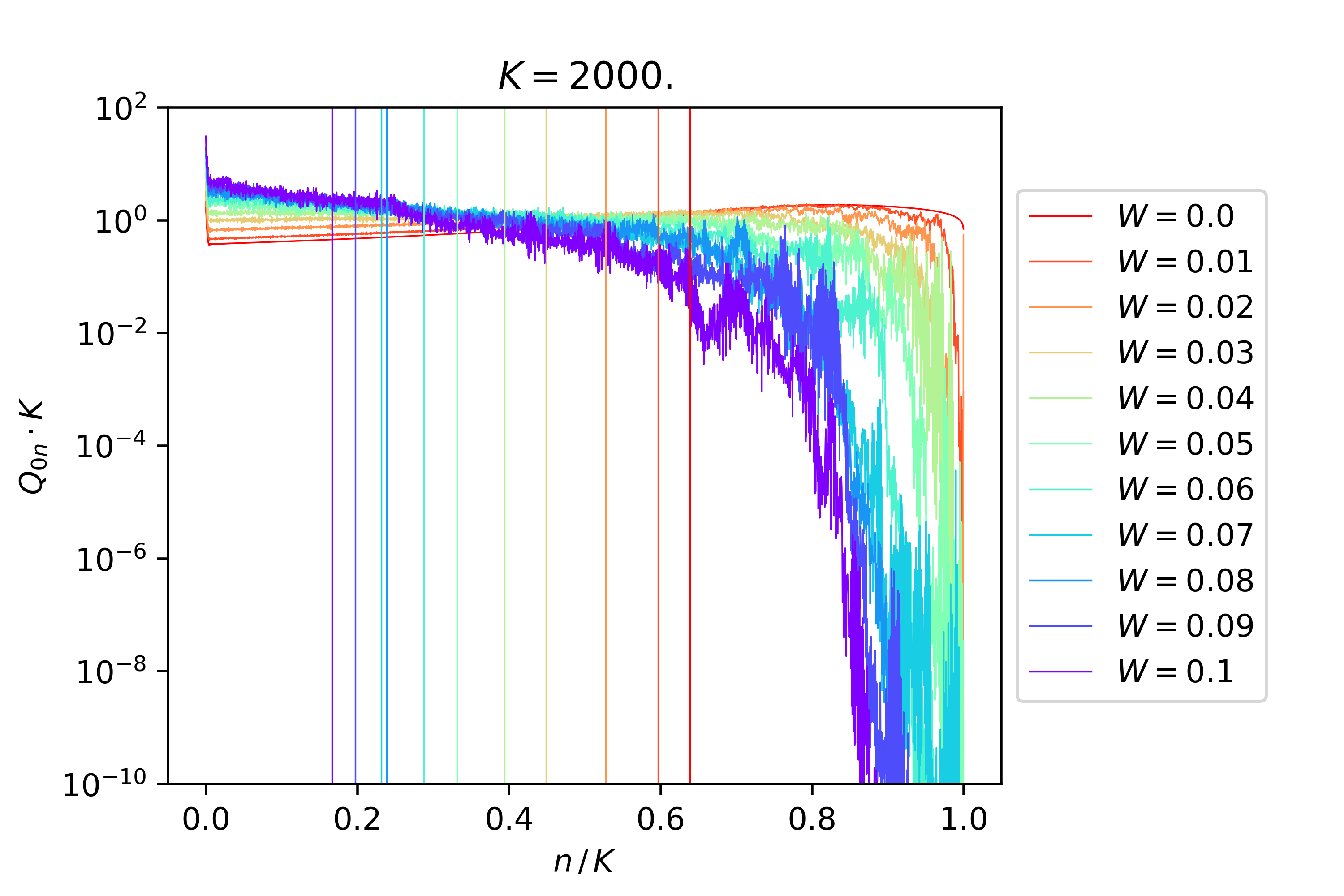} \includegraphics[width = 7.5cm]{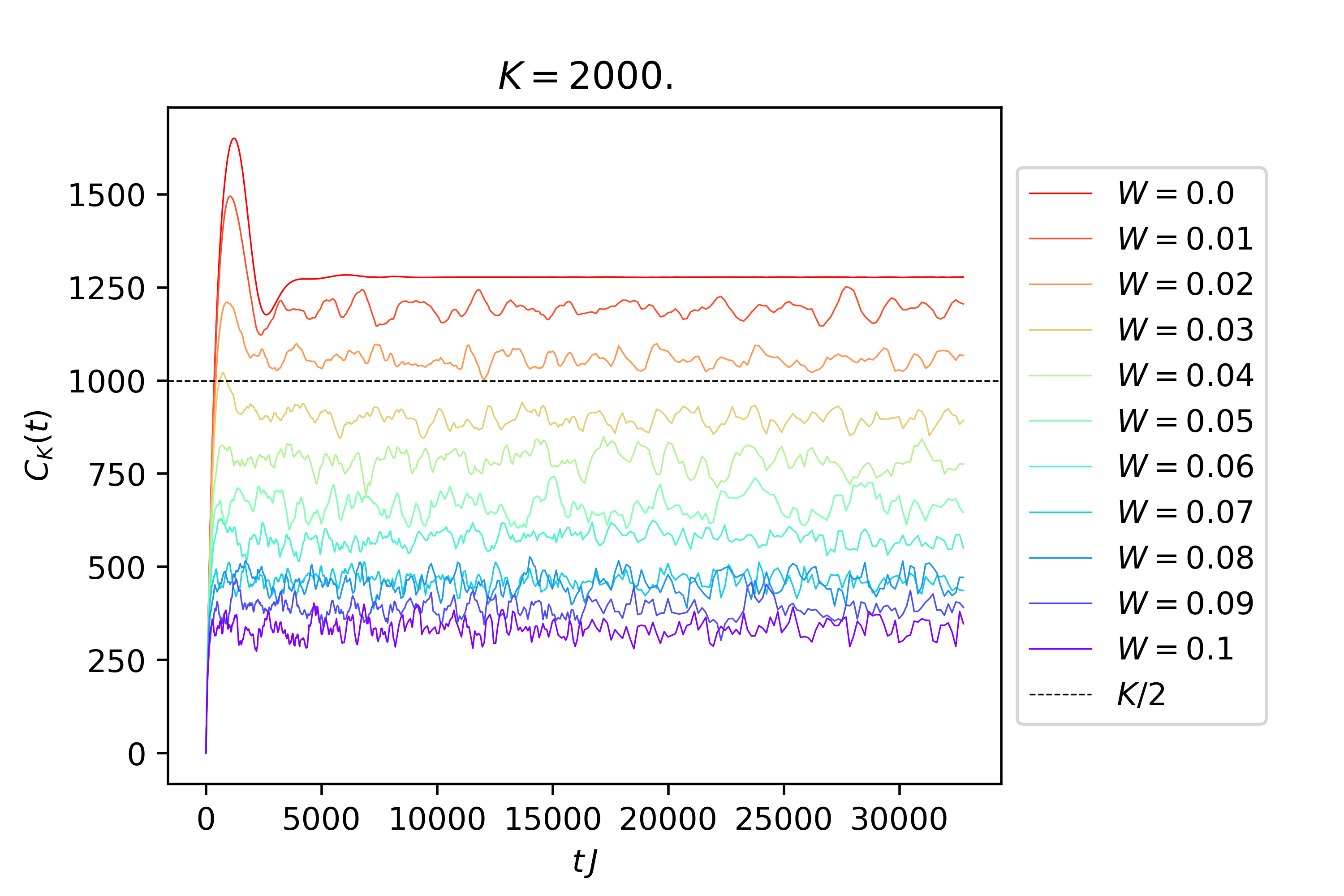}
    \caption{\textbf{Top left:} $b$-sequence for the phenomenological model (\ref{Toy_LIN}). One random realization for different values of the disorder strength is depicted. \textbf{Top right:} Same sequence, plotted with logarithmic scale along the horizontal axis. \textbf{Bottom left:} Time-averaged transition probabilities $Q_{0n}$ computed out of each sequence. As disorder increases they start to follow a profile that decreases with $n$, signaling localization. \textbf{Bottom right:} K-complexity as a function of time for each disorder strength. We observe that the late-time saturation value decreases monotonously with $W$.}
    \label{fig:Toy_LIN}
\end{figure}

This time we add i.i.d. Gaussian fluctuations on top of a Lanczos sequence $b_n$ that increases linearly up to $b_*$ at $n_*\sim \log K$ and then decays linearly to (almost) zero at $n=K-1$. That is:
\begin{equation}
    \centering
    \label{Toy_LIN}
    b_n =  \begin{cases}
        b_0 + \frac{b_*-b_0}{n_*}n + W_n\,J &  0< n < n_*\\
        b_*-\frac{b_*}{K-n_*}(n-n_*) + W_n\,J & n_*\leq n < K
    \end{cases} 
\end{equation}
where, as announced, $W_n$ are i.i.d. Gaussian random variables with zero mean and standard deviation given by the dimensionless parameter $W$. Sequences generated with different disorder strengths as well as the transition probabilities and K-complexities computed out of them are depicted in Figure \ref{fig:Toy_LIN}. The numerical values of $b_0$ and $b_*$ were chosen so that the resulting profiles reassembled those of XXZ\footnote{In fact, for adequately normalized Hamiltonians, $b_*$ should not depend strongly on system size.}.

In the absence of disorder, we observe that K-complexity saturates above $\frac{K}{2}$. This is due to the asymmetric shape of the Lanczos sequence: Since the hopping amplitudes are smaller on the right side of the chain, the packet tends to spend more time in that region on average, as reflected in the transition probability $Q_{0n}$ being slightly biased towards the right. However, as soon as disorder is turned on, the late-time saturation value of complexity drops below $\frac{K}{2}$ to some finite fraction of $K$. This localization phenomenon is also reflected in the fact that the transition probability decays with $n$ at a seemingly exponential rate (even though with a small exponent), for the disorder strengths explored.

One might naively think that under-saturation of K-complexity in the XXZ results is due to the descent in the Lanczos sequence. The model presented here contradicts that statement: We have observed that, in complete absence of disorder, the ascent-descent profile results, conversely, in ``over-saturation'' (in the sense that the late-time average is bigger than half the Krylov dimension $\frac{K}{2}$). It is thanks to the addition of disorder that the late-time complexity value can drop down, as we have explicitly demonstrated (c.f. Figure \ref{fig:Toy_LIN}).

There is one slightly unsatisfactory feature of this phenomenological model: The profile of the descent on top of which disorder is added is linearly decaying, but the variance of the fluctuations doesn't depend on $n$. Hence, the fluctuations become stronger with respect to their mean value as $n$ increases, amplifying the decay of the transition probability. Furthermore, some coefficients towards the end of the sequence can incidentally become very close to zero, effectively ``breaking'' the chain. This all produces a big drop in the transition probability, as can be observed in Figure \ref{fig:Toy_LIN} (bottom left). Localization is, however, still operating throughout the chain, as $Q_{0n}$ decreases with $n$ even on the left side of the chain when disorder is sufficiently strong. A further refinement of the model, to be presented next, tries to get rid of the above-mentioned edge effect by adding some convexity towards the end of the descent.

\subsection{Disordered sequence with ascent and quasi-linear convex descent}

If the descent of the Lanczos sequence had a slightly convex profile, the region where the strength of the fluctuations becomes comparable to the mean value on top of which they are added would be pushed towards the right. In fact, the analytical expression for the Lanczos sequence in RMT at large size has this feature \cite{Kar:2021nbm}: the quasi-linear descent gets eventually modified by a square-root behavior. With this motivation, we use the following Ansatz for the toy Lanczos sequence:

\begin{equation}
    \centering
    \label{Toy_SQRT}
    b_n =  \begin{cases}
        b_0 + \frac{b_*-b_0}{n_*}n + W_n\,J & 0< n < n_*\\
        b_*\sqrt{1 - \frac{n-n_*}{K-n_*}} + W_n\,J & n_*\leq n < K
    \end{cases}~.
\end{equation}
For $n_*<n\ll K$ the descent is linear, but towards the edge $n\sim K$ the square-root behavior becomes dominant.

\begin{figure}
    \centering
    \includegraphics[width=7.5cm]{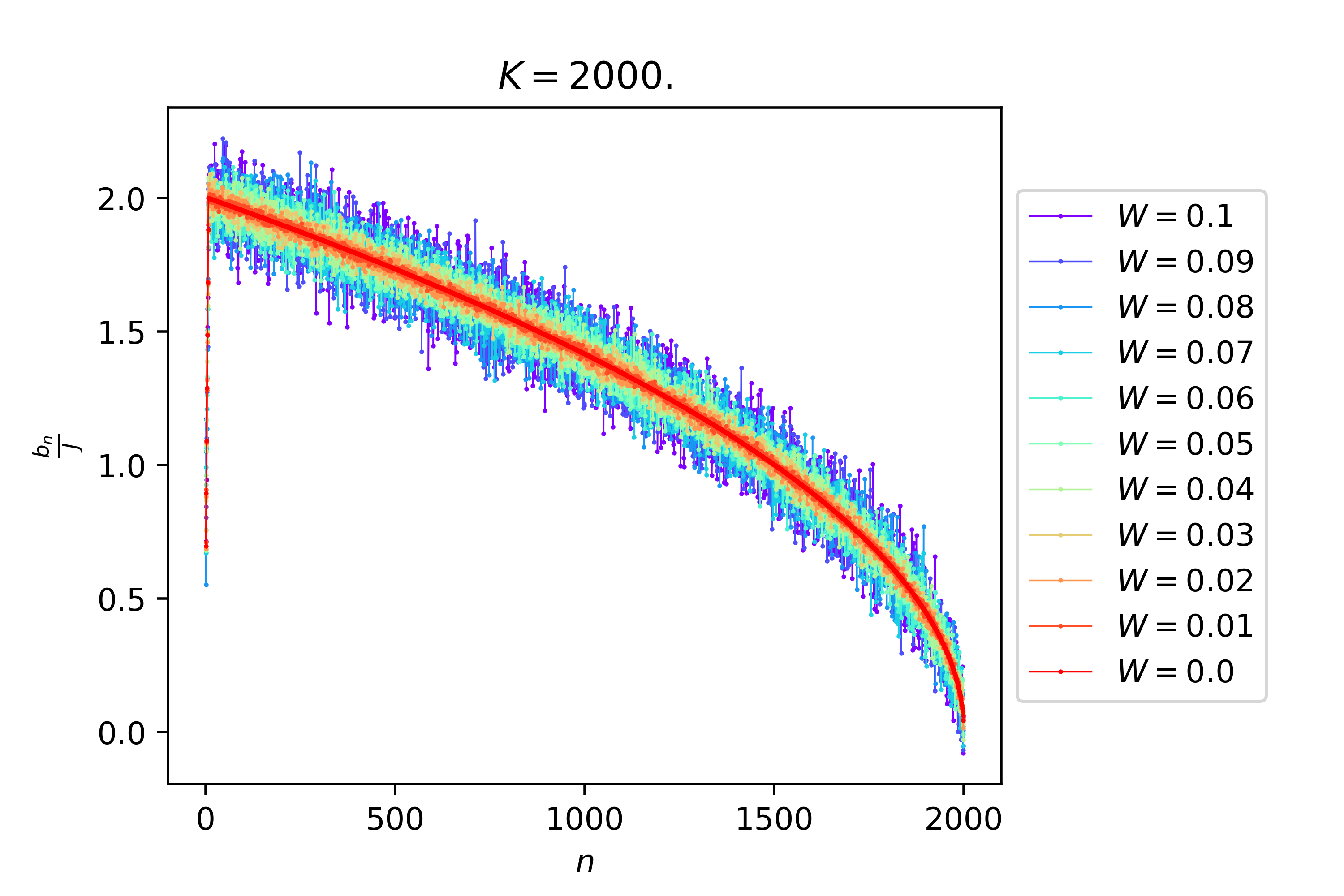} \includegraphics[width=7.5cm]{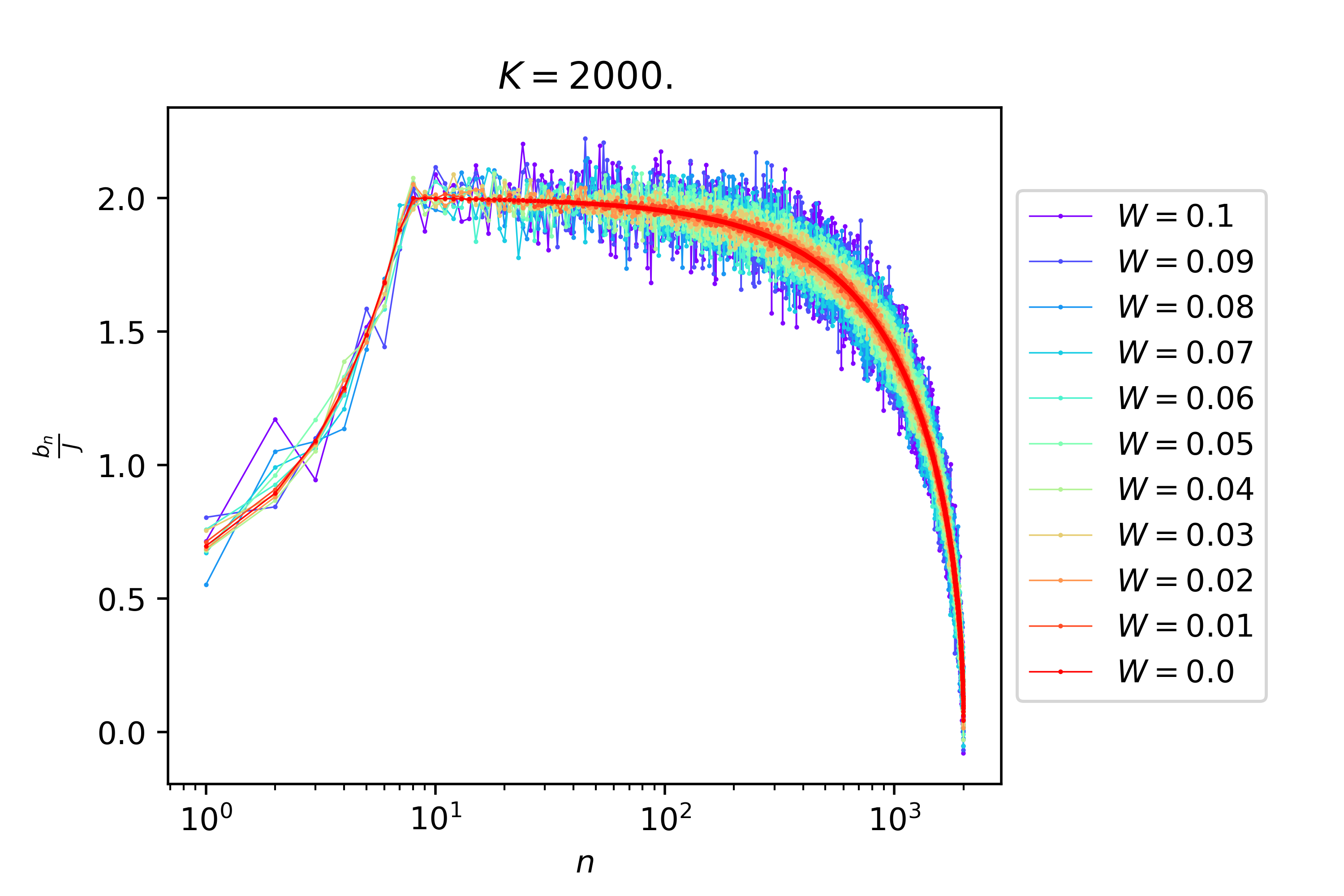} \\
    \includegraphics[width = 7.5cm]{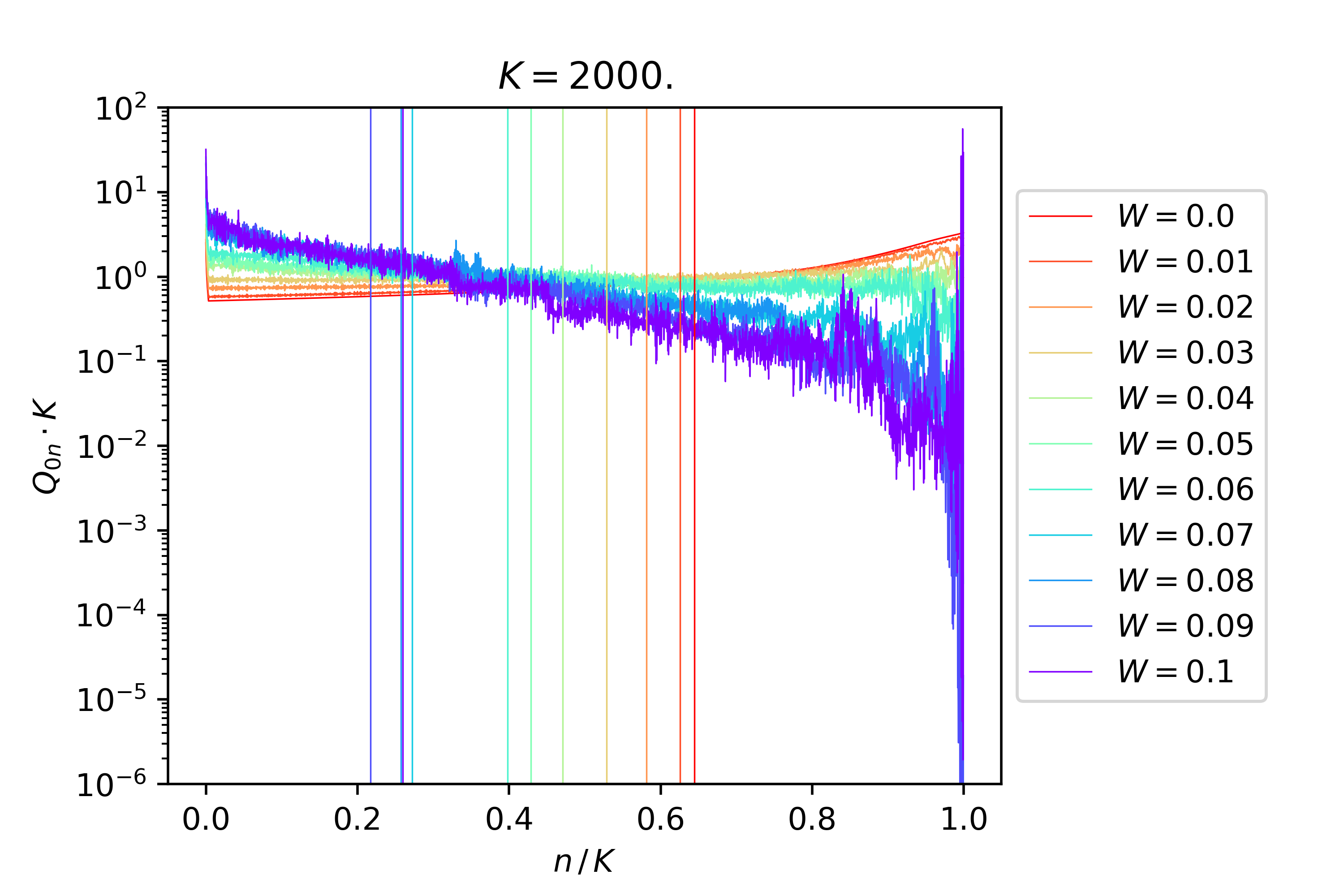} \includegraphics[width = 7.5cm]{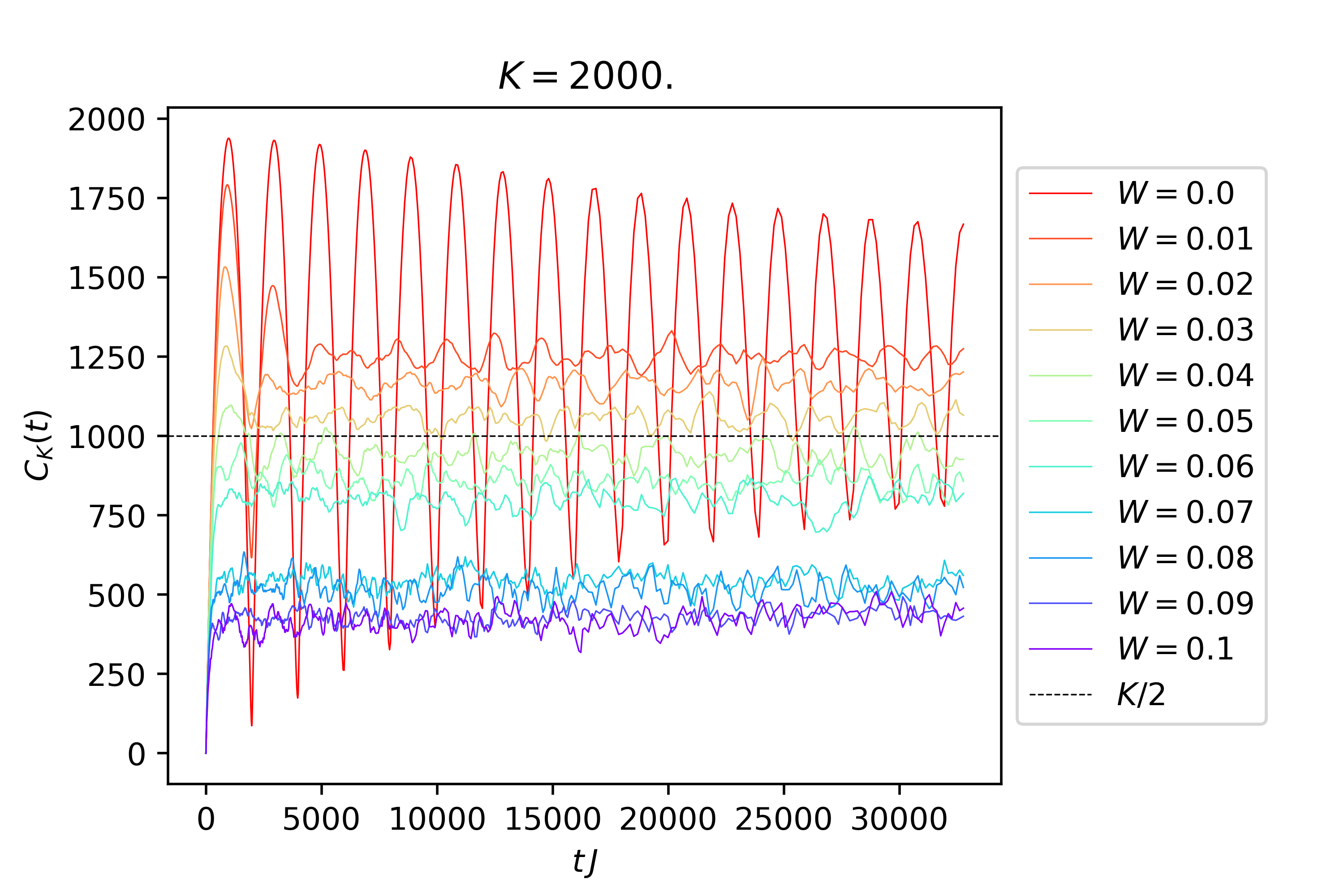}
    \caption{\textbf{Top left:} Lanczos sequences with a slightly convex shape towards the end of the descent, with fluctuations controlled by $W$. \textbf{Top right:} Idem, with logarithmic scale along the horizontal axis. \textbf{Bottom left:} Transition probabilities $Q_{0n}$ computed out of each $b$-sequence. We note the similarity to the results depicted in Figure \ref{fig:Qn0_XXZ_vs_SYK}. \textbf{Bottom right:} K-complexity as a function of time, $C_K(t)$, for each disorder strength. Large oscillations disappear when disordered is turned on, and the complexity saturation value decreases as disorder strength increases.}
    \label{fig:Toy_SQRT}
\end{figure}

Figure \ref{fig:Toy_SQRT} depicts the generated $b$-sequences as well as $Q_{0n}$ and $C_K(t)$. Their striking similarity to the XXZ results presented earlier in this paper backs up the claim that disorder is responsible for the obtained transition probabilities and the lower saturation value of K-complexity. A weak disorder $W=0.1$ is able to reduce the complexity saturation value from above $\frac{K}{2}$ to $\sim\frac{K}{4}$.

It is, again, interesting to note that the total absence of disorder would imply long standing coherent oscillations of complexity and a saturation value slightly above $\frac{K}{2}$ due to the asymmetric profile of the $b$-sequence that has already been discussed. This feature should remain there at larger system sizes, even though less pronounced. Hence, the fact that in a single random realization of SYK K-complexity does saturate around $\frac{K}{2}$ (in fact, even a bit below) and doesn't feature strong oscillations (c.f. \cite{Rabinovici:2020ryf}) indicates that even in this system the small disorder of its Lanczos sequence is operating. From this perspective, the difference between SYK and XXZ is more quantitative than qualitative: The $b$-sequence of the latter is more disordered than that of the former, but they both display localization in the Krylov chain to some extent. This is consistent with usual discussions in Anderson localization: in one spatial dimension there is no critical disorder strength, and this phenomenon is always present for any value of the variance of the fluctuations, but the localization length can vary drastically.

\section{Conclusions}\label{Sect_conclusions}

In this work we studied the behavior of K-complexity for local operators in the XXZ spin chain, a strongly-interacting integrable system solvable via the Bethe Ansatz. We found that the associated Krylov space dimension is exponential in system size, however the disorder in the sequence of Lanczos coefficients induces a localization effect that prevents maximally efficient exploration of the Krylov chain by the operator wave function, resulting in a saturation value of K-complexity at late times below $\frac{K}{2}$. It should be noted that the starting Hamiltonian, i.e. an XXZ spin chain with constant coupling strength, was not defined through a set of disordered couplings: disorder appears in the Lanczos sequence and it is enhanced by the Poissonian statistics of its spectrum; localization operates therefore in Krylov space rather than in direct space. Results also indicate that the difference between XXZ and SYK is more quantitative than qualitative: in both cases there is some disorder in the Lanczos sequence, but it is stronger in the integrable model; since the dynamics on the Krylov chain are described by a one-dimensional hopping problem, where there is no critical value for the disorder strength, localization should be operating in both cases, to a different extent in each of them.  In turn, the enhanced disorder in the Lanczos sequences for XXZ causes a `tilt' in the transition probability from the initial operator $|\mathcal{O}_0)$ to $|\mathcal{O}_n)$ as can be seen both in our numerical results for XXZ (Figure \ref{fig:Qn0_XXZ_vs_SYK}) as well as in our phenomenological models, thus reducing the value of the long-time average of K-complexity. This is also manifest in the presence of Liouvillian eigenstates with smaller K-complexities in XXZ, especially around the band center (Figures \ref{fig:KC_KCi} and \ref{fig:KC_KCi_band_center}), which is again due to stronger localization as a result of the larger disorder in the Lanczos sequences.

More work needs to be done in order to determine how generic this phenomenon is. In particular, it is necessary to formalize more strictly the relation between Poissonian statistics (and hence, integrability) and the erratic structure of the Lanczos coefficients, in a way that allows to predict the strength of the fluctuations of the $b$-sequence from data of the initial Hamiltonian (and perhaps the operator). Eventually, it would also be enlightening to push this towards an analytical estimate for the saturation value of K-complexity, which so far can only be accessed numerically. It should be noted, however, that this last point translates to a generically complicated problem in the conventional framework of Anderson localization.

Last but not least let us comment on the relationship between our findings and what role complexity may play in holography, the very question that has been the source of motivation to study the phenomenology of Krylov complexity in detail in this work. As we mentioned in the introduction it is of course of great interest to push ahead in the direction of identifying precise notions of complexity on either side of the duality, in an attempt to solidify the corresponding dictionary entries. It was our goal to add to the detailed understanding of K-complexity and its associated physics, as has been summarized above. Let us now venture into more speculative territory, by addressing a few puzzles and potential resolutions raised by our results.

Going beyond the necessity to juxtapose integrable and chaotic behavior in order to understand what properties of complexity can be associated unambiguously to chaotic operator spreading, and which ones cannot, general holographic wisdom would suggest that we are only interested in the behavior of K-complexity in chaotic systems. This is in accordance with the general expectation that semi-classical black holes are thermodynamic systems and the dynamics of the theory is (maximally) chaotic, as evidenced, for example by the maximal value of the OTOC-Lyapunov exponent \cite{Shenker:2013pqa}, which is determined by the red-shift factor of the classical black hole. On the other hand, there is ample evidence that the field-theory side of the duality (e.g. $\mathcal{N}=4$ SYM in $d=3+1$) is quantum integrable -- at least in the planar sector, and for certain classes of observables \cite{Beisert:2010jr}. Accepting, for argument's sake, that this integrability extends to the quantities of interest in this paper, this raises a conundrum: how can both of these statements be correct at the same time? One potential route out of this impasse lies in timescales. Our methods and results have targeted very large timescales, exponential in the entropy, where we have found quantitative and qualitative differences between integrable and chaotic behavior. The semiclassically chaotic behavior of the bulk black hole is manifest at much earlier timescales, around the scrambling time, so that our results (leaving aside the obvious fact that we are talking about the XXZ spin chain, and not $\mathcal{N}=4$ SYM) are not in immediate conflict with the early growth of complexity corresponding to the black hole geometry. It would therefore be of great interest to study further the issue of earlier ($\sim$ scrambling-) time complexity dynamics in order to see whether any significant changes can be discerned between chaotic and integrable cases. In this regard it is interesting to remark that \cite{Dymarsky:2021bjq} presents evidence that the early growth phase of K-complexity is not in fact a reliable indicator of quantum chaos. On the bulk side it would then be highly interesting to elucidate what other geometries, besides the black holes contribute at later time scales \cite{Maldacena:2001kr, Barbon:2003aq, Saad:2018bqo}, and whether those geometries can account for the difference between chaotic and integrable cases observed in this work. Recent work \cite{Iliesiu:2021ari} has investigated a bulk proposal of the late-time non-perturbative behavior of complexity in chaotic two-dimensional gravitational systems. It would be very interesting to see if a similar analysis can be made in integrable cases and whether our suppressed complexity saturation could be seen from the bulk.\\

\textit{\underline{Note}: during the final stages of preparation of this paper, the related article \cite{Trigueros:2021rwj} appeared on arXiv. It studies localization on the Krylov chain of disordered systems featuring many-mody localization, in the thermodynamic limit. They find that, in the MBL phase, the operator wave function is localized on the Krylov chain and that the growth of K-complexity is suppressed. While they studied the thermodynamic limit of systems with disordered coupling strengths, we have considered late-time dynamics of systems with a fixed, spatially uniform coupling strength, where the Lanczos coefficients appear to be disordered due solely to their integrable nature. We thank the authors of \cite{Trigueros:2021rwj} for drawing our attention to their work.} 

%These findings for integrable systems in the thermodynamic limit are complementary to our results for finite integrable systems.  

\acknowledgments
We would like to thank Dmitry Abanin, Michael Ben-Or, Pietro Pelliconi and Tom\'{a}s Reis for enlightening discussion. The numerical computations were performed on the Landau cluster at the Hebrew University and on the Baobab HPC cluster at the University of Geneva. This work has been partially supported by the SNF through Project Grants 200020 182513, as well as the NCCR 51NF40-141869 The Mathematics of Physics (SwissMAP). The work of ER and RS is partially supported by the Israeli Science Foundation Center of Excellence.

\appendix

\section{Moments and Hankel determinants} \label{Appx_Hankel_Det}
In this appendix we show how to arrive at (\ref{Dn}) starting with (\ref{Hankel_Det}).  Firstly, let us look at 
\begin{eqnarray}
    D_{K-1} &=& \begin{vmatrix} 
    \mu_0 & \mu_1 & \mu_2 & \dots & \mu_{K-1} \\
    \mu_1 & \mu_2 & \mu_3 & \dots & \mu_{K}  \\
    \vdots & \vdots & \vdots & \dots & \vdots \\
    \mu_{K-1} & \mu_K & \mu_{K+1} & \dots & \mu_{2K-2}
    \end{vmatrix} ~.
\end{eqnarray}
Given (\ref{moments}), the matrix in the determinant can be written as
\begin{eqnarray} \label{DK-1}
    \begin{pmatrix} 
    1 & \sum_i |O_i|^2 \omega_i & \sum_i |O_i|^2 \omega_i^2 & \dots & \sum_i |O_i|^2 \omega_i^{K-1} \\
    \sum_i |O_i|^2 \omega_i & \sum_i |O_i|^2 \omega_i^2 & \sum_i |O_i|^2 \omega_i^3 & \dots & \sum_i |O_i|^2 \omega_i^K  \\
    \vdots & \vdots & \vdots & \dots & \vdots \\
    \sum_i |O_i|^2 \omega_i^{K-1} & \sum_i |O_i|^2 \omega_i^K & \sum_i |O_i|^2 \omega_i^{K+1} & \dots & \sum_i |O_i|^2 \omega_i^{2K-2}
    \end{pmatrix}\,,
\end{eqnarray}
where we have assumed the operator is normalized $\sum_{i=0}^{K-1}|O_i|^2=1$. (\ref{DK-1}) can be decomposed as
\small{\begin{eqnarray}
    \begin{pmatrix} 
    1 & 1  &  1 & \dots & 1 \\
    \omega_0 &  \omega_1 & \omega_2 & \dots & \omega_{K-1}  \\
    \vdots & \vdots & \vdots & \dots & \vdots \\
     \omega_0^{K-1} &  \omega_1^{K-1} &   \omega_2^{K+1} & \dots & \omega_{K-1}^{K-1}
    \end{pmatrix}
    \begin{pmatrix}
    |O_0|^2 & & & \\
     & |O_1|^2 & & \\
     & & \ddots & \\
     & & & |O_{K-1}|^2
    \end{pmatrix}
    \begin{pmatrix} 
    1 & \omega_0  &  \omega_0^2 & \dots & \omega_0^{K-1} \\
    1 &  \omega_1 &  \omega_1^2 & \dots &  \omega_1^{K-1} \\
    \vdots & \vdots & \vdots & \dots & \vdots \\
     1 & \omega_{K-1} &  \omega_{K-1}^2 & \dots &  \omega_{K-1}^{K-1}
    \end{pmatrix} \nonumber\,.
\end{eqnarray}}
In this form the determinant is immediate 
\begin{equation}
    D_{K-1} = \prod_{i=0}^{K-1} |O_i|^2 \prod_{0\leq i < j \leq K-1} (\omega_j-\omega_i)^2\,,
\end{equation} 
where the final product is the square of the Vandermonde determinant of the Liouvillian frequencies $\{\omega_i\}_{i=0}^{K-1}$.

Hankel determinants $D_n$ with $n<K-1$ are a bit more complicated since the Vandermonde matrices in this case are not square. The matrix in the determinant of $D_n$, is the product of matrices with sizes $(n+1\times K)$ and $(K \times n+1)$, namely
\small{\begin{eqnarray}
    \begin{pmatrix} 
    |O_0|^2 & |O_1|^2  &  |O_2|^2 & \dots & |O_{K-1}|^2 \\
    |O_0|^2\omega_0 &  |O_1|^2\omega_1 & |O_2|^2\omega_2 & \dots & |O_{K-1}|^2\omega_{K-1}  \\
    \vdots & \vdots & \vdots & \dots & \vdots \\
     |O_0|^2\omega_0^{n} &  |O_1|^2\omega_1^n &   |O_2|^2\omega_2^{n} & \dots & |O_{K-1}|^2\omega_{K-1}^{n}
    \end{pmatrix}
    \begin{pmatrix} 
    1 & \omega_0  &  \omega_0^2 & \dots & \omega_0^{n} \\
    1 &  \omega_1 &  \omega_1^2 & \dots &  \omega_1^{n} \\
    \vdots & \vdots & \vdots & \dots & \vdots \\
     1 & \omega_{K-1} &  \omega_{K-1}^2 & \dots &  \omega_{K-1}^{n}
    \end{pmatrix}\,.
\end{eqnarray}}
The determinant of a matrix which is a product of two rectangular matrices, can be reduced via the Cauchy-Binet formula (see for example \cite{enwiki:1030584213}) to the form (\ref{Dn}).

\section{Hilbert space dimension of XXZ sectors}\label{Appx_Sectors}

This Appendix presents the details of the computation of the dimensions of Hilbert space sectors of XXZ with fixed magnetization, parity and $R$-charge (whenever the latter is also a conserved quantum number within a given sector).

\subsection{Parity sectors at fixed magnetization}\label{Appx_Sectors_Parity}

We start by considering the sector $\mathcal{H}_M$ of fixed magnetization with $M$ spins up, whose dimension is given by:

\begin{equation}
    \centering
    \label{Dim-M}
    D_M = \binom{N}{M}\,,
\end{equation}
and we shall proceed to split it in sectors of fixed parity, $\mathcal{H}_M=\mathcal{H}_M^+\oplus \mathcal{H}_M^{-}$, whose dimensions we denote by $D_M^+$ and $D_M^{-}$, respectively. In order to study these parity sectors, it is convenient to use the natural tensor product basis of the spins in the XXZ chain. For the sake of language economics, we shall refer to it as the ``computational basis'', identifying ones with spins up and zeroes with spins down. If we examine individually each element $\ket{\psi}$ of this basis, we find the two mutually exclusive possibilities:

\begin{itemize}
    \item[a)] $P\ket{\psi}=\ket{\psi}$, i.e. $\ket{\psi}$ is already invariant under parity, hence it already belongs to $\mathcal{H}_M^+$.
    \item[b)] $\bra{\psi}P \ket{\psi}=0$, i.e. $\ket{\psi}$ is not invariant, so the action of parity gives a different element in the computational basis, and therefore their overlap vanishes due to orthogonality. For the states in this class, it immediately follows that $\ket{\psi}+P\ket{\psi}$ has positive parity while $\ket{\psi}-P \ket{\psi}$ has negative parity.
\end{itemize}

We denote $A,B$ as the number of states in the computational basis that satisfy (a), (b), respectively. $B$ needs to be always even because if $\ket{\psi}$ belongs to (b) then so does $P \ket{\psi}$; the fact that $P^2=\mathds{1}$ groups the states in the class (b) pairwise and would lead to a contradiction if $B$ is odd. We note that the $\mathcal{H}_M^+$ sector receives $A$ states from class (a) and $\frac{B}{2}$ states from class (b), where the factor of $\frac{1}{2}$ comes from the fact that half of the independent linear combinations of pairs constructed in (b) have positive parity and half have negative parity. To summarize:

\begin{equation}
    \centering
    \label{Parity-dims-1}
    \begin{split}
        & D_M^+ = A+\frac{B}{2} \\
        &D_M^{-} = \frac{B}{2} ~.
    \end{split} 
\end{equation}
Since (a) and (b) are mutually exclusive, we have that:

\begin{equation}
    \centering
    \label{a-b-exclusive}
    A+B = D_M\,,
\end{equation}
and therefore:

\begin{equation}
    \centering
    \label{DplusDminus_A}
    \begin{split}
        &D_M^+ = \frac{D_M+A}{2}\\
        &D_M^{-}=\frac{D_M-A}{2}\,.
    \end{split}
\end{equation}
So all that's left is to determine $A$, i.e. the number of states in the computational basis that are invariant under parity. Since parity performs a mirror transformation with respect to the chain center, the general philosophy is to count the number of states in (a) by finding the number of ways to arrange half of the ones in the left half of the chain, as the requirement of invariance under parity will determine the position of the remaining ones in the other half. However, depending on whether $N$ and $M$ are even or odd, some subtleties need to be taken into account.

\begin{itemize}
    \item[1.] \textbf{$N$ odd:} In this case there is always a site at the chain center that is fixed under parity transformations. This resource shall be conveniently exploited.
    \begin{itemize}
        \item[1.1.] \textbf{$M$ odd:} We need to put a one on the chain center, and then find all possible ways to arrange $\frac{M-1}{2}$ ones on the $\frac{N-1}{2}$ positions of the left side of the chain. That is:
        
        \begin{equation}
            \centering
            \label{A-1-1}
            A = \binom{\frac{N-1}{2}}{\frac{M-1}{2}}\,,
        \end{equation}
        
        and using (\ref{DplusDminus_A}) this yields:
        
        \begin{equation}
            \centering
            \label{DplusDminus_1_1}
            D_M^{\pm} = \frac{1}{2}\binom{N}{M}\pm\frac{1}{2}\binom{\frac{N-1}{2}}{\frac{M-1}{2}}\,.            \end{equation}
        \item[1.2.] \textbf{$M$ even:} In this case we can't put a one on the chain center. We thus find all possibilities to arrange $\frac{M}{2}$ ones on $\frac{N-1}{2}$ positions:
        
        \begin{equation}
            \centering
            \label{A-1-2}
            A = \binom{\frac{N-1}{2}}{\frac{M}{2}}\,,
        \end{equation}
        
        which implies:
        
        \begin{equation}
            \centering
            \label{DplusDminus_1_2}
            D_M^{\pm}=\frac{1}{2}\binom{N}{M}\pm \frac{1}{2}\binom{\frac{N-1}{2}}{\frac{M}{2}} ~.
        \end{equation}
    \end{itemize}
    \item[2.] \textbf{$N$ even:} This time there is no fixed point under parity, which constrains the possibility of having states in class (a) at all.
    \begin{itemize}
        \item[2.1.] \textbf{$M$ odd:} In this case there is no way to arrange half of the ones in half of the chain, so simply:
        
        \begin{equation}
            \centering
            \label{A-2-1}
            A=0\,,
        \end{equation}
        and thus:
        \begin{equation}
            \centering
            \label{DplusDminus_2_1}
            D_M^+=D_M^{-}=\frac{1}{2}\binom{N}{M} ~.       \end{equation}
            \item[2.2.] \textbf{$M$ even:} We just need to find all possible ways to arange $\frac{M}{2}$ ones on $\frac{N}{2}$ sites:
            \begin{equation}
                \centering
                \label{A_2_2}
                A = \binom{\frac{N}{2}}{\frac{M}{2}}\,,
            \end{equation}
            so that:
            \begin{equation} \label{DplusDminus_2_2}
                \centering
                D_M^{\pm}=\frac{1}{2}\binom{N}{M}\pm\frac{1}{2}\binom{\frac{N}{2}}{\frac{M}{2}}\,.
                \end{equation}
    \end{itemize}
\end{itemize}
We have therefore succeeded in computing the dimension of all Hilbert space sectors of fixed magnetization and parity. The results are summarized in Table \ref{Table_Dim_ParitySectors}.

\subsection{$R$-subsectors}\label{Appx_Sectors_R}

For XXZ chains with an even number of sites $N$ there exists a zero-magnetization sector with $M=\frac{N}{2}$ in which both $P$ and $R$ are conserved quantum numbers due to the symmetry enhancement discussed in Section \ref{Section_XXZ}. Therefore, we can further split the parity sectors into subsectors of fixed $R=\pm 1$, $\mathcal{H}_{\frac{N}{2}}^P=\mathcal{H}_{\frac{N}{2}}^{P,+}\oplus \mathcal{H}_{\frac{N}{2}}^{P,-}$. Their dimensions $D_{\frac{N}{2}}^{PR}$ can again be computed making use of the computational basis for the starting sector $\mathcal{H}_{\frac{N}{2}}$. Again splitting the elements of this basis into classes (a) and (b) defined earlier in \ref{Appx_Sectors_Parity}, we can make the following observations:

\begin{itemize}
    \item No state $\ket{\psi}$ in the computational basis is invariant under $R$. In particular, no state in (a) is invariant under $R$, even though they are all parity eigenstates with positive eigenvalue.
    \item There is a subset of the computational basis states $\ket{\psi}$ in (b) that satisfy:
    
    \begin{equation}
        \centering
        \label{States_P_R}
        P\ket{\psi} = R\ket{\psi}.
    \end{equation}
    
    For a state fulfilling this property, we can build two orthonormal combinations $\ket{\psi_+}$ and $\ket{\psi_-}$ as:
    
    \begin{equation}
        \centering
        \label{R-combinations}
        \ket{\psi_\pm}=\frac{1}{\sqrt{2}}\big(\ket{\psi}\pm P\ket{\psi}\big)=\frac{1}{\sqrt{2}}\big(\ket{\psi}\pm R\ket{\psi}\big). 
    \end{equation}
    
    And we note that $\ket{\psi_{\pm}}$ verifies simultaneously $P=R=\pm 1$. The states of the form of $\ket{\psi_+}$ introduce a bias towards $R=+1$ in the $P=+1$ sector, while those of the form $\ket{\psi_-}$ introduce a bias towards $R=-1$ in the $P=-1$ sector.
    
\end{itemize}

We denote the number of states fulfilling (\ref{States_P_R}) as $B^\prime<B$. In order to find this number, we need to count the total number of possible configurations of the left half of the spin chain: Once half the chain is determined, the other half is fixed by reflecting and ``flipping''. Note that the requirement of $M=\frac{N}{2}$ is baked in, since if there are $m$ 1's in one half of the chain, there will be $\frac{N}{2}-m$ 1's in the other half. This yields $B^\prime = 2^{\frac{N}{2}}$, and with this we are in conditions to proceed with the counting of the dimension of the $R$-subsectors within each parity sector:
    
    \begin{itemize}
        \item $\mathbf{P=+1.}$ We can enumerate the following contributions:
        
        \begin{itemize}
            \item[1.] $\frac{B^\prime}{2}$ out of the $\frac{B}{2}$ positive-parity states coming from the class (b) are of the form $\ket{\psi_+}$ in (\ref{R-combinations}) and thus have directly $R=+1$.
            \item[2.] The $\frac{B-B^\prime}{2}$ remaining positive-parity states coming from class (b) are not invariant under $R$.
            \item[3.] The $A$ states in class (a), which are all positive-parity states, are not invariant under $R$.
        \end{itemize}
        
        Hence, points $2$ and $3$ above give $A+\frac{B-B^\prime}{2}=D_{\frac{N}{2}}^+-\frac{B^\prime}{2}$ positive-parity states that are not invariant under $R$. Since $R^2=1$, we proceed analogously as we did in (\ref{Appx_Sectors_Parity}) in order to construct parity eigenstates out of states in (b): We group them pairwise in linear combinations that give as many states with $R+1$ as states with $R=-1$. This yields:
        
        \begin{equation}
            \centering
            \label{Sectors_Pplus_R}
            \begin{split}
                &D_{\frac{N}{2}}^{++} = \frac{B^\prime}{2}+\frac{1}{2}\left(D_{\frac{N}{2}}^+-\frac{B^\prime}{2}\right) = \frac{1}{2}D_{\frac{N}{2}}^++2^{\frac{N-4}{2}} \\
                &D_{\frac{N}{2}}^{+-} = \frac{1}{2}\left(D_{\frac{N}{2}}^+-\frac{B^\prime}{2}\right) =\frac{1}{2}D_{\frac{N}{2}}^+-2^{\frac{N-4}{2}}\,.
            \end{split}
        \end{equation}
        
        \item $\mathbf{P=-1.}$ In this case we have the contributions:
        
        \begin{itemize}
            \item[1.] $\frac{B^\prime}{2}$ out of the $\frac{B}{2}$ negative-parity states coming from the class (b) are of the form $\ket{\psi_-}$ in (\ref{R-combinations}) and thus have directly $R=-1$.
            \item[2.] The remaining $\frac{B-B^\prime}{2}=D_{\frac{N}{2}}^--\frac{B^\prime}{2}$ negative-parity states coming from (b) are not $R$-invariant, and thus have to be grouped pairwise in linear combinations giving as many $R=+1$ as $R=-1$ eigenstates.
        \end{itemize}
        
        This yields the expressions:
        
        \begin{equation}
            \centering
            \label{Sectors_Pminus_R}
            \begin{split}
                & D_{\frac{N}{2}}^{-+} = \frac{1}{2}\left(D_{\frac{N}{2}}^--\frac{B^\prime}{2}\right)=\frac{1}{2}D_{\frac{N}{2}}^--2^{\frac{N-4}{2}} \\
                & D_{\frac{N}{2}}^{--}=\frac{B^\prime}{2}+\frac{1}{2}\left(D_{\frac{N}{2}}^--\frac{B^\prime}{2}\right)=\frac{1}{2}D_{\frac{N}{2}}^-+2^{\frac{N-4}{2}}\,. 
            \end{split}
        \end{equation}
        
    \end{itemize}
    Happily, equations (\ref{Sectors_Pplus_R}) and (\ref{Sectors_Pminus_R}) can be merged together into a single expression:
    
    \begin{equation}
        \centering
        \label{Sectors_P_R_appx}
        D_{\frac{N}{2}}^{PR}=\frac{1}{2}D_{\frac{N}{2}}^P+P\cdot R \cdot 2^{\frac{N-4}{2}},\quad\quad\quad\text{for }N \text{ even.}
    \end{equation}

\subsubsection{Selection rules and Krylov dimension for operators charged under $R$}\label{Appx_selection_rules_Krylov}
Because of the $R$-symmetry enhancement, operators charged under $R$ are constrained by selection rules that enforce nullity of a part of their matrix elements in the energy basis, which has the effect of reducing the dimension of their associated Krylov space. The reason why these selection rules exist is that eigenstates of the Hamiltonian are themselves $R$-eigenstates, rather than being degenerate, as proposed in \cite{Doikou:1998jh} and as we have confirmed performing numerical checks. We shall now prove explicitly those selection rules and give the expression for the Krylov dimension that they imply in each case.

\begin{itemize}
    \item \textbf{Negatively charged operator.} We consider an observable $\mathcal{O}$ acting on the sector $\mathcal{H}_{\frac{N}{2}}^P$ that verifies:
    
    \begin{equation}
        \centering
        \label{Op_R_neg_charged}
        R \mathcal{O} R^{\dagger} = -\mathcal{O}\quad\Longleftrightarrow \quad\left[R,\mathcal{O}\right]=-2\mathcal{O}R  ~.
    \end{equation}
    
    Since eigenstates in the sector of zero magnetization and fixed parity $P$ of this chain of even length are also eigenstates of $R$ with eigenvalue $\pm 1$, it follows that the action of $\mathcal{O}$ on these states has the effect of changing the sign of the $R$-eigenvalue. To show this, we shall label eigenstates in the sector explicitly with their energy and $R$-eigenvalues, $\ket{E,r}$ with $r=\pm 1$. Using (\ref{Op_R_neg_charged}) it is possible to show that:

\begin{equation}
    \centering
    \label{R-flip}
    R \mathcal{O}\ket{E,r}=-r\,\mathcal{O}\ket{E,r} ~.
\end{equation}
That is: While $\ket{E,r}$ is an eigenstate of $R$ with eigenvalue $r$, $\mathcal{O}\ket{E,r}$ is an eigenstate of $R$ with eigenvalue $-r$. This in turn implies the selection rule:

\begin{equation}
    \centering
    \label{Selection-rule}
    \langle E^\prime,r^\prime | \mathcal{O}|E,r\rangle=0 \quad\text{if}\quad r^\prime = r ~.
\end{equation}
But (\ref{Selection-rule}) are the matrix elements of the observable in the energy basis, so they are direcly related to the Krylov dimension, and the fact that many of them vanish identically because of the selection rule has the effect of lowering the Krylov dimension with respect to its upper bound $D^2-D+1$ (where we use $D$ instead of $D_{\frac{N}{2}}^P$ for the sake of notational simplicity). To give the upper bound for the Krylov dimension in this case, let's assume that all matrix elements that are not forced to vanish by the selection rule are non-zero. Then, the number of non-zero matrix elements $\langle E^\prime,r^\prime | \mathcal{O}|E,r\rangle$ (and hence the Krylov space dimension, as there are no degeneracies in the spectrum of the Hamiltonian in this sector) is given by:

\begin{equation}
    \centering
    \label{K-selection}
    K = 2R^+R^-\,,
\end{equation}
where, again for notational simplicity, $R^\pm \equiv D_{\frac{N}{2}}^{P,R=\pm}$ denote schematically the number of eigenstates with $R$-eigenvalues $r=\pm1$ in the sector $\mathcal{H}_{\frac{N}{2}}^P$.

\item $\mathbf{R}$\textbf{-invariant operator.} In this case, we consider an observable $\mathcal{O}$ acting on the sector $\mathcal{H}_{\frac{N}{2}}^P$ that verifies:

    \begin{equation}
        \centering
        \label{Op_R_pos_charged}
        R \mathcal{O} R^{\dagger} = \mathcal{O}\quad\Longleftrightarrow \quad\left[R,\mathcal{O}\right]=0 ~.
    \end{equation}
    
    The selection rule in this case is: 
    
    \begin{equation}
    \centering
    \label{TwoSiteCenter_Selection}
    \langle E^\prime,r^\prime|\mathcal{O}|E,r\rangle = 0 \quad \text{if}\quad r^\prime = -r ~.
\end{equation}
Again, assuming that all matrix elements unconstrained by (\ref{TwoSiteCenter_Selection}) are non-zero, and noting that some of them are diagonal elements, we reach the prediction:

\begin{eqnarray}
    \text{Number of non zero matrix elements:} \quad (R^+)^2+(R^-)^2 \\
    K = R^+(R^+-1)+R^-(R^--1)+1=(R^+)^2+(R^-)^2-D+1\,,
    \label{K_R_inv}
\end{eqnarray}
where again $D\equiv D_{\frac{N}{2}}^P$ and $R^\pm \equiv D_{\frac{N}{2}}^{P,R=\pm}$ for simplicity. In fact, since $\mathcal{O}$ is invariant under $R$, we could have just considered separately its restrictions on $\mathcal{H}_{\frac{N}{2}}^{P,+}$ and $\mathcal{H}_{\frac{N}{2}}^{P,-}$. It is apparent from this perspective that (\ref{K_R_inv}) is nothing but the combination of the upper bounds of the form (\ref{K_bound}) applied separately to each $R$-sector, where the operator is dense in the energy basis.
\end{itemize}

\begin{figure}[ht]
     \centering
     \begin{subfigure}[b]{0.4\textwidth}
         \centering
         \includegraphics[width=\textwidth]{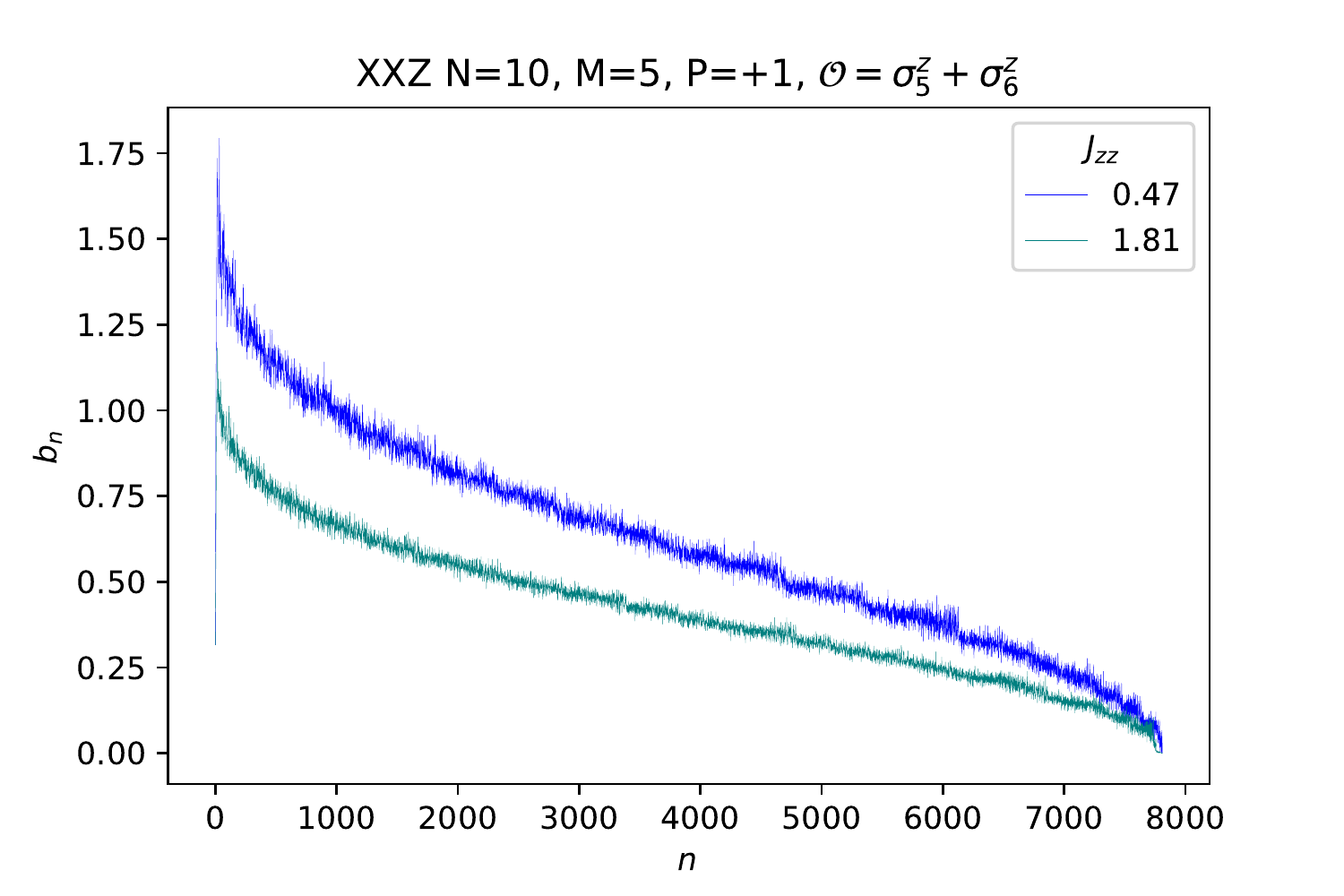}
         \caption{Lanczos sequences}
         \label{fig:Lanczos_N10}
     \end{subfigure}
     \begin{subfigure}[b]{0.25\textwidth}
         \centering
         \includegraphics[width=\textwidth]{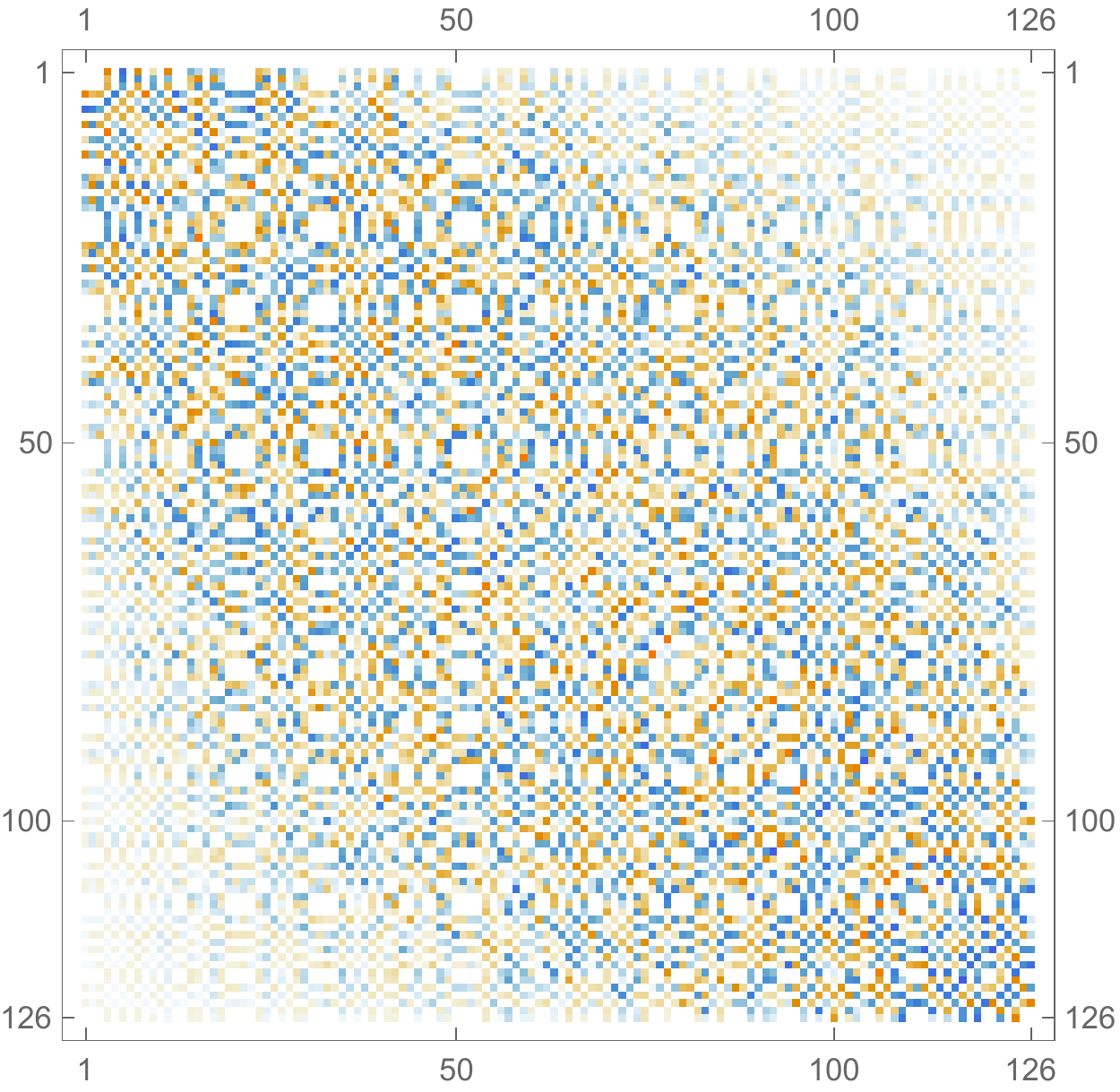}
         \caption{$J_{zz}=0.47$}
         \label{fig:Jzz047}
     \end{subfigure}
     \begin{subfigure}[b]{0.25\textwidth}
         \centering
         \includegraphics[width=\textwidth]{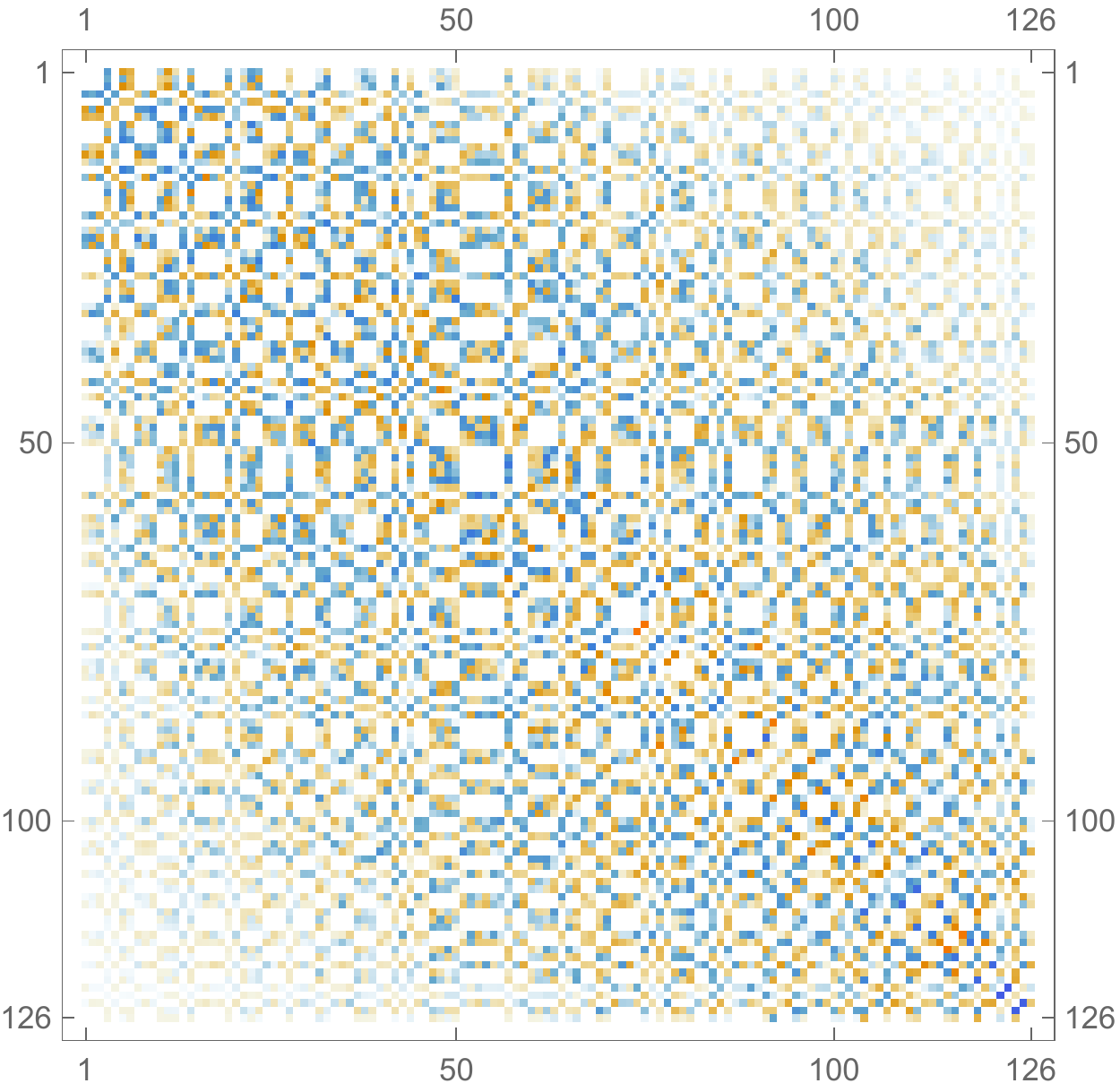}
         \caption{$J_{zz}=1.81$}
         \label{fig:Jzz181}
     \end{subfigure}
        \caption{For XXZ with $N=10$ in the sector $M=5,\, P=+1$ for which the Hilbert space dimension is $D=126$ according to Table \ref{Table_Dim_ParitySectors}, the operator $\mathcal{O}=\sigma_5^z+\sigma_6^z$ is not dense in the energy basis.  The left panel shows the Lanczos sequences which terminate at $K=7810$, and the middle and right panels show matrix plots of the operator in the energy basis for the two $J_{zz}$ couplings for which the Lanczos sequences were computed.}
        \label{fig:XXZ_N10_M5}
\end{figure}

Consider the operator we study, $\mathcal{O}=\sigma_m^z+\sigma_{N-m+1}^z$  in an XXZ system with an even number of spins $N$ in the zero magnetization sector $M=N/2$ and $P=+1$. As discussed in Section \ref{Section_XXZ}, in this sector both $P$ and $R$ are symmetries of the Hamiltonian. It can be shown, using the property $\sigma^j\sigma^k = \delta^{jk}+i\varepsilon^{jkl}\sigma^l$ of the Pauli matrices at a fixed site, that the operator $\mathcal{O}$ chosen here satisfies (\ref{Op_R_neg_charged}), thus being negatively charged under $R$. We therefore expect it to span (in the absence of other degeneracies in the spectrum) a Krylov space of dimension (\ref{K-selection}). We confirm this expectation for the case of XXZ with $N=10, M=5$ in the $P=+1$ sector with the operator $\mathcal{O}=\sigma^z_5+\sigma^z_6$, where we find the Krylov space dimension to be $K=7810$, as indeed predicted by (\ref{K-selection}), once we work with high enough accuracy.  In Figure \ref{fig:XXZ_N10_M5} we show the Lanczos sequences as well as matrix plots of the operator in the energy basis, for two different $J_{zz}$ coefficients.

\section{Analytical results for a constant $b$-sequence}\label{appx:Constant_b_analytics}
To contrast with the disordered systems we study in the main text, we discuss here a flat, smooth $b$-sequence with no disorder. Such systems were considered in the context of Anderson localization, for example in \cite{luck:cea-01485001}.  

Consider the eigenvalue problem (\ref{EV_problem}) with $b_n=1$ (in some units) for all $n$, i.e.
\begin{equation}
    \omega \psi_n = \psi_{n+1}+\psi_{n-1}, \quad n=0,1,\dots, K-1\,,
\end{equation}
where we want to find $\psi_n$ and $\omega$ which solve this equation. 
Assuming boundary conditions $\psi_{-1}=\psi_{K}=0$, it can be solved via the transfer matrix recurrence equation:
\begin{equation}
    \begin{pmatrix}
        \psi_{n+1}\\ \psi_{n} 
    \end{pmatrix} = 
    \begin{pmatrix}
        \omega & -1 \\ 1 & 0
    \end{pmatrix}
    \begin{pmatrix}
        \psi_n\\ \psi_{n-1}
    \end{pmatrix}~.
\end{equation}
The solution in terms of the transfer matrix and boundary conditions is
\begin{equation}
    \begin{pmatrix}
        \psi_{n}\\ \psi_{n-1} 
    \end{pmatrix} = 
    \begin{pmatrix}
        \omega & -1 \\ 1 & 0
    \end{pmatrix}^n
    \begin{pmatrix}
        \psi_0\\ 0
    \end{pmatrix}~.
\end{equation}
Finding the eigenvalues and eigenvectors of the transfer matrix, it can be written as
\begin{equation}
    \begin{pmatrix}
        \omega & -1 \\ 1 & 0
    \end{pmatrix} = \frac{1}{\lambda_1-\lambda_2}\begin{pmatrix}
        \lambda_1 & \lambda_2 \\ 1 & 1
    \end{pmatrix} \begin{pmatrix}
        \lambda_1 & 0 \\ 0 & \lambda_2
    \end{pmatrix}\begin{pmatrix}
        1 & -\lambda_2 \\ -1 & \lambda_1
    \end{pmatrix}\,,
\end{equation}
where $\lambda_{1,2}=\frac{1}{2}\left(\omega\pm \sqrt{\omega^2-4} \right)$.  This allows us to write the solution for any $n$ in terms of $\psi_0$ and $\lambda_{1,2}$, as follows:
\begin{equation} \label{eigvec_eq}
    \begin{pmatrix}
        \psi_{n}\\ \psi_{n-1} 
    \end{pmatrix} = 
    \frac{1}{\lambda_1-\lambda_2}\begin{pmatrix}
        \lambda_1 & \lambda_2 \\ 1 & 1
    \end{pmatrix} \begin{pmatrix}
        \lambda_1^n & 0 \\ 0 & \lambda_2^n
    \end{pmatrix}\begin{pmatrix}
        1 & -\lambda_2 \\ -1 & \lambda_1
    \end{pmatrix}\begin{pmatrix}
        \psi_0\\ 0
    \end{pmatrix} = \frac{\psi_0}{\lambda_1-\lambda_2} \begin{pmatrix}
        \lambda_1^{n+1}-\lambda_2^{n+1}\\ \lambda_1^n -\lambda_2^n
    \end{pmatrix}~.
\end{equation}
Now we can use the other boundary condition, namely $\psi_{K}=0$:
\begin{equation}
    \begin{pmatrix}
        0\\ \psi_{K-1} 
    \end{pmatrix} = \frac{\psi_0}{\lambda_1-\lambda_2} \begin{pmatrix}
        \lambda_1^{K+1}-\lambda_2^{K+1}\\ \lambda_1^K -\lambda_2^K
    \end{pmatrix}\,,
\end{equation}
from which we extract the condition $\lambda_1^{K+1}-\lambda_2^{K+1}=0$, which in terms of $\omega$ reads
\begin{equation}\label{E_condition}
    \left(\omega+\sqrt{\omega^2-4}\right)^{K+1} - \left(\omega-\sqrt{\omega^2-4}\right)^{K+1} = 0 ~.
\end{equation}
A family of solutions is
\begin{equation}\label{E_no_disorder}
    \omega_i = 2 \cos\left( \frac{\pi (i+1)}{K+1}\right), \quad i=0,1,2,\dots, K-1 ~.
\end{equation}
It is also clear now from (\ref{eigvec_eq}), that the eigenvector element $\psi_{ni}$ for eigenvalue $\omega_i$ is:
\begin{equation}
    \centering
    \label{eigenvector-disorder-free}
    \psi_{ni} = \psi_{0i}\frac{\sin \left(\frac{(n+1)(i+1)}{K+1}\pi\right)}{\sin \left(\frac{i+1}{K+1}\pi\right)}\equiv C_i \sin \left(\frac{(n+1)(i+1)}{K+1}\pi\right),\;\;n,i=0,1,2,...,K-1\,,
\end{equation}
where we have absorbed the denominator in the normalization constant because it doesn't depend on $n$ (recall that the label $i$ designates the eigenstate). $C_i$ can now be fixed from normalization, and the normalized eigenstates are
\begin{eqnarray}
   \begin{pmatrix}
       \psi_{0i}\\
       \psi_{1i}\\
       \vdots \\
       \psi_{ni}\\
       \vdots \\
       \psi_{K-1,i}
   \end{pmatrix} =
   \sqrt{\frac{2}{K+1}} \begin{pmatrix}
        \sin \frac{ (i+1) \pi}{K+1} \\
        \sin \frac{2 (i+1) \pi}{K+1} \\
        \vdots \\
        \sin \frac{(n+1) (i+1) \pi}{K+1} \\
        \vdots \\
        \sin \frac{K (i+1) \pi}{K+1}
   \end{pmatrix}, \quad i = 0,1,2, \dots, K-1~.
\end{eqnarray}
In terms of Krylov elements, we found that
\begin{eqnarray}
    ( \mathcal{O}_n |\omega_i) = \psi_{ni}= \sqrt{\frac{2}{K+1}} \sin \frac{(n+1) (i+1) \pi}{K+1} \quad  i,n=0,1,\dots, K-1\,, 
\end{eqnarray}
from which we can compute K-complexity for the pure no-disorder system:
\begin{eqnarray}
    \overline{C_K} = \sum_{i=0}^{K-1} |( \mathcal{O}_0 |\omega_i)|^2   \sum_{n=0}^{K-1} n\,  |(\mathcal{O}_n |\omega_i)|^2 = \sum_{n=0}^{K-1} n \sum_{i=0}^{K-1} \psi_{0i}^2 \psi_{ni}^2 ~.
\end{eqnarray}
The transition probability (\ref{Transition_Probability}) is\footnote{To arrive at this result we used $
    \sum_{n=0}^{N}e^\frac{2i\pi m n}{N+1} = \frac{1-e^{2i\pi m}}{1-e^{\frac{2i\pi m }{N+1}}} = \begin{cases}
        0, & (N+1) \not |m\\
        N+1, & (N+1) | m 
    \end{cases}$. See \cite{luck:cea-01485001}.}
\begin{eqnarray}
    Q_{0n} &=&
    % \sum_{i=0}^{K-1} \psi_{0i}^2 \psi_{ni}^2 =
    \left(\frac{2}{K+1}\right)^2 \sum_{i=0}^{K-1} \sin^2\frac{ (i+1)\pi}{K+1} \sin^2 \frac{(n+1) (i+1) \pi}{K+1} \nonumber\\
    &=& \frac{1}{K+1}\left(1+\frac{1}{2} \delta_{n0}+\frac{1}{2}\delta_{n,K-1} \right)
\end{eqnarray}
and the time-averaged K-complexity is given by,
\begin{eqnarray}
    \overline{C_K} &=&\frac{1}{K+1} \sum_{n=0}^{K-1} n \left(1+\frac{1}{2} \delta_{n0}+\frac{1}{2}\delta_{n,K-1} \right)\nonumber\\ &=&\frac{K(K-1)}{2(K+1)} +0+  \frac{K-1}{2(K+1)}   = \frac{K}{2}\times O(1) +O(1)\sim \frac{K}{2} ~.
\end{eqnarray}
This analytical result shows that for a flat and smooth Lanczos-sequence K-complexity will saturate at $K/2$.

\bibliographystyle{JHEP}
\bibliography{references}
\end{document}